\documentclass[twocolumn,rmp,aps,nofootinbib]{revtex4}

\usepackage{amsmath}
\usepackage{amssymb}
\usepackage{graphicx}

\newcommand{\beq}{\begin{equation}}
\newcommand{\eeq}{\end{equation}}
\newcommand{\beqa}{\begin{eqnarray}}
\newcommand{\eeqa}{\end{eqnarray}}

\def\A{\mathbb A}
\def\C{\mathbb C}
\def\G{\mathbb G}
\def\H{\mathbb H}
\def\I{\mathbb I}
\def\P{\mathbb P}
\def\Q{\mathbb Q}
\def\S{\mathbb S}
\def\T{\mathbb T}
\def\V{\mathbb V}

\def\k{\kappa}
\def\kb{k_B}

\def\px{p_{\ss \bfX}}
\def\qx{q_{\ss \bfX}}
\def\pz{p_{\ss \bfZ}}

\def\cov{\text{cov}}
\def\E{\text{E}}

\def\cz{c_{\ss \bfZ}}
\def\muz{\mu_{\ss \bfZ}}
\def\pxi{p_{\ss \bf \Xi}}

\def\bfmux{\bfmu_{\ss \bfX}}
\def\bfmuz{\bfmu_{\ss \bfZ}}
\def\hatcz{\widehat{c}_{\ss \bfZ}}
\def\hatmux{\widehat{\mu}_{\ss \bfX}}
\def\hatmuz{\widehat{\mu}_{\ss \bfZ}}

\def\bfe{\pmb e}
\def\bfg{\pmb g}
\def\bfl{\pmb l}
\def\bfp{\pmb p}
\def\bfq{\pmb q}
\def\bfr{\pmb r}
\def\bfs{\pmb s}
\def\bfu{\pmb u}
\def\bfx{\pmb x}
\def\bfy{\pmb y}
\def\bfz{\pmb z}

\def\bfW{\pmb W}
\def\bfX{\pmb X}
\def\bfZ{\pmb Z}

\def\bfchi{\pmb \chi}
\def\bfmu{\pmb \mu}
\def\bfxi{\pmb \xi}
\def\bfzeta{\pmb \zeta}

\def\bfPhi{\pmb \Phi}
\def\bfXi{\pmb \Xi}

\def\sA{\mathcal A}
\def\sB{\mathcal B}
\def\sC{\mathcal C}
\def\sD{\mathcal D}
\def\sE{\mathcal E}
\def\sG{\mathcal G}
\def\sI{\mathcal I}
\def\sK{\mathcal K}
\def\sM{\mathcal M}
\def\sN{\mathcal N}
\def\sO{\mathcal O}
\def\sP{\mathcal P}

\def\sW{\mathcal W}
\def\sX{\mathcal X}
\def\sZ{\mathcal Z}

\def\ith{i\text{-th}}
\def\jth{j\text{-th}}
\def\kth{k\text{-th}}
\def\lth{l\text{-th}}
\def\mth{m\text{-th}}
\def\nth{n\text{-th}}

\def\ss{\scriptscriptstyle}
\def\nn{\nonumber}

\begin{document}
\title{Markovian Dynamics on Complex Reaction Networks}

\author{John Goutsias}
\email{goutsias@jhu.edu}
\affiliation{Whitaker Biomedical Engineering Institute,
The Johns Hopkins University,
Baltimore, MD 21218}
\author{Garrett Jenkinson}
\affiliation{Whitaker Biomedical Engineering Institute,
The Johns Hopkins University,
Baltimore, MD 21218}

%
%

\begin{abstract}
Complex networks, comprised of individual elements that interact
with each other through reaction channels, are ubiquitous across
many scientific and engineering disciplines. Examples include
biochemical, pharmacokinetic, epidemiological, ecological, social,
neural, and multi-agent networks. A common approach to modeling
such networks is by a master equation that governs the dynamic
evolution of the joint probability mass function of the underling
population process and naturally leads to Markovian dynamics for
such process. Due however to the nonlinear nature of most reactions,
the computation and analysis of the resulting stochastic population
dynamics is a difficult task. This review article provides a coherent
and comprehensive coverage of recently developed approaches and
methods to tackle this problem. After reviewing a general
framework for modeling Markovian reaction networks and giving
specific examples, the authors present numerical and computational
techniques capable of evaluating or approximating the solution
of the master equation, discuss a recently developed approach
for studying the stationary behavior of Markovian reaction
networks using a potential energy landscape perspective, and
provide an introduction to the emerging theory of thermodynamic
analysis of such networks. Three representative problems of
opinion formation, transcription regulation, and neural network
dynamics are used as illustrative examples.
\end{abstract}

\date{\today}
\maketitle
\tableofcontents

%
%

\section{Introduction}

\vspace{-6pt}

Complex interaction networks are at the core of many problems of
scientific and engineering interest, and this realization has
caused the interdisciplinary study of networks to burgeon over
the past decade. Example applications include (but are not limited
to): chemical reaction networks~\cite{HeinSchu:96,Newm:03,Newm:10},
cellular (signaling, transcriptional and metabolic)
networks~\cite{Newm:03,BaraOltv:04,Newm:10}, pharmacokinetic
networks used to study the absorbtion, distribution, metabolism,
and elimination of chemicals and drugs by the human
body~\cite{BoisZeisToze:90}, epidemiological (disease-spreading)
networks~\cite{Heth:00,Newm:10}, ecological networks~\cite{Newm:03,%
Basc:09,PoweBola:09,Basc:10,Newm:10,ThebFont:10,BlacMcKa:12},
social networks~\cite{Newm:03,Free:04,Weid:06,BorgMehrBras-etal:09,%
HillRandNowa-etal:10,MasuGibeRedn:10,Newm:10}, neural
networks~\cite{Newm:03,BenaCowaDron-etal:10,Newm:10}, multi-agent
networks comprised of intelligent agents that observe and act
upon each other to achieve a certain objective~\cite{XiTanBara:06},
and evolutionary game theory networks~\cite{SzabFath:07}.

A common approach to modeling the dynamic behavior of complex
interaction networks is by a master equation that governs the
time evolution of the joint probability mass function of the
underling population processes and naturally leads to Markovian
dynamics. Due however to the nonlinear nature of most interaction
networks, computing the exact solution of the master equation is
not possible in general. As a consequence, the analysis of nonlinear
Markovian interaction networks is a formidable task. Deterministic
approximations of the master equation have been developed to address
this problem, but these approximations may \textit{fail} to predict
important system behavior~\cite{McQuJachRuss:64,ThakRescLisi:78,%
LeonReic:90,ZhenRoss:91,RaoArki:03,Gout:07,UribVerg:07,VellQian:07,%
BuicCowaChow:10}. For example, deterministic approximations cannot
predict the emergence of noise-induced behavior, a fundamental
property of nonlinear interaction networks with stochastic
dynamics~\cite{ArtyDasKard-etal:07,ArtyMathSamo-etal:09,%
QianShiXing:09,BishQian:10,ZhanGeQian:10,Qian:10,Qian:11}.

The earliest Markovian interaction network model proposed in the
literature seems to be that of~\textcite{Delb:40} who developed
it to study statistical fluctuations in an autocatalytic reaction
mechanism of chemical kinetics. This approach was subsequently
adopted by several investigators who focused on models of simple
reaction mechanisms in small systems that exhibit large fluctuations
and developed methods for their analysis~\cite{Sing:53,Bart:58,%
Bart:59,Ishi:60,Bart:62,McQu:63,Ishi:64,McQuJachRuss:64,DarvStaf:66,%
McQu:67,Kurt:72,NicoPrig:77,Hake:75,Schn:76}. Parallel to these
developments, the pioneering work of N.\ G.\ van Kampen and
D.\ T.\ Gillespie provided fundamental analytical and computational
methods for dealing with stochasticity in nonlinear chemical
reaction networks through approximations of the master equation
or Monte Carlo sampling ~\cite{Kamp:61,Gill:76,Kamp:76,Gill:77,%
Gill:92,Gill:96,Gill:00,Gill:01,Kamp:07}. These methods however
were largely overlooked by the chemical modeling community
which, for many decades, concentrated its main effort on
developing system-based and control-theoretic methods for the
analysis of chemical reaction networks using deterministic rate
equations~\cite{HeinSchu:96}. It turns out that the deterministic
approach is theoretically and computationally much easier to
handle than the stochastic approach. Successful application
to numerous chemical modeling and analysis problems is one of
the main reasons why deterministic approaches have garnered
wide-spread popularity.

Strong experimental evidence has recently revealed that
stochasticity plays a fundamental role in cell
regulation~\cite{RossBrowHume:94,McAdArki:97,HastPradDoln-etal:00,%
KeplElst:01,ThatOude:01,ElowLeviSigg-etal:02,BlakKaerCant-etal:03,%
MunsNeueOude:12}. This evidence has catalyzed a new effort
on modeling biochemical reaction networks using stochastic
(mainly Markovian) approaches, resulting in the development
of novel mathematical, computational and experimental tools
for quantitatively understanding the dynamic interplay between
stochastic fluctuations and system function. In addition to
refining previously
suggested algorithms and developing new numerical and
computational techniques for estimating or approximating
the solution of the master equation, two important and related
methodologies are emerging as fundamental to the analysis of
nonlinear biochemical reaction networks. The first is based
on a potential energy landscape perspective~\cite{Ao:04,%
AoKwonQian:07,HanWang:07,KimWang:07,LapiHanWang:08,%
WangXuWang:08,WangLiEang:10,WangXuWang-etal:10,%
WangZhanWang:10,ZhouQian:11,WangZhanXu-etal:11} and leads to
a powerful approach for conceptualizing and quantifying
emergent complex behavior in nonlinear biochemical reaction
networks with stochastic dynamics. The second methodology is
based on non-equilibrium stochastic thermodynamics~\cite{Schn:76,%
Schl:80,MouLuoNico:86,LuoZhaoHu:02,AndrGasp:04,JianQianQian:04,%
Qian:06,AndrGasp:07,SchmSeif:07,HanWang:08,Ross:08,Seif:08,Ge:09,%
Qian:09,VellQian:09,BroeEspo:10,Demi:10,EspoBroe:10b,GeQian:10,%
PuglPigoRond-etal:10,Qian:10,RossVill:10,RaoXiaoHou:11,SantQian:11,%
GeQianQian:12b,ZhanQianQian:12a} and can be effectively used
to study the macroscopic behavior of Markovian biochemical
reaction networks and, in particular, properties related to
their self-organization, functional stability, robustness and
evolutionary behavior~\cite{Hake:75,Prig:78,HanWang:08}.

In parallel to the previous developments, substantial effort has
been independently focused on modeling and analyzing stochastic
behavior in problems of epidemiology~\cite{Bart:49,Bail:50,%
Hask:54,Bail:57,Bart:57,Bart:60,Bail:63,HillSeve:69,Kamp:73,%
Kamp:76,ChenBokk:05,KeelRoss:08,KeelRoss:09,BlacMcKa:10,%
YousScog:11,JenkGout:12}, ecology~\cite{Bart:60,DilaDomi:00,%
DattDeliLaw:10,LiWangWang:11,BlacMcKa:12}, sociology~\cite{Weid:72,%
Hake:75,WeidHaag:83,Weid:91,Weid:06}, and theoretical
neuroscience~\cite{Hake:75,Cowa:91,OhirCowa:93,BuicCowa:07,%
SoulChow:07,BousDest:09,Bres:09,BenaCowaDron-etal:10,Bres:10,%
BuicCowaChow:10}. The main premise underlying this effort is
the realization that environmental, demographic, behavioral,
and biological factors fluctuate randomly and that the
resulting stochasticity can cause dramatic deviation from
what is predicted by deterministic approaches.

A common theme of most works cited above is the representation
of stochasticity by a master equation that naturally leads to
Markovian dynamics. This provides a direct mathematical and
computational link with the techniques developed in stochastic
chemical kinetics. As a matter of fact, there is a growing
consensus among network researchers in diverse scientific
disciplines that most mathematical, numerical, and computational
tools developed for solving problems in stochastic chemical
kinetics can also be used to solve problems within seemingly
disparate fields of scientific inquiry. It turns out that
Markovian reaction networks provide a unified mathematical
framework for studying stochastic dynamics on networks in a
variety of scientific and engineering applications.

Our main goal in this article is to provide a comprehensive
and coherent coverage of recently developed approaches and
methods to model complex nonlinear Markovian reaction networks
and analyze their dynamic behavior. To achieve this, we first
review in Section~II a general framework for modeling Markovian
reaction networks and subsequently discuss specific examples
of this framework in Section~III. In Section~IV, we provide
a comprehensive review of the main numerical and computational
techniques available for estimating or approximating the
solution of the master equation. Moreover, in Section~V,
we focus on multiscale methods for approximately computing
the solution of stiff master equations. In addition, we
review in Section~VI several mathematical facts pertaining to
the mesoscopic (probabilistic) behavior of the master equation.
These facts are well-known from the theory of Markov processes,
but we recast them here in the more  specific form dictated
by the framework of Markovian reaction networks. In Section~VII,
we discuss a recently developed approach for studying the
stationary behavior of Markovian reaction networks using a
potential energy landscape perspective, whereas, in Section~VIII
we present an introduction to the emerging theory of thermodynamic
analysis of Markovian reaction networks. Finally, we provide
in Section~IX a general outlook of what we believe lies ahead
in this very fundamental and exciting area of research and
summarize our conclusions in Section~X. To illustrate key
concepts, we employ three representative examples dealing
with opinion formation in social networks, transcriptional
control in cell regulation, and avalanche formation in neural
networks. The MATLAB software used to implement these examples
is available on line and can be freely downloaded from
www.cis.jhu.edu/$\sim$goutsias/CSS\%20lab/software.html.

\newpage

With such a rich and diverse subject matter, the authors
regret that realistic limitations forbid an exhaustive
treatise on the history and present state of the field. The
references provided in this review can serve as a starting
point to more in depth or diverse coverage. We sincerely
apologize to the authors whose works do not receive recognition,
but hope that the listed citations can provide a ``path of
least resistance'' to early-stage investigators who may
feel lost in the vast sea of publications available in the
area of complex interaction networks.

%
%

\vspace{-6pt}

\section{Reaction networks}

\vspace{-6pt}

\subsection{Chemical systems and reaction networks}

\vspace{-6pt}

Networks of chemical reactions are used extensively to model
biochemical activity in cells. It turns out that many physical
and man-made systems of interest to science and engineering
can be viewed as special cases of chemical reaction networks
when it comes to mathematical and computational analysis. For
this reason, chemical reaction networks can serve as archetypal
systems when studying dynamics on complex networks.

A chemical reaction system is comprised of a (usually) large
number of molecular species and chemical reactions. A~group of
molecular species, known as \textit{reactants}, interact through
a chemical reaction to create a new set of molecular species,
known as \textit{products}. In general, we can think of a set
of chemical reactions as a system that consists of $N$ molecular
species $X_1,X_2,\ldots,X_{\! N}$ that interact through $M$
coupled reactions of the form:
\beq
\sum_{n \in \sN} \nu_{nm} X_n \rightarrow \sum_{n \in \sN}
\nu'_{nm} X_n, ~~ m \in \sM ,
\label{eq-II-1}
\eeq
where $\sN :=\{1,2,\ldots,N\}$ and $\sM :=\{1,2,\ldots,M\}$. The
quantities $\nu_{nm} \geq 0$ and $\nu'_{nm} \geq 0$ are known as
the \textit{stoichiometric coefficients} of the reactants and
products, respectively. These coefficients tell us how many
molecules of the $\nth$ species are consumed or produced by the
$\mth$ reaction. In particular, the notation used in (\ref{eq-II-1})
implies that occurrence of the $\mth$ reaction changes the
molecular count of species $X_n$ by $s_{nm} := \nu'_{nm} - \nu_{nm}$,
where~$s_{nm}$ is known as the \emph{net} stoichiometric coefficient.

The inter-connectivity between components in a chemical reaction
system can be graphically represented as a network~\cite{KlamHausThei:09,%
Newm:10} and, more specifically, by means of a directed, weighted,
bipartite graph. Since molecular species react with each other to
produce other molecular species, we can refer to this network in
more general terms as a \textit{reaction network}.

To illustrate how we can map a chemical reaction system to a
network, let us consider the following reactions that correspond
to a quadratic autocatalator with positive feedback~\cite{Gout:07}:

\newpage

$~~~~~~~~~~~~~~~~~~~~~~~$

\begin{figure}
\includegraphics[width=3.35in]{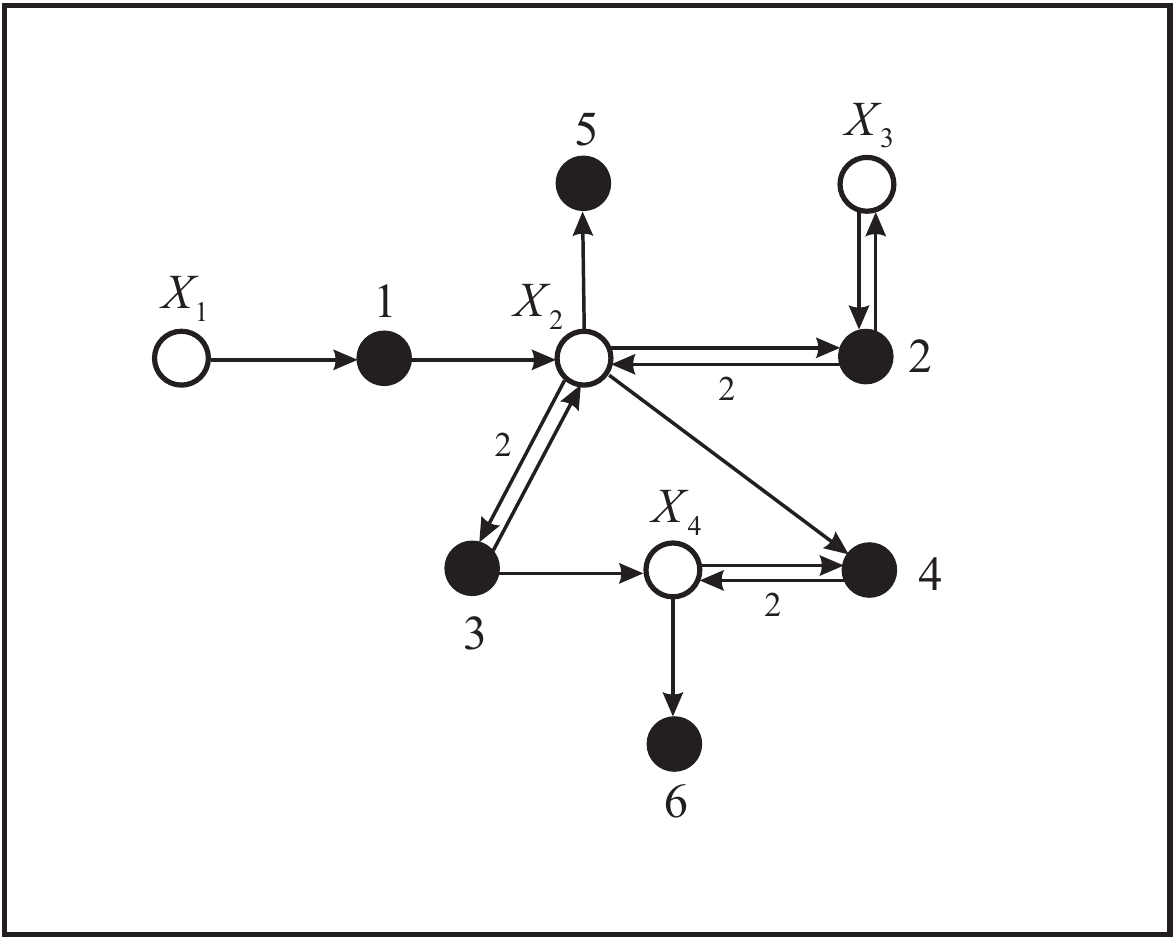}
\caption{A directed, weighted, bipartite graphical representation
of the chemical reaction system given by~(\ref{eq-II-2}). The
molecular species are represented by the white nodes, whereas, the
reactions are represented by the black nodes. Edges emanating from
white nodes and incident to black nodes correspond to the reactants
associated with a particular reaction, whereas, edges emanating from
black nodes and incident to white nodes correspond to the products
of that reaction.}
\label{fig-1}
\vspace{-9pt}
\end{figure}

\vspace{-30pt}

\beq
\begin{array}{rcl}
\text{S}            & \rightarrow   & \text{P}              \\ [2pt]
\text{D} + \text{P} & \rightarrow   & \text{D} + 2\text{P}  \\ [2pt]
2\text{P}           & \rightarrow   & \text{P} + \text{Q}   \\ [2pt]
\text{P} + \text{Q} & \rightarrow   & 2\text{Q}             \\ [2pt]
\text{P}            & \rightarrow   & \emptyset             \\ [2pt]
\text{Q}            & \rightarrow   & \emptyset,
\end{array}
\label{eq-II-2}
\eeq
where the last two reactions indicate the degradation of
molecules~$\text{P}$ and~$\text{Q}$. This chemical reaction
system is comprised of $N=4$ molecular species that interact
through the $M=6$ reactions given by~(\ref{eq-II-2}). We
can (arbitrarily) label the molecular species as $X_1=\text{S}$,
$X_2=\text{P}$, $X_3=\text{D}$, $X_4=\text{Q}$, and the reactions
as $1,2,\ldots,6$. We can now represent the system by the network
of interactions depicted in Fig.~\ref{fig-1}. This network
consists of two types of nodes: those representing the molecular
species (white circles) and those representing the reactions
(black circles). The directed edges represent interactions
between molecular species and reactions and, naturally,
connect only white nodes with black nodes. Edges emanating
from white nodes and incident to
black nodes correspond to the reactants associated with a
particular reaction, whereas, edges emanating from black
nodes and incident to white nodes correspond to the products
of that reaction. Edges are labeled by their weights, which
correspond to the stoichiometric coefficients associated with
the molecular species represented by the white nodes and the
reactions represented by the corresponding black nodes. For
simplicity, an edge is not labeled when the value of the
associated stoichiometric coefficient is one.

An alternative representation of a reaction network is by means of the
two $N \times M$ \textit{stoichiometric matrices} $\V$ and~$\V'$ with
elements $\nu_{nm}$ and $\nu'_{nm}$, respectively. These matrices play
a similar role as the adjacency matrix of a simple graph~\cite{Newm:10}.
For the reaction network depicted in Fig.~\ref{fig-1},
we have that
\beq
\V = \left [
\begin{array}{cccccc}
1 &  0 & 0 & 0 & 0 & 0 \\
0 &  1 & 2 & 1 & 1 & 0 \\
0 &  1 & 0 & 0 & 0 & 0 \\
0 &  0 & 0 & 1 & 0 & 1
\end{array}
\right ]  ~~ \mbox{and} ~~
\V' = \left [
\begin{array}{cccccc}
0 &  0 & 0 & 0 & 0 & 0 \\
1 &  2 & 1 & 0 & 0 & 0 \\
0 &  1 & 0 & 0 & 0 & 0 \\
0 &  0 & 1 & 2 & 0 & 0
\end{array}
\right ]. \nn
\label{eq-II-3}
\eeq
It is not difficult to see that, given the two stoichiometric matrices
$\V$ and $\V'$, we can uniquely construct the chemical reaction system
given by~(\ref{eq-II-2}) and, therefore, the network depicted in
Fig.~\ref{fig-1}. Hence, knowledge of the two stoichiometric matrices
completely specifies the network topology. Note that a quick glance
of these matrices may allow us to make some interesting observations
about the chemical reaction system under consideration. For example,
the fact that all but one of the elements of the first row of matrix
$\V$ are zero indicates that the molecular species $X_1$ is a reactant
only in one reaction, whereas, the fact that the first row of matrix
$\V'$ is zero indicates that this species is not produced by any
reaction. Moreover, the last two zero columns of matrix $\V'$ indicate
that reactions 5 and 6 do not result in any products (i.e., they act
as sink nodes).

Although the mathematical study of the topological structure of a
reaction network is an important topic of research, we will not
consider this problem here. Moreover, we will not consider
situations in which the topology of the network varies with
time. The reader is referred to~\textcite{Newm:10} and the
references therein for such topological considerations.
Instead, our objective is to discuss mathematical methods and
computational techniques for the modeling and analysis of the
dynamic behavior of reaction networks.

\vspace{-12pt}

\subsection{Stochastic dynamics on reaction networks}

\vspace{-6pt}

In many reaction networks of interest, the underlying reactions may
occur at random times. If $Z_m(t)$ denotes the number of times that
the $\mth$ reaction occurs within the time interval $[0,t)$, then
$\{Z_m(t), t \geq 0\}$ will be a random counting process~\cite{Ross:96}.
By convention, we set $Z_m(0)=0$ (i.e., the reaction never occurs
before the initial time $t=0$). We can employ the $M \times 1$
random vector $\bfZ(t)$ with elements $Z_m(t)$, $m=1,2,\ldots,M$, to
characterize the state of the system at time $t > 0$. $Z_m(t)$ is
usually referred to as the \textit{degree of advancement} (DA) of
the $\mth$ reaction~\cite{Kamp:07}. For this reason, we refer to
the multivariate counting process $\{\bfZ(t), t > 0\}$ as the
\textit{DA process}.

An alternative way to characterize a reaction network is by using
the $N \times 1$ random state vector
\beq
\bfX(t) := \bfx_0 + \S \bfZ(t),
\label{eq-II-4}
\eeq
for $t \geq 0$, where $\S := \V'-\V$ is the \textit{net}
stoichiometric matrix of the reaction network and $\bfx_0$
is some known value of $\bfX(t)$ at time $t=0$. Usually, the
$\nth$ element~$X_n(t)$ of $\bfX(t)$ represents the population
number of the $\nth$ species present in the system at time~$t$,
although this may not be true in certain problems (see the
examples discussed in Sections~III-D and III-E). We will be
referring to the multivariate stochastic process
$\{\bfX(t), t > 0\}$ as the \textit{population process}. For
a given initial population vector $\bfx_0$, Eq.~(\ref{eq-II-4})
allows us to uniquely determine the random population vector
$\bfX(t)$ from the DAs $\bfZ(t)$, provided than $\bfZ(t)$ is
finite with probability
one.

\vspace{-12pt}

\subsubsection{Markovian dynamics}

\vspace{-6pt}

A large class of reaction networks can be characterized by Markovian
dynamics, in which case we refer to them as \textit{Markovian
reaction networks}. Markovian reaction networks are based on the
fundamental premise that, for a sufficiently small~$dt$, the
probability of one reaction to occur within the time interval
$[t,t+dt)$ is proportional to $dt$, with proportionality factor
that depends only on the species population present in the system
at time $t$. Specifically, we have that $\Pr {\bigl [}
\mbox{one reaction}~m~\mbox{occurs within}~[t,t+dt)
\mid \bfX(t)=\bfx {\bigr ]} $ $=$ $ \pi_m(\bfx) dt + o(dt)$,
for some function $\pi_m(\bfx)$ of the population, known as the
\textit{propensity function}~\cite{Gill:00}, where $o(dt)$ is a
term that goes to zero faster than $dt$. Under these assumptions,
$\{Z_m(t), t > 0\}$ is a Markovian counting process with intensity
$\pi_m(\bfX(t))$. In particular, the probability  $\pz(\bfz;t)
:= \Pr[\bfZ(t) = \bfz \mid \bfZ(0)={\bf 0}]$ associated with this
process satisfies the following partial differential
equation~\cite{HaseRawl:02,Gout:05,Gout:06}:
\beqa
&& \!\!\!\!\!\!\!\!\!\!\!\!\!\!
\frac{\partial \pz(\bfz;t)}{\partial t} \nn \\ [6pt]
&& \!\!\!\!\!\!\!\!\!\!\! = \!\!\! \sum_{m\in\sM} \!\! {\Bigl \{}
\alpha_m (\bfz - {\bf e}_m) \pz(\bfz - {\bf e}_m;t) -
\alpha_m(\bfz) \pz(\bfz;t) {\Bigr \}},
\label{eq-II-6}
\eeqa
for $t > 0$, where
\beq
\alpha_m(\bfz) := \left \{
\begin{array}{ll}
\pi_m(\bfx_0+\S \bfz),  & \mbox{if~~$\bfz \geq 0$} \\ [6pt]
0,                      & \mbox{otherwise} \: ,
\end{array}
\right . \nn
\label{eq-II-7}
\eeq
and ${\bf e}_m$ is the $\mth$ column of the $M\times M$ identity
matrix. This equation is initialized by setting $\pz(\bfz;0)
= \Delta(\bfz)$, where $\Delta(\bfz)$ is the Kronecker delta
function. It turns out that the population process $\{\bfX(t), t > 0\}$
is a Markov process as well with probability $\px(\bfx;t) :=
\Pr[\bfX(t) = \bfx \mid \bfX(0)=\bfx_0]$ that satisfies the
following partial differential equation~\cite{Gill:92}:
\beqa
&& \!\!\!\!\!\!\!\!\!\!\!\!\!\!
\frac{\partial \px(\bfx;t)}{\partial t} \nn \\ [6pt]
&& \!\!\!\!\!\!\!\!\!\!\! = \!\!\! \sum_{m\in\sM} \!\! {\Bigl \{}
\pi_m (\bfx - {\bf s}_m ) \px(\bfx - {\bf s}_m;t) -  \pi_m(\bfx)
\px(\bfx;t) {\Bigr \}},
\label{eq-II-8}
\eeqa
for $t > 0$, initialized by $\px(\bfx;0)=\Delta(\bfx-\bfx_0)$, where
$\bfs_m$ is the $\mth$ column of the net stoichiometric matrix
$\S$.\footnote{The solution $\qx(\bfx;t)$ of Eq.~(\ref{eq-II-8}),
initialized with an arbitrary probability mass function $q(\bfx)$,
is related to the solution $\px(\bfx;\bfx_0,t)$ of Eq.~(\ref{eq-II-8}),
initialized with $\Delta(\bfx-\bfx_0)$, by $\qx(\bfx;t)
= \sum_{\bfx_0} \px(\bfx;\bfx_0,t) q(\bfx_0)$. Therefore, it
suffices to only calculate $\px(\bfx;\bfx_0,t)$, for every $\bfx_0$
such that $q(\bfx_0) \not= 0$. For this reason, we focus our
discussion on solving Eq.~(\ref{eq-II-8}) initialized with
$\Delta(\bfx-\bfx_0)$.} For notational simplicity, we hide the
dependency of $\px(\bfx;t)$ on $\bfx_0$. Most often, Eqs.~(\ref{eq-II-6})
and~(\ref{eq-II-8}) are referred to as \textit{master equations}
although they are both special cases of the well-known forward
Kolmogorov equations in the theory of Markov processes~\cite{Kamp:07}.

The previous master equations provide a suggestive interpretation
on how the probabilities $\pz(\bfz;t)$ and $\px(\bfx;t)$ evolve as a
function of time. For example, Eq.~(\ref{eq-II-8}) implies that
the probability $\px(\bfx;t)$ of the population process $\bfX(t)$
taking value $\bfx$ increases during the time interval $[t,t+dt)$
by an amount $dt \sum_{m\in\sM} \pi_m (\bfx - {\bf s}_m )
\px(\bfx - {\bf s}_m;t)$ due to possible transitions from states
$\bfx-\bfs_m$, $m \in \sM$, at time $t$, to state $\bfx$ at time
$t+dt$. However, during the same time period the probability
$\px(\bfx;t)$ also decreases by an amount $dt \sum_{m\in\sM}
\pi_m (\bfx) \px(\bfx;t)$ due to possible transitions from state
$\bfx$ at time $t$ to states $\bfx+\bfs_m$, $m \in \sM$, at time
$t+dt$. Note finally that, in most practical situations, the elements
of $\bfx$ are limited to being not larger than some finite value.
As a consequence, we assume that $\px(\bfx;t) = \pi_m(\bfx)=0$,
when at least one element of~$\bfx$ is greater than that value.

\vspace{-12pt}

\subsubsection{Hidden Markov models}

\vspace{-6pt}

Although the DA process uniquely determines the population process via
Eq.~(\ref{eq-II-4}), the opposite is not true in general. This is due
to the fact that the matrix $\S^T\S$ may not invertible. Invertibility
of $\S^T\S$ is only possible when the nullity of $\S$ is zero, in which
case $\bfZ(t)=(\S^T\S)^{-1}\S^T[\bfX(t)-\bfx_0]$ and the DA process
can be uniquely determined from the population process. Therefore, we
can consider the DA process to be more informative in general than the
population process. Note that, if the solution $\pz(\bfz;t)$ of the
master equation~(\ref{eq-II-6}) is known, then we can calculate the
probability mass function $\px(\bfx;t)$ without having to solve
Eq.~(\ref{eq-II-8}). Since we are dealing with discrete random
variables, we have that
\beq
\px(\bfx;t) \: = \!\! \sum_{\bfz \in \sB(\bfx)} \!\! \pz(\bfz;t) ,
\label{eq-II-9}
\eeq
for $t \geq 0$, where $\sB(\bfx) := \{\bfz \! :~ \bfx = \bfx_0 + \S \bfz\}$.

In many reaction networks, it is much easier to observe the population
process than the DA process, which is usually very difficult or
impossible to measure. Thus, we can consider the elements of $\bfZ(t)$
as being the \textit{hidden} state variables of the system under
consideration and the elements of $\bfX(t)$ as being the \textit{observed}
state variables. If we choose to model the population process by
Eq.~(\ref{eq-II-4}), then we would be using what is known
as a \textit{hidden Markov model} (HMM) for our system~\cite{Gout:06}.
This opens the possibility of employing well-known techniques for
the statistical analysis and stochastic control of HMMs to
mathematically and computationally study stochastic dynamics
on reaction networks.

\vspace{-12pt}

\subsubsection{Topological structure and propensity functions}

\vspace{-6pt}

At a first glance, Eqs.~(\ref{eq-II-6}) and~(\ref{eq-II-8})
may give the impression that the probability distributions $\pz(\bfz;t)$
and $\px(\bfx;t)$ of the DA and population processes associated with
a reaction network do not depend on a detailed knowledge of the
topological structure of the network. This is due to the fact that
the previous master equations seem to depend only on the difference
$\S = \V'-\V$  between the stoichiometric matrices $\V$ and $\V'$ and
not on the individual matrices. This however is not true. It
turns out that, for all reaction networks encountered in practice,
the propensity function $\pi_m(\bfx)$ associated with the $\mth$
reaction node in the network does not depend on all elements of the
state vector $\bfx$ but only on those elements associated with the
adjacent reactant nodes, as specified by the stoichiometric matrix
$\V$. In other words, the propensity function does not depend on
terms involving variables on non-adjacent nodes. As a consequence,
the topological structure of a reaction network directly affects
its dynamics through this mathematical property of the propensity
functions.

%
%

\vspace{-12pt}

\section{Examples}

\vspace{-6pt}

We now provide a few examples which clearly demonstrate that the
previously discussed general framework for reaction networks, based
on~(\ref{eq-II-1}), is sufficiently general to characterize
Markovian dynamics on many other important networks. Each example
is associated with a set of ``species'' that affect each other's
population by interacting through well-defined ``reactions.''
To determine the DA and population dynamics, we only need to specify
the mathematical form of the underlying propensity functions -- from
these, the dynamics follow by solving Eq.~(\ref{eq-II-6})
or Eq.~(\ref{eq-II-8}) for $\pz(\bfz;t)$ and $\px(\bfx;t)$, respectively.

\vspace{-12pt}

\subsection{Biochemical networks}

\vspace{-6pt}

When dealing with biochemical reactions, we usually assume that the
system is well-stirred and in thermal equilibrium at fixed volume.
It can be shown in this case that the probability of a randomly
selected combination of reactant molecules at time $t$ to react
through the $\mth$ reaction during the infinitesimally small time
interval $[t,t+dt)$ is proportional to~$dt$, with a proportionality
factor $\kappa_m$ known as the \textit{specific probability rate
constant} of the reaction~\cite{Gill:92}. As a consequence,
$\Pr {\bigl [} \mbox{one reaction}~m~\mbox{occurs within}~[t,t+dt)
\mid \bfX(t)=\bfx {\bigr ]}$ $=$ $\kappa_m \gamma_m(\bfx) dt + o(dt)$,
where $\gamma_m(\bfx)$ is the number of distinct subsets of
molecules that can form a reaction complex at time $t$, given by
\vspace{-3pt}
\beqa
\gamma_m(\bfx) &\!\!=\!\!\!& \prod_{n\in\sN} \binom{x_n}{\nu_{nm}} \nn \\
&\!\!=\!\!\!&  \prod_{n\in\sN} \left [ x_n \geq \nu_{nm} \right ]
\frac{x_n!}{\nu_{nm}!(x_n-\nu_{nm})!}~, \nn
\label{eq-III-2}
\vspace{-3pt}
\eeqa
with $[a_1 \geq a_2]$ being the Iverson bracket.\footnote{$[a_1
\geq a_2]=1$, if $a_1 \geq a_2$, and $0$ otherwise.} Note that the Iverson
bracket guarantees that a reaction will proceed only if all reactants
are present in the system. Moreover, we use the convention $0!=1$,
so $\binom{x_n}{0}=1$, indicating that the rate of a reaction
is only determined by the state of the reactants. As a consequence,
we obtain the following propensity functions:
\beq
\pi_m(\bfx) = \kappa_m \!\! \prod_{n\in\sN} \binom{x_n}{\nu_{nm}}~,
\quad \mbox{for}~~ m \in \sM, \nn
\label{eq-III-3}
\eeq
which are said to follow the \textit{mass-action law}.

We should note here that certain reactions cannot be adequately
characterized by propensity functions that follow the mass-action
law. For example, let us consider a reaction $X_1 + X_2 \rightarrow
X_3$ that can occur only when a molecule $X_1$ is bound by at least
one molecule $X_2$ at two independent binding sites with the same
affinity $\theta$. It can be shown [e.g., see~\textcite{DillBrom:11}]
that the fraction of molecules $X_1$ bound by $X_2$ is given by
$\theta x_1/(1+\theta x_1)$. This leads to the following
\textit{hyperbolic} propensity function for the reaction:
\beq
\pi(x_1,x_2) = \frac{\kappa \theta x_1 x_2}{1+\theta x_1}, \nn
\label{eq-III-4}
\eeq
where $\kappa$ is the associated specific probability rate constant.
Clearly, the mathematical form of the propensity function of
a given reaction depends on the underlying molecular mechanism.

\vspace{-12pt}

\subsection{Pharmacokinetic networks}

\vspace{-6pt}

Physiological pharmacokinetic models are used extensively to study
the absorption, distribution, metabolism, and elimination of chemicals
and drugs by the body of animals and humans. As a consequence, they
are of crucial importance for drug dosing in clinical
pharmacology~\cite{HardLimb:01}. A large class of pharmacokinetic
models is based on the notion of compartmentalization~\cite{MachIlia:06}.
These models assume the existence of a central compartment (e.g., heart,
lungs, brain, etc.), which serves as a site for drug administration
to peripheral compartments (e.g., fat, muscles, central nervous system,
and liver).

To illustrate the connection between pharmacokinetic models and
Markovian reaction networks, we consider here a model for studying
the effect of tetrachloroethylene, a widely used solvent, on
carcinogenesis~\cite{BoisZeisToze:90}. This model assumes a division
of the human body into the lungs, which serve as the central compartment,
and four peripheral compartments, namely fat tissue, poorly perfused
tissue (muscles and skin), richly perfused tissue (central nervous
system and viscera, except liver), and liver. To model this system,
we denote by $X_n$ the solvent present in the $\nth$ compartment.
Then, we can represent the system by $N=5$ species interacting by
the following $M=10$ reactions:
\beq
\begin{array}{lrcl}
\text{reaction}~1:~~    & \emptyset & \rightarrow  & X_1           \\ [2pt]
\text{reaction}~2:~~    & X_1       & \rightarrow  & X_2           \\ [2pt]
\text{reaction}~3:~~    & X_2       & \rightarrow  & X_1           \\ [2pt]
\text{reaction}~4:~~    & X_1       & \rightarrow  & X_3           \\ [2pt]
\text{reaction}~5:~~    & X_3       & \rightarrow  & X_1           \\ [2pt]
\text{reaction}~6:~~    & X_1       & \rightarrow  & X_4           \\ [2pt]
\text{reaction}~7:~~    & X_4       & \rightarrow  & X_1           \\ [2pt]
\text{reaction}~8:~~    & X_1       & \rightarrow  & X_5           \\ [2pt]
\text{reaction}~9:~~    & X_5       & \rightarrow  & X_1           \\ [2pt]
\text{reaction}~10:~~   & X_5       & \rightarrow  & \emptyset ~.  \\ [2pt]
\end{array} \nn
\label{eq-III-5}
\eeq
The underlying reactions model the injection of solvent into
lung blood (reaction~1), the exchange of one molecule of
solvent between the lung blood and fat tissue (reactions~2~\&~3),
poorly perfused tissue (reactions~4~\&~5), richly perfused
tissue (reactions~6~\&~7), and liver tissue (reactions~8~\&~9),
as well as the metabolic clearance of the solvent by the
liver (reaction~10).

If we assume that all compartments are homogeneous, that the
injection of solvent into the lung blood takes place at a
constant rate $\kappa_1$, and that the probability of a randomly
selected solvent molecule to move from compartment $n$ to
compartment $n'$ within an infinitesimally small time interval
$[t,t+dt)$ is proportional to $dt$ with proportionality
constant~$\kappa_{nn'}$, then we can model the previous
pharmacokinetic system as a Markovian reaction network
with \textit{linear} mass-action propensity functions
\beq
\begin{array}{lll}
\pi_1(\bfx) = \kappa_1, & \pi_2(\bfx) = \kappa_{12} x_1, &
\pi_3(\bfx) = \kappa_{21} x_2, \\ [4pt]
\pi_4(\bfx) = \kappa_{13} x_1, &
\pi_5(\bfx) = \kappa_{31} x_3, &
\pi_6(\bfx) = \kappa_{14} x_1, \\ [4pt]
\pi_7(\bfx) = \kappa_{41} x_4, &
\pi_8(\bfx) = \kappa_{15} x_1, &
\pi_9(\bfx) = \kappa_{51} x_5,
\end{array} \nn
\label{eq-III-6}
\eeq
where the $\nth$ element $x_n$ of vector $\bfx$ denotes the population
of tetrachloroethylene in the $\nth$ compartment. Moreover, if we
assume that tetrachloroethylene metabolism in the liver is saturable
according to the Michaelis-Menten relationship of enzyme
kinetics~\cite{BoisZeisToze:90}, then the propensity function of the
last reaction will be given by the following \textit{nonlinear}
(hyperbolic) expression~\cite{SanfGillPetz:11}:
\beq
\pi_{10}(\bfx) = \frac{V x_5}{K + x_5}, \nn
\label{eq-III-7}
\eeq
where $V,K$ are two parameters associated with the underlying metabolic
mechanism.

\vspace{-12pt}

\subsection{Epidemiological networks}

\vspace{-6pt}

Epidemiological networks study the spread of infectious diseases
or agents through a population of individuals. Although numerous
publications can be found on the subject, we refer the reader
to~\textcite{Newm:10} for an elementary introduction to
epidemiological networks. For a mathematical review of
\textit{deterministic} epidemiological models, see~\textcite{Heth:00},
whereas, for a \textit{stochastic} modeling approach to epidemiological
modeling, see~\textcite{ChenBokk:05}.

To illustrate the connection between epidemiological networks and
Markovian reaction networks, we consider the simplest and most
widely used model, known as the SIR epidemic model. In this model,
an individual in a population can be in one of three states with
respect to a disease: susceptible~(S), infected~(I), or resistant~(R).
According to this model, there are two types of interactions that an
individual may undergo: (a)~if a susceptible individual comes into
contact with an infectious individual, the susceptible person can be
infected, and (b)~an infected individual may become resistant if his
immune system fights off the infection and confers resistance, or if
the individual dies by the infection. These interactions can be
modeled by a reaction network comprised of $N=3$ species (S, I, and R)
that interact through the following $M=2$ reactions:
\beq
\begin{array}{rcl}
X_1 + X_2 & \rightarrow   & 2X_2    \\ [2pt]
X_2       & \rightarrow   & X_3 ~,
\end{array}
\label{eq-III-8}
\eeq
where $X_1=\text{S}$, $X_2=\text{I}$ and $X_3=\text{R}$. In this case,
\beq
\V = \left [
\begin{array}{cc}
 1 & 0 \\
 1 & 1 \\
 0 & 0
\end{array}
\right ] \!\!, ~
\V' = \left [
\begin{array}{cc}
0 & 0 \\
2 & 0\\
0 & 1
\end{array}
\right ] \!\!, ~ \mbox{and} ~
\S = \left [
\begin{array}{rr}
-1 &  0 \\
 1 & -1\\
 0 &  1
\end{array}
\right ] \! . \nn
\label{eq-III-9}
\eeq

We can now assume that the probability of a randomly selected
susceptible individual at time $t$ to become infected by a randomly
selected infectious individual during an infinitesimally small time
interval $[t,t+dt)$ is proportional to~$dt$, with proportionality
factor $\kappa_1$ that does not depend on the particular individuals
involved. Moreover, we can assume that the probability of a randomly
selected infected individual at time $t$ to recover or die from the disease
during $[t,t+dt)$ is also proportional to $dt$, with proportionality
factor~$\kappa_2$ that does not depend on the particular infected individual.
Then, the previous interactions lead to a Markovian reaction network
with mass-action propensity functions given by~\cite{ChenBokk:05}
\beq
\pi_1(x_1,x_2,x_3) = \kappa_1 x_1 x_2 \quad \mbox{and} \quad
\pi_2(x_1,x_2,x_3) = \kappa_2 x_2, \nn
\label{eq-III-10}
\eeq
where $x_1,x_2,x_3$ are the populations of susceptible, infectious,
and resistant individuals, respectively.

We can use the previous 3-species/2-reactions motif, given
by~(\ref{eq-III-8}), to construct more complex Markovian
reaction networks that model the spread of an infectious
disease in a population of individuals grouped into classes
(e.g., households, work spaces, cities, etc.);
see~\textcite{ZionCoheShne:10}. We may group, for example,
individuals into two classes, those living in Baltimore and
Philadelphia, and give each class its own distinct set of variables,
namely $X_1,X_2,X_3$, for susceptible, infected, and resistant
individuals in Baltimore, as well as $X_4,X_5,X_6$, for susceptible,
infected, and resistant individuals in Philadelphia. Each class
will be characterized by the previous 3-species/2-reactions motif,
resulting in the following four reactions:
\beq
\begin{array}{rcl}
X_1 + X_2 & \rightarrow & 2X_2 \\ [2pt]
X_2       & \rightarrow & X_3  \\ [2pt]
X_4 + X_5 & \rightarrow & 2X_5 \\ [2pt]
X_5       & \rightarrow & X_6~. \nn
\label{eq-III-11}
\end{array}
\eeq
In this case however there is also a flow (by air, road, or rail) of
individuals between the two different cities, which we can model by
using the following six reactions:
\beq
\begin{array}{rcl}
X_1 & \rightarrow  & X_4 \\ [2pt]
X_4 & \rightarrow  & X_1 \\ [2pt]
X_2 & \rightarrow  & X_5 \\ [2pt]
X_5 & \rightarrow  & X_2 \\ [2pt]
X_3 & \rightarrow  & X_6 \\ [2pt]
X_6 & \rightarrow  & X_3 \: . \nn
\label{eq-III-12}
\end{array}
\eeq
The propensity functions associated with these new reactions will
be proportional to the population of the input species, with the
proportionality factor being the specific probability rate constant
of an individual traveling from one city to the other. In this
fashion, we can build complex Markovian reaction network models
for epidemiological dynamics that are more realistic and more
predictive than traditional deterministic models.

Likewise, new reactions may be incorporated into the epidemiological
network to account for additional transitions between states. For
instance, if we assume that a vaccine is available, then we must
include the reaction $X_1 \rightarrow X_3$ in the formulation. Vital
dynamics (i.e., births and deaths) may also be included in this
fashion. For example, if infants born at a fixed rate are always
susceptible, then the reaction $\emptyset \rightarrow X_1$ must be
included in the system. Finally, one may consider social networks
on which epidemiological networks reside. Specifically, age
stratification in the population~\cite{Heth:00}, or the scale-free
structure of social/sexual networks~\cite{Newm:10}, may be handled
in a manner similar -- albeit not identical -- to the aforementioned
geographic considerations.

\vspace{-12pt}

\subsection{Ecological networks}

\vspace{-6pt}

Ecological networks aim to study interactions among organisms
living in a particular area as well as between these organisms
and nonliving physical components of the environment, such as
air, soil, water, and sunlight. The main objective of this
type of network is to model how mass and energy are transferred
from primary producers (or autotrophs), who generate their own
energy from the sun's rays, up to the apex predators who gather
their energy and body mass through the consumption of prey
lower in the food chain. We illustrate here the fact that
ecological networks can also be modeled as Markovian reaction
networks using a simple example.

Consider a food web comprised of grass~($X_1$), rabbits~($X_2$)
and wolves~($X_3$), whose net mass at time $t$ is given by $X_1(t)$,
$X_2(t)$ and $X_3(t)$, respectively. These states can take non-integer
values. In particular, $X_1(t)=x$ means that, at time~$t$, the mass
of grass equals $x$-times some reference value, and likewise for
rabbits and wolves. More advanced models may also choose to keep
track of the number of individuals~\cite{DattDeliLaw:10}. Here
however we consider a common situation in which the net mass
of each species is sufficient to describe the system.

We can assume that changes in mass distribution are caused by discrete
steps in body size as predators eat prey as well as by the mortality
that comes with this process. In particular, we can model the predation
of grass by rabbits and rabbits by wolves with the following two
reactions~\cite{DilaDomi:00}:
\beq
\begin{array}{rcl}
X_1 + X_2   & \rightarrow   & (1+a_1) X_2 \\ [6pt]
X_2 + X_3   & \rightarrow   & (1+a_2) X_3  \: , \nn
\end{array}
\label{eq-III-13}
\eeq
where $a_1,a_2 > 0$ are constants representing the conversion factor
of mass. Moreover, when rabbits or wolves die for reasons other
than predation they fertilize the grass. We can model this conversion
by~\cite{DilaDomi:00}
\beq
\begin{array}{rcl}
X_2 & \rightarrow & b_1 X_1 \\ [6pt]
X_3 & \rightarrow & b_2 X_1 \: , \nn
\end{array}
\label{eq-III-14}
\eeq
where $b_1,b_2 >0$ are appropriately chosen recycling constants.
As a consequence, the stoichiometric matrices of the resulting
reaction network, comprised of the $N=3$ species and the $M=4$
reactions above, are given by
\beq
\V = \left [
\begin{array}{cccc}
 1 & 0 & 0 & 0 \\
 1 & 1 & 1 & 0 \\
 0 & 1 & 0 & 1
\end{array}
\right ] \!\!, \quad
\V' = \left [
\begin{array}{cccc}
0     & 0     & b_1 &~~ b_2 \\
1+a_1 & 0     & 0 &~~ 0 \\
0     & 1+a_2 & 0 &~~ 0
\end{array}
\right ] \!\!, \nn
\eeq
\vspace{-9pt}
\beq
\S = \left [
\begin{array}{rrrr}
-1   &  0   &  b_1 & b_2 \\
a_1  & -1   & -1   & 0   \\
 0   &  a_2 &  0   & -1
\end{array}
\right ] \!\!. \nn
\label{eq-III-15}
\eeq

Under appropriate assumptions, similar to the ones made before, the
previous interactions lead to a Markovian reaction network with
mass-action propensity functions given by~\cite{DilaDomi:00}
\beq
\begin{array}{ll}
\pi_1(\bfx) = \kappa_1 [x_1,x_2 \geq 1] x_1 x_2, & \!\!\!\!\!\!\!\!\!\!\!\!
\pi_2(\bfx) = \kappa_2 [x_2,x_3 \geq 1]x_2 x_3 \\ [6pt]
\pi_3(\bfx) = \kappa_3 [x_2 \geq 1] x_2, \hspace{.6in} & \!\!\!\!\!\!\!\!\!\!\!\!
\pi_4(\bfx) = \kappa_4 [x_3 \geq 1] x_3,
\end{array} \nn
\label{eq-III-16}
\eeq
where the Iverson brackets are used to make sure that the reactions
occur only when the net mass of a reactant species is at least as
large as the corresponding reference value. Here, $\kappa_1$ is the
specific probability rate constant of rabbits eating grass, $\kappa_2$
is the specific probability rate constant of wolves eating rabbits,
and $\kappa_3,\kappa_4$ are the specific probability rate constant
of natural deaths of rabbits and wolves, respectively.

More complicated ecological reaction network models can include
geographic considerations, direct competition, mutualism, and
more complex food chains~\cite{LassBastManr-etal:01,PoweBola:09,%
ThebFont:10}. In addition, epidemiological networks can be
combined with ecological networks to study the effects of a disease
on a given ecosystem~\cite{AugeMchiChow-etal:09}.

\vspace{-12pt}

\subsection{Social networks}

\vspace{-6pt}

Recently, interest has emerged in developing mathematical models
for social networks that can be used to better understand human
behavior. In particular, much effort has been devoted to studying
dynamics on social networks~\cite{MoreNekoPach:04,AntaKrapRedn:06,%
Weid:06,ZaneGil:06,HillRandNowa-etal:10,MasuGibeRedn:10}, a problem
that has been investigated by the physics community many decades
ago~\cite{Hake:75}. Several models for dealing with dynamic processes
on social networks are currently available, with many fitting
nicely into the Markovian reaction framework discussed in this review.
As an example, we focus on a model of opinion formation in social
networks, a process that is of political, marketing, and general
sociological interest.

The critical behavior of a society moving from a liberal to a
totalitarian political system can be evaluated when individuals
are endowed with two separate opinions: a publicly pronounced and
a privately held opinion for/against the ideology of the ruling
party. The public and private opinions of an individual can be
different when, for example, public dissent against the ruling
ideology is a punishable offence. Along these lines, let us
consider a fixed homogeneous group of $2L$ individuals who
react in the same manner to a given situation. An individual
simultaneously holds a public and a private opinion that each
takes values $1/2$ or $-1/2$ if it is for or against the ruling
ideology, respectively. Let us denote by $X_1$ the net public
opinion, which corresponds to the sum of the publicly held
opinions of all $2L$ individuals. Likewise, let us denote by
$X_2$ the net private opinion. We are now dealing with $N=2$
species interacting through the following $M=4$ reactions:
\beq
\begin{array}{ll}
\text{reaction}~1:~~ X_1 + X_2 \rightarrow 2X_1 + X_2    \\ [4pt]
\text{reaction}~2:~~ X_1 + X_2 \rightarrow X_2           \\ [4pt]
\text{reaction}~3:~~ X_1 + X_2 \rightarrow X_1 + 2X_2    \\ [4pt]
\text{reaction}~4:~~ X_1 + X_2 \rightarrow X_1 .
\end{array}
\label{eq-III-17}
\eeq

The first two reactions model the influence of net private opinion
$X_2$ on the net public opinion $X_1$ that results in a single
individual changing her public opinion in support of (reaction~1)
or against (reaction~2) the ruling ideology. In this case, the
net private opinion remains unchanged, whereas, the net public
opinion is increased by one in reaction~1 [due to a value change
from $-1/2$ (against) to $1/2$ (for)] and decreased by one in
reaction~2 [due to a value change from $1/2$ (for) to $-1/2$
(against)]. Likewise, the subsequent two reactions model the
influence of net public opinion $X_1$ on the net private
opinion~$X_2$ that results in a single individual changing her
private opinion in support of (reaction~3) or against (reaction~4)
the ruling ideology. These reactions are governed by the
following propensity functions~\cite{Weid:06}:
\beq
\begin{array}{l}
\pi_1(\bfx) = \k_1(L-x_1)\exp(a_1 x_1 + a_2 x_2)  \\ [4pt]
\pi_2(\bfx) = \k_1(L+x_1)\exp(-a_1 x_1 - a_2 x_2) \\ [4pt]
\pi_3(\bfx) = \k_2(L-x_2)\exp(a_3 x_1)          \\ [4pt]
\pi_4(\bfx) = \k_2(L+x_2)\exp(-a_3 x_1),
\end{array}
\label{eq-III-18}
\eeq
where $x_1$, $x_2$ represent the net values of all publicly and
privately held opinions, respectively, $\k_1,\k_2 > 0$ are two
specific probability rate constants associated with the four
reactions, and $a_1 \geq 0$, $a_2 > 0$, $a_3$ are three model
parameters. Note that $x_1$ and $x_2$ are integer-valued with
$-L \leq x_1,x_2 \leq L$, where $-L$ represents total disapproval
and $L$ represents total approval of the ruling ideology.

Parameter $a_1 \geq 0$ controls pressure inflicted on public
opinion due, for example, to oppression of this opinion by the
ruling party (the value of this parameter is zero in the U.S.\
where free speech is protected, but strictly positive in countries
where public dissidence has consequences). On the other hand,
parameter $a_2 > 0$ controls the influence of privately held
beliefs on publicly stated opinions, whereas, parameter $a_3$
controls how affirmative (for $a_3 > 0$) or dissident (for $a_3 < 0$)
the private opinion is towards the ruling ideology. When the
values of $a_1$ and $a_3$ vary, an abrupt change from a liberal
to a totalitarian political system can be observed~\cite{Weid:06}.
This critical social behavior predicted by the model is reminiscent
to the well-known phenomenon of phase transition in statistical
mechanics and provides a crucial focus of study when dealing with
opinion spreading in social networks.

\vspace{-12pt}

\subsection{Neural networks}

\vspace{-6pt}

A discussion on reaction networks cannot be complete without
mentioning biological neural networks. With 100 billion or
more neurons in the human brain connected by 100-500 trillion
synapses, there is no other reaction network that can compete
in size and complexity.

There is a large body of literature surrounding the modeling
and analysis of biological neural networks. As an example,
we consider a Markovian reaction model for neural networks
recently proposed by~\textcite{BenaCowaDron-etal:10} that is
intuitive enough for novices in neurobiology to comprehend
and yet rich enough to be a viable candidate for understanding
many features of this preeminent reaction network. The model
consists of $L$ neurons, with each neuron being in either a
quiescent or an active state. Let $X_{2l-1}$ and $X_{2l}$ denote
a quiescent or active neuron $l$, respectively. We can assign
the following two reactions to the $\lth$ neuron in the network:
\beq
\begin{array}{rcl}
X_{2l-1} + {\displaystyle \sum_{l'\neq \: l} }
~ \nu_{l'l} X_{2l'} & \rightarrow & X_{2l}
+ {\displaystyle \sum_{l' \neq \: l}}
~ \nu_{l'l} X_{2l'} \\ [6pt]
X_{2l} & \rightarrow & X_{2l-1} ,
\end{array}
\label{eq-III-19}
\eeq
where $\nu_{ij}$ measures the synaptic weight between neurons $i$
and $j$, with a positive value indicating an excitatory synapsis
and a negative value indicating an inhibitory synapsis. Note that
the first reaction models transition of the $\lth$ neuron from
the quiescent to the active state, which is assumed to be
influenced by appropriately weighted active neurons $X_{2l'}$,
$l' \not= l$, in the network [see Eq.~(\ref{eq-III-20}) below]
that act as ``catalysts.'' On the other hand, the second reaction
models transition of the neuron from the active to the quiescent
state, which is assumed to occur constitutively. As a
consequence, we obtain a reaction network with $N=2L$ species
and $M=2L$ reactions.

We can describe this system by a $2L \times 1$ state vector $\bfx$ with
binary-valued $0/1$ elements $x_{2l-1}$, $x_{2l}$ indicating the state
of the $\lth$ neuron (with $0$ being quiescent and $1$ being active).
Due to the fact that a neuron must be either quiescent or active, the
state variables must satisfy the mass conservation relationships
$x_{2l-1} + x_{2l} = 1$, for $l=1,2,\ldots,L$. It has been suggested
by~\textcite{BenaCowaDron-etal:10} that the probability of the $\lth$
neuron becoming active during an infinitesimally small time interval
$[t,t+dt)$, given that the neuron is quiescent at time $t$, can be
taken to be $x_{2l-1} [\phi_l(\bfx)>0] \tanh[\phi_l(\bfx)]dt+o(dt)$,
where $[a>0]$ is the Iverson bracket and $\phi_l$ is the net synaptic
input to the~$\lth$ neuron, given by
\beq
\phi_l(\bfx) = \sum_{l' \neq \: l} \nu_{l'l} x_{2l'} + h_l,
\label{eq-III-20}
\eeq
with $h_l$ being an external input to the neuron. The term $x_{2l-1}$
ensures that the neuron becomes active within $[t,t+dt)$ only when it
is quiescent at time $t$. As a consequence, the propensity of the first
reaction in~(\ref{eq-III-19}) will be given by
\beq
\pi_{2l-1}(\bfx) = x_{2l-1} [\phi_l(\bfx)>0] \tanh[\phi_l(\bfx)],
\label{eq-III-21}
\eeq
and therefore depends on the synaptic inputs from neurons connected to
the $\lth$ neuron and any external input to that neuron. On the other
hand, if we assume that the $\lth$ neuron decays from an active to a
quiescent state at a constant rate~$\gamma_l$, then the propensity of the
second reaction will be given by
\beq
\pi_{2l}(\bfx) = \gamma_l x_{2l} ,
\label{eq-III-22}
\eeq
where the term $x_{2l}$ ensures that the neuron becomes inactive within
$[t,t+dt)$ only when it is active at time $t$.

\vspace{-12pt}

\subsection{Multi-agent networks}

\vspace{-6pt}

The study of multi-agent networks focuses on systems in which many
intelligent agents, such as autonomous vehicles that observe and
act upon their environment, interact with each other to achieve
a certain goal. To illustrate the fact that multi-agent systems
can also be modeled as Markovian processes on reaction networks,
we consider here a system comprised of $L$ autonomous unmanned
vehicles (AUVs) that can move over a two-dimensional bounded
rectangular space in a discrete fashion~\cite{XiTanBara:06}.
For simplicity, we assume that, at each step, an AUV located at
a discrete point $(i,j)$ in space can move towards one of four
possible directions, namely east to point $(i+1,j)$, west to
point $(i-1,j)$, north to point $(i,j+1)$, or south to point
$(i,j-1)$. We want to develop a mathematical approach that can
be used to describe vehicular motion so that the AUVs reach a
spatial configuration $\bfx$ at steady-state with desired
probability $\rho(\bfx)$ which assigns high probability over
configurations that maximize a given design objective and low
or zero probability over the remaining configurations. The
construction of such probability can be thought of as an inverse
problem that can be solved using a statistical mechanics
approach, as the one proposed by~\textcite{CohnKuma:09}.

In the following, we employ two species $X_{2l-1}$ and $X_{2l}$
whose populations $x_{2l-1}$ and $x_{2l}$ denote the position of
the~$\lth$ AUV on the two-dimensional rectangular grid. For
example, if the $\lth$ vehicle is located at point $(i,j)$ on
the grid, then $x_{2l-1}=i$ and $x_{2l}=j$. We can now characterize
the motion of all AUVs in the multi-agent network under consideration
by $N=2L$ species interacting through the following $M=4L$ reactions:
\beqa
&& \!\!\!\!\!\!\!\!\!\!\!\!\!\!\!\!\!\!\!\!
X_{2l-1} + X_{2l} + {\displaystyle \sum_{l' \neq \: l}(X_{2l'-1} + X_{2l'})}
\rightarrow \nn \\ [-4pt]
&& ~~~~~~ 2X_{2l-1}+X_{2l} + {\displaystyle  \sum_{l' \neq \: l}
(X_{2l'-1} + X_{2l'})} \nn \\
&& \!\!\!\!\!\!\!\!\!\!\!\!\!\!\!\!\!\!\!\!
X_{2l-1}+X_{2l} + {\displaystyle  \sum_{l' \neq \: l}
(X_{2l'-1} + X_{2l'})} \rightarrow \nn \\ [-4pt]
&& ~~~~~~ X_{2l} + {\displaystyle \sum_{l' \neq \: l}(X_{2l'-1}
+ X_{2l'})} \nn \\
&& \!\!\!\!\!\!\!\!\!\!\!\!\!\!\!\!\!\!\!\!
X_{2l-1} + X_{2l} + {\displaystyle \sum_{l' \neq \: l}(X_{2l'-1} + X_{2l'})}
\rightarrow \nn \\ [-4pt]
&& ~~~~~~ X_{2l-1}+2X_{2l} + {\displaystyle  \sum_{l' \neq \: l}
(X_{2l'-1} + X_{2l'})} \nn \\
&& \!\!\!\!\!\!\!\!\!\!\!\!\!\!\!\!\!\!\!\!
X_{2l-1}+X_{2l} + {\displaystyle  \sum_{l' \neq \: l}
(X_{2l'-1} + X_{2l'})} \rightarrow \nn \\ [-4pt]
&& ~~~~~~ X_{2l-1} + {\displaystyle \sum_{l' \neq \: l}
(X_{2l'-1} + X_{2l'})} \:.
\label{eq-III-23}
\eeqa
The first two reactions model one-step motion of the $\lth$ AUV towards
east/west, whereas, the other two reactions model one-step motion towards
north/south. Note that, when the first reaction occurs, the horizontal
position~$i$ of the $\lth$ AUV is increased by one (transition from
$X_{2l-1}$ to $2X_{2l-1}$), whereas its vertical position~$j$ remains
unchanged (transition from $X_{2l}$ to itself). Moreover, this is done
by using the positions $X_{2l'-1}$, $X_{2l'}$, $l' \not= l$, of the
remaining vehicles [see Eq.~(\ref{eq-III-25}) below], which act as
``catalysts'' of the reaction. Similar remarks apply for the other
three reactions as well.

Let us now define the potential energy $V(\bfx)$ of the reaction system
being in configuration $\bfx$ at steady-state by
\beq
V(\bfx) := \left \{
\begin{array}{cl}
- \ln {\displaystyle \frac{\rho(\bfx)}{\rho(\bfx_0)}},
& \quad \mbox{for~~$\bfx \in \sD$} \\ [12pt]
\infty,
& \quad \mbox{otherwise},
\end{array}
\right .
\label{eq-III-24}
\eeq
where $\sD$ is a set that contains all permissible vehicle configurations
(e.g., $\bfx$ should not allow two vehicles to occupy the same grid
position or positions occupied by obstacles, thus avoiding collisions
or assignment of vehicles to grid positions outside the bounded rectangular
region). Moreover, $\bfx_0 \in \sD$ is an appropriately chosen reference
configuration of zero potential energy. Given that $\bfX(t)=\bfx$, we
can assume that, during the infinitesimally small time interval $[t,t+dt)$,
the $\lth$ AUV can move one step towards east if two events take place:
(a)~during $[t,t+dt)$, the AUV initiates motion with probability
that is proportional to $dt$, with proportionality factor $\kappa_l$,
and (b)~given that the AUV initiates motion during $[t,t+dt)$, it
moves with probability $\exp \left \{- V(\bfx+\bfs_{4l-3}) \right \}$,
where $\bfs_m$ denotes the $\mth$ column of the net stoichiometric
matrix of the reaction network given by~(\ref{eq-III-23}). As a
consequence, the AUV will be moving east with higher probability if
the motion produces a larger reduction in potential energy. Note that
parameter $\kappa_l$ controls the speed of the $\lth$ vehicle, with
higher values of $\kappa_l$ resulting in faster motion.

By making similar assumptions for vehicle motion towards the
other three directions, the dynamics on the reaction network
given by~(\ref{eq-III-23}) will be Markovian with propensity
functions
\beq
\pi_m(\bfx) = \kappa_l e^{-V(\bfx+\bfs_m) },
\label{eq-III-25}
\eeq
for $m=4l-3,4l-2,4l-1,4l$, $l=1,2,\ldots,L$.
Note that $\bfs_{4l-3}$, $\bfs_{4l-2}$, $\bfs_{4l-1}$ and $\bfs_{4l}$
equal ${\textbf e}_{2l-1}$, $-{\textbf e}_{2l-1}$, ${\textbf e}_{2l}$,
and $-{\textbf e}_{2l}$, respectively, where ${\textbf e}_m$ is the
$\mth$ column of the $2L \times 2L$ identity matrix. It turns out that
the resulting master equation governing the population process $\bfX(t)$
has a unique stationary distribution $\overline{p}_{\ss \bf X} (\bfx) :=
\lim_{t \rightarrow \infty} \px(\bfx;t)$, given by the Gibbs distribution
\beq
\overline{p}_{\ss \bf X} (\bfx) = \frac{1}{\zeta} \: e^ {- V(\bfx) },
\label{eq-III-26}
\eeq
where
\beq
\zeta := \sum_{\bfx} e^{ - V(\bfx)}
\label{eq-III-27}
\eeq
is the associated partition function. As a consequence of
Eqs.~(\ref{eq-III-24}), (\ref{eq-III-26}) and (\ref{eq-III-27}), we
have that $\overline{p}_{\ss \bf X} (\bfx)=\rho(\bfx)$. Therefore, the
AUVs will asymptotically position themselves in the two-dimensional
space at locations $\bfx$ with probability~$\rho(\bfx)$, as desired.

\vspace{-12pt}

\subsection{Evolutionary game theory}

\vspace{-6pt}

Game theory deals with mathematical models of conflict and cooperation
among intelligent and rational individuals. Evolutionary game theory
extends the paradigm of classical game theory by removing some stringent
assumptions and by naturally incorporating the dynamic aspects of
learning and experimentation into the problem.

As an example of how evolutionary game theory can fit within the
current context, suppose that a population of~$L$ individuals play
with each other a game with $N$ possible strategies $X_1,X_2,\ldots,X_N$.
Let $X_n(t)$ be the number of individuals playing strategy $X_n$ at
time~$t$. Here, we consider a simple situation in which each individual
competes with all other individuals. However, the framework presented
in this paper is also capable of handling more general situations,
such as those discussed by~\textcite{SzabFath:07}. Given that
$\bfX(t)=\bfx$, let $P_n(\bfx)$ be the payoff to an individual
playing the $\nth$ strategy at time $t$. Based on the current payoff,
this individual may decide at a random time to follow a new strategy
$X_{n'}$ in an attempt to improve his payoff. This can be modeled by
the following $M=N(N-1)$ reactions:
\beq
X_{n} + X_{n'} + \!\!\!\! \sum_{n'' \not= \: n,n'}\!\!\!\!\!
X_{n''} ~\rightarrow~ 2X_{n'} + \!\!\!\!\sum_{n''\not= \: n,n' }
\!\!\!\!\! X_{n''}, \quad n' \neq n. \nn
\label{eq-III-28}
\eeq
Note that, in this case, the number of individuals $X_{n''}$ that
follow strategies other than $n$ and $n'$ affect the transition
of an individual from strategy $n$ to strategy $n'$ [see
Eq.~(\ref{eq-III-29}) below] without changing their own
strategies and, therefore, act as ``catalysts."

There are many alternative propensity functions that can be chosen
to dictate when players will change their strategy, with each
corresponding to different learning techniques or update
rules~\cite{SzabFath:07}. A common choice however is given
by the imitation rule of the Moran process~\cite{Mora:62}:
\beq
\pi(\bfx) = \k \: \frac{x_{n} }{L} \: \frac{ x_{n'} P_n(\bfx)}
{\sum_{n'' \in \sN} x_{n''}P_{n''}(\bfx)} \:,
\label{eq-III-29}
\eeq
where $\k>0$ is a specific probability rate constant detailing how often
individuals choose to update their strategies. The second term in
Eq.~(\ref{eq-III-29}) is the fraction of individuals playing strategy $X_n$,
whereas, the third term is the fraction of the net payoff paid to
individuals who play strategy $X_{n'}$. These propensity functions have
been originally developed to model natural selection
and genetic drift in an asexually reproducing population of $N$ genetically
distinct individuals, where each genotype represents a strategy and the
payoffs provide measures of reproductive fitness.

\vspace{-12pt}

\subsection{Petri nets}

\vspace{-6pt}

Petri nets have been extensively used to describe discrete-event
distributed systems, a class of systems that are of particular
interest in computer science applications~\cite{Diaz:09}. A Petri
net is a weighted, directed, bipartite graph, in which the nodes
represent \textit{places} and \textit{transitions}. Places model
passive system components, whereas, transitions correspond to
events that inter-convert places. Directed arcs join places
to transitions (connect places that can be converted during a
transition) and transitions to places (connect a transition with
the corresponding products). Weights associated with arcs indicate
the multiplicity of the arc. Each place is associated with
\textit{tokens}, indicating the number of existing places.
Whether or not a transition takes place is described by a rule,
which may be deterministic or stochastic~\cite{Haas:02,Diaz:09},
that depends on the number of tokens available in the places
connecting to the transition by incoming arcs. The occurrence
of a transition results in removing a token from the input
places and adding a token to the output places of the transition.

The flow of tokens on a Petri net can be used to model the dynamics
on a reaction network. As a matter of fact, a number of investigators
(including Petri himself) have already proposed using Petri nets
for modeling biochemical reaction systems~\cite{ReddLiebMavr:96,%
GossPecc:98,Chao:07,HeinGilbDona:08}. This approach however is
very similar to traditional methods for modeling biochemical
reaction systems based on first-order differential equations or
the chemical master equation, which have been extensively studied
in the literature~\cite{Gill:92,HeinSchu:96}. In particular,
Markovian Petri nets are identical to the Markovian reaction
networks considered in this review, with the places playing the
role of species and the transitions representing reactions. It is
however important to carefully study the theory of stochastic
Petri nets~\cite{Haas:02}, since many results derived in that
theory will likely prove very useful for the analysis of the
Markovian reaction networks reviewed in this paper.

%
%

\vspace{-12pt}

\section{Solving the master equation}

\vspace{-6pt}

\begin{figure}
\includegraphics[width=3.35in]{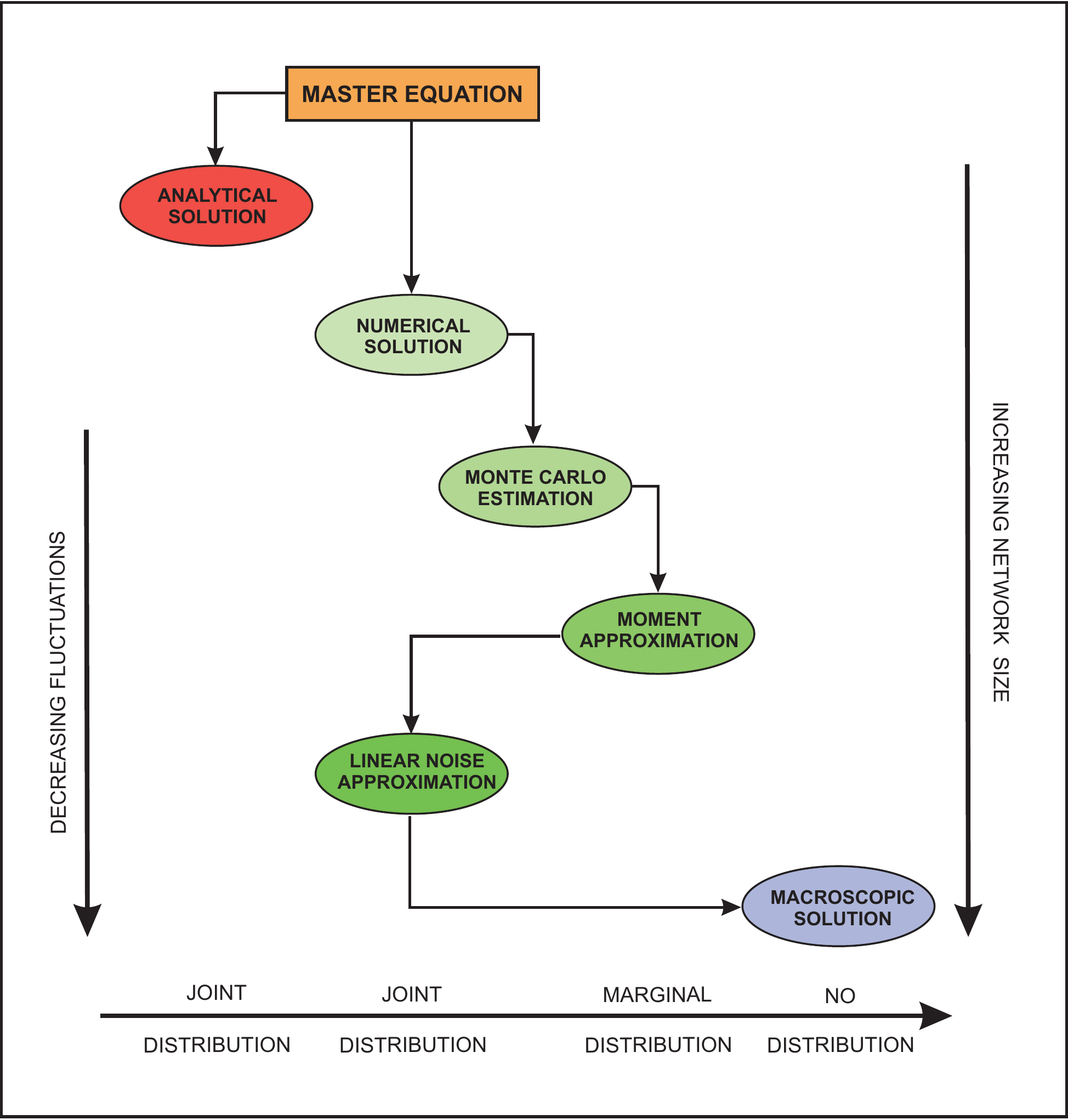}
\caption{Six methods for solving the master equation. Some methods can
be used to approximate the \textit{joint} probability distributions of
the DA and population processes while other methods can only be used
to approximate \textit{marginal} distributions. Analytical solutions
can be obtained only in special cases. Numerical solutions are currently
limited to small reaction networks. Large networks require use of a
moment approximation scheme or adoption of linear noise approximation
method as opposed to Monte Carlo sampling. For large reaction networks,
computing the macroscopic solution may be the only feasible choice.
This solution however can only be trusted at low fluctuation levels.}
\label{fig-2}
\vspace{-6pt}
\end{figure}

Although the algebraic form of the master equations~(\ref{eq-II-6})
and~(\ref{eq-II-8}) is simple, solving these equations [i.e., calculating
the probabilities $\pz(\bfz;t)$ and $\px(\bfx;t)$ at each time $t > 0$]
is a very difficult task in general. Many methods have been proposed in
the literature to address this problem, which can be grouped into the six
general categories depicted in Fig.~\ref{fig-2}. In the following, we
discuss the most prominent techniques available to date. Whether a
technique can be applied to a particular problem depends on the size
and complexity of the reaction network at hand.

\vspace{-12pt}

\subsection{Exact analytical solution}

\vspace{-6pt}

Deriving exact analytical solutions for $\pz(\bfz;t)$ and $\px(\bfx;t)$
is possible only in simple cases [e.g., see~\textcite{McQu:63,%
McQuJachRuss:64,DarvStaf:66,LeonReic:90,Laur:00,GadgLeeOthm:05,%
ZhanCockBuga-etal:05,HeueQian:06,JahnHuis:07,Gard:10}]. For example,
an analytical solution for the master equation~(\ref{eq-II-8}) can
be derived in the case of a \textit{linear} reaction network (i.e.,
a network with linear propensity functions). It has been shown
by~\textcite{GadgLeeOthm:05} that, for \textit{closed} linear
reaction networks (i.e., linear reaction networks with fixed
net population), the solution of the master equation~(\ref{eq-II-8})
is a multinomial distribution, provided that the initial joint
distribution is also multinomial. Moreover, for \textit{open}
linear reaction networks (i.e., linear reaction networks with
varying net population), the solution of the master
equation~(\ref{eq-II-8}) is a product Poisson distribution,
provided that the initial joint distribution is also product
Poisson [see also~\textcite{HeueQian:06}]. These results are
special cases of a more general result derived by~\textcite{JahnHuis:07}
who have shown that the probability distribution $\px(\bfx;t)$
of the population process in a linear reaction network with
initial state $\bfx_0$ can be expressed as the convolution of
multinomial and product Poisson distributions with time-dependent
parameters that evolve according to well-defined systems of
first-order linear differential equations [see
also~\cite{ZhanCockBuga-etal:05}].

\newpage

\vspace{-6pt}

\subsection{Numerical solution}

\vspace{-6pt}

A substantial research effort has recently been focused on approximately
solving the master equation~(\ref{eq-II-8}) using numerical techniques.
Although the methods developed so far show promise for addressing this
problem, they are mostly limited to relatively small reaction networks.
For this reason, we only provide a brief discussion here. The interested
reader can find details in the references.

The master equation~(\ref{eq-II-8}) can be expressed as a linear
system of \textit{coupled} first-order differential equations,
given by
\beq
\frac{d\bfp(t)}{dt} = \P \: \bfp(t),
\label{eq-IV-1}
\eeq
for $t > 0$, where $\bfp(t)$ is a $K \times 1$ vector that
contains the nonzero probabilities $\px(\bfx;t)$, $\bfx \in \sX$,
of the population process~$\bfX(t)$ and $\P$ is a large $K \times K$
\textit{sparse} matrix whose structure can be inferred directly from
the master equation. For example, when the columns of the stoichiometric
matrix $\S$ are all different from each other, the only nonzero elements
of the $\ith$ column of $\P$ are the $M$ off-diagonal elements,
whose values are given by~$\pi_m(\bfx_i)$, and the diagonal element,
whose value is given by $-\sum_{m=1}^M \pi_m(\bfx_i)$, where $M \ll K$
is the number of reactions. If we assume that the cardinality $K$ of
the state-space $\sX$ is finite, then we can calculate the
probabilities~$\px(\bfx;t)$ by solving Eq.~(\ref{eq-IV-1}),
in which case
\beq
\bfp(t) = \exp(t\P) \: \bfp(0),
\label{eq-IV-2}
\eeq
for $t > 0$. This simple idea has led to a numerical technique,
proposed by~\textcite{MunsKham:06}, for approximately solving
the master equation known as \textit{finite state projection}
(FSP). This method requires an appropriate truncation of the
state-space to determine the smallest possible set $\sX$ and
development of a computationally feasible algorithm for
calculating the matrix exponential in Eq.~(\ref{eq-IV-2}).

Although a number of methods are available for computing matrix
exponentials [e.g., see~\textcite{MoleLoan:03}], we briefly
discuss here a popular technique known as  \textit{Krylov
subspace approximation} (KSA) method~\cite{Sidj:98,SidjStew:99}.
For a sufficiently small time step $\tau > 0$, this is the best
available method for approximating the vector $\bfp(t+\tau) =
\exp(\tau\P) \: \bfp(t)$, when $\P$ is a large and sparse matrix.
This is done by using a polynomial series expansion of the form:
\beq
\widehat{\bfp}(t+\tau) = c_0 \bfp(t) + c_1 \tau\P \bfp(t)
+ \cdots + c_{\ss K_0-1}(\tau\P)^{K_0-1} \bfp(t), \nn
\label{eq-IV-3}
\eeq
where the coefficients $c_0,c_1,\ldots,c_{\ss K_0-1}$ are estimated
by minimizing the least-squares error $||\bfp(t+\tau)-\widehat{\bfp}
(t+\tau)||_2^2$. It turns out that the optimal $K_0$-th order polynomial
approximation of $\bfp(t+\tau)$ is a point in the $K_0$-dimensional
Krylov subspace $\sK(t) = \text{span} \! \left \{ \bfp(t),
\tau\P\bfp(t),\ldots,(\tau\P)^{K_0-1}\bfp(t) \right \}$. This
element can be approximated by
\beq
\widehat{\bfp}(t+\tau) := || \bfp(t) ||_2 \V(t)
\exp\{\tau \H(t)\} \: \bfe_1 , \nn
\label{eq-IV-4}
\eeq
where $\V(t)$ is a $K \times K_0$ matrix whose columns form an
orthonormal basis for the Krylov subspace $\sK(t)$ and $\H(t)$
is a $K_0 \times K_0$ Hessenberg matrix (upper triangular with
an extra subdiagonal), both computed by the well-known Arnoldi
procedure~\cite{SidjStew:99}. Finally, $\bfe_1$ is the first
column of the $K_0 \times K_0$ identity matrix.

The KSA method reduces the problem of calculating the exponential of a
large and sparse $K \times K$ matrix~$\P$ to the problem of calculating
the exponential of the much smaller and dense $K_0 \times K_0$ matrix
$\H$ ($K_0 \ll K$, with $K_0=30$--$50$ being sufficient for many
applications). Computation of the reduced size problem can be done by
standard methods, such as a Chebyshev or Pad{\' e}
approximation~\cite{Sidj:98,SidjStew:99,MoleLoan:03}. Note that we can
recursively estimate the solution $\bfp(t)$ in Eq.~(\ref{eq-IV-2})
at some time $t_j$ by
\beqa
\widehat{\bfp}(t_j) &\!\!=\!\!& \exp\{(t_j-t_{j-1})\P\} \:
\widehat{\bfp}(t_{j-1}) \nn \\ [4pt]
&\!\!=\!\!& || \widehat{\bfp}(t_{j-1}) ||_2 \V(t_{j-1})
\exp\{(t_j-t_{j-1}) \H(t_{j-1})\}
\bfe_1 , \nn
\label{eq-IV-5}
\eeqa
for $j=1,2,\ldots$, where $\widehat{\bfp}(0)=\bfp(0)$
and $0=t_0 < t_1 < t_2 < \cdots$ is an increasing sequence of (not
necessarily uniformly spaced) time points. These points are selected
automatically, in conjunction with an appropriately designed error
estimation procedure, to ensure stability and accuracy of the overall
algorithm~\cite{Sidj:98}.

Unfortunately, and for most realistic reaction networks, $\sX$ contains
a very large number of states with non-negligible probability, thus
making the practical implementation of FSP difficult. This is a direct
consequence of the fact that $\sX$ contains $R_1 \times R_2 \times
\cdots \times R_N$ distinct elements, where $R_n$ is an assumed
maximum copy number of the $\nth$ species. A number of approaches have been
proposed in the literature to address this problem~\cite{PeleMunsKham:06,%
HeglBurdSant-etal:07,MunsKham:07,DeufHuisJahn-etal:08,HeglHellLots:08,%
JahnHuis:08,MacnBurrSidj:08,WolfGoelMate-etal:10,ZhanWatsCao:10}.
Although some approaches perform well, most are limited
to small reaction networks. It turns out that the most difficult issue
associated with these methods is solving the resulting system of
differential equations, which is usually prohibitively large.

We should point out here that another numerical approach has been
recently proposed in the literature that also attempts to address the
previous problem~\cite{Jahn:10,JahnUdre:10}. The method is based on
representing the probability mass function of the population process
by an appropriately chosen wavelet decomposition scheme whose basis
elements and the associated wavelet coefficients are being adaptively
updated in time by solving a much smaller system of linear equations.
Although preliminary results indicate that the method works well,
it is not clear at this point whether it can be efficiently used to
evaluate population probabilities in reaction networks containing
more than a few reactions and species.

The KSA method is based on several approximations, whose cumulative
effect may appreciably affect its accuracy, numerical stability and
computational efficiency. These drawbacks can be addressed by solving
the master equation~(\ref{eq-II-6}) associated with the DA process,
instead of Eq.~(\ref{eq-II-8}). This leads to a recently developed
numerical technique for solving the master equation known as
\textit{implicit Euler} (IE) method~\cite{JenkGout:12}. Similarly
to the KSA technique, derivation of the IE method starts by
expressing the master equation~(\ref{eq-II-6}) as a linear system
of \textit{coupled} first-order differential equations, given by
\beq
\frac{d\bfq(t)}{dt} = \Q \: \bfq(t), \nn
\label{eq-IV-6}
\eeq
for $t > 0$, where $\bfq(t)$ is a $Q \times 1$ vector that contains the
nonzero probabilities $\pz(\bfz;t)$, $\bfz \in \sZ$, of the DA
process $\bfZ(t)$ and~$\Q$ is a large $Q \times Q$ \textit{sparse}
matrix whose structure can be inferred directly from the master
equation (each column of $\Q$ contains $M+1$ nonzero elements
that sum to zero, where $M \ll Q$ is the number of reactions).
Ordering the elements in $\sZ$ lexicographically results in
a matrix $\Q$ that is lower triangular. As a consequence, and
for a given time step $\tau > 0$, we  can use the implicit Euler
method for solving differential equations~\cite{PresTeukVett-etal:07}
to estimate $\bfq(t)$ at discrete time points $t_j := j\tau$,
$j=1,2,\ldots$. Thus, given an estimate $\widehat{\bfq}(t_{j-1})$
of $\bfq(t_{j-1})$, an estimate $\widehat{\bfq}(t_j)$ of
$\bfq(t_j)$ can be obtained by solving the following system of
linear equations:
\beq
(\I-\tau \Q) \widehat{\bfq}(t_j) = \widehat{\bfq}(t_{j-1}), \nn
\label{eq-IV-7}
\eeq
where $\I$ is the $Q \times Q$ identity matrix. It has been shown
by~\textcite{JenkGout:12} that this is possible for any value of
$\tau$ and can be efficiently done by a standard forward substitution
scheme~\cite{PresTeukVett-etal:07}. Moreover, the resulting method
is always stable, producing a valid probability vector at each
iteration, whereas, its accuracy can be controlled by a single
parameter, the step-size $\tau$. Finally, we can use Eq.~(\ref{eq-II-9})
to obtain an estimate $\widehat{\bfp}(t)$ of the probabilities
$\px(\bfx;t)$ from $\widehat{\bfq}(t)$.

The IE method is computationally superior to KSA when
the cardinality of the state-space $\sZ$ is not appreciably larger
than the cardinality of the state-space $\sX$. This however is not
always possible, since the DAs are non-decreasing, whereas, the
population numbers can either increase or decrease in a way that
their values remain within a fixed and bounded domain. As a
consequence, this method can only be used when the number
of reaction events are sufficiently constrained or remain small
during a time interval of interest. The IE method has been used
by~\textcite{JenkGout:12} to numerically approximate the solution
of the SIR epidemic model discussed in Section~III-C with remarkable
success compared to the KSA method. In this case, the nullity of
the stoichiometric matrix $\S$ is zero and, therefore, there is
one-to-one correspondence between $\bfZ(t)$ and $\bfX(t)$, which
implies that the state-spaces $\sZ$ and $\sX$ are isomorphic.

\vspace{-12pt}

\subsection{Monte Carlo estimation}

\vspace{-6pt}

Numerical approaches for solving the master equation are not practical
when the reaction network contains many reactions and species. In this
case, Monte Carlo sampling~\cite{Liu:01} can be used to evaluate the
statistical behavior of the network. If, by simulation, we generate
$L$ sample trajectories $\{\bfz^{(l)}(t),t >0 \}$, $l=1,2,\ldots,L$,
of the DA process $\{\bfZ(t), t >0 \}$, then we can estimate the
dynamics of its moments, such as of the means $\{\muz(m;t) :=
\E[Z_m(t)], t >0 \}$ and covariances $\{\cz(m,m';t) :=
\cov[Z_m(t),Z_{m'}(t)], t >0 \}$, by using the following Monte
Carlo estimators:
\beqa
\hatmuz(m;t) &\!\!=\!\!& \frac{1}{L} \sum_{l=1}^L z^{(l)}_m(t), \nn \\ [6pt]
&& \hspace{-50pt} \hatcz(m,m';t) \nn \\
&& \hspace{-40pt} = \frac{1}{L-1} \sum_{l=1}^L
\left [ z^{(l)}_m(t) - \hatmuz(m;t) \right ] \left [ z^{(l)}_{m'}(t)
- \hatmuz(m';t) \right ] \! . \nn
\label{eq-IV-8}
\eeqa
Moreover, we can estimate the probability distribution $\pz(\bfz;t)$
by using
\beq
\widehat{p}_{\ss \bfZ}(\bfz;t) = \frac{1}{L} \sum_{l=1}^L \Delta
(\bfz^{(l)}(t)-\bfz), \nn
\label{eq-IV-9}
\eeq
for $t >0$, where $\Delta(\bfz)$ is the Kronecker delta
function. Due to the simple relationship between the DA and
population processes given by Eq.~(\ref{eq-II-4}), we can use
similar estimators to approximate the dynamic evolution of the
corresponding population statistics.

Unfortunately, to obtain sufficiently accurate Monte Carlo estimates,
we need a large number of sample trajectories, which is computationally
inefficient, especially when estimating high-order moments or
probability distributions.\footnote{When estimating probability
distributions, the issue of efficiently sampling low probability
events is crucial and becomes the main bottleneck for
deriving accurate and computationally efficient Monte Carlo
estimators.} This problem can be addressed by developing
computationally efficient approaches for sampling the master
equation~(\ref{eq-II-6}). In the following, we discuss a number
of methods available in the literature.

\vspace{-12pt}

\subsubsection{Exact sampling}

\vspace{-6pt}

The simplest way to draw samples from the master equation (\ref{eq-II-6})
is by using the Gillespie algorithm~\cite{Gill:76,Gill:77,Gill:92}.
This method can generate a trajectory $\{\bfz(t), t >0\}$
of the DA process by following two steps. First, given that the system
is at state $\bfz(t)$ at time $t$, the time $t+\tau$ of the next
reaction to occur can be determined by drawing a sample~$\tau$ from
the exponential distribution:
\beq
e_t(\tau) = \left \{ \sum_{m \in \sM} \!\! \alpha_m(\bfz(t)) \right \}
\exp \! \left \{ -\tau \!\!\! \sum_{m \in \sM} \!\! \alpha_m(\bfz(t)) \right \},
\label{eq-IV-10}
\eeq
for $\tau > 0$. Then, which reaction occurs at time $t+\tau$ can be specified
by drawing a sample from the probability mass function
\beq
r_t(m) = \frac{\alpha_m(\bfz(t))}{{\displaystyle
\sum_{m' \in \sM} \!\! \alpha_{m'}(\bfz(t))}} \:,
\label{eq-IV-11}
\eeq
and by increasing the corresponding value of $\bfz$ by one.

Unfortunately, this algorithm is computationally demanding,
especially when applied to large and highly reactive systems,
due to the fact that every single reaction event must be
simulated. Attempts by \textcite{GibsBruc:00},
\textcite{CaoLiPetz:04}, and \textcite{CollPeteCox-etal:06}
to improve the computational efficiency of the Gillespie
algorithm have produced sampling methods that significantly
increase computational speed for large reaction networks.
However, despite these efforts, the previous methods are
still inefficient, especially when used in conjunction with Monte
Carlo estimation. For this reason, work has focused on developing
sampling techniques that appreciably reduce computational complexity
by trading-off accuracy. We discuss some of these methods next.

\vspace{-12pt}

\subsubsection{Langevin approximation}

\vspace{-6pt}

We can obtain a useful approximation to the master equation~(\ref{eq-II-6})
by assuming that there exists a time step~$\tau$ such that, for
every time $t$, two conditions are satisfied: (a)~occurrence of
reactions within the time interval $[t,t+\tau)$ does not appreciably
affect the propensity functions $\alpha_m(\bfz(t))$, $m \in \sM$,
and~(b)~the expected number of occurrences of each reaction during
$[t,t+\tau)$ is much larger than one. In this case, we can approximate
the DA process $\bfZ(t)$ by another process $\widehat{\bfZ}(t)$ that
satisfies the following equations~\cite{Gill:00,Gill:76,Gill:77,Gill:92}:
\beq
\widehat{Z}_m((j+1)\tau) \!=\! \widehat{Z}_m(j\tau)
+ \alpha_m(\widehat{\bfZ}(j \tau)) \tau
+ \sqrt{\alpha_m(\widehat{\bfZ}(j\tau)) \tau} \: G_m,
\label{eq-IV-12}
\eeq
for $j=0,1,\ldots, \quad m \in \sM$, initialized by
$\widehat{Z}_m(0)=0$, for every $m \in \sM$, where
$\{G_m, m \in \sM\}$ are mutually independent standard
normal random variables that are statistically independent
of $\widehat{\bfZ}$.

We can use Eq.~(\ref{eq-IV-12}) to approximately sample
the master equation in an iterative fashion. Starting with
zero DA values at time zero, we can approximate the DA
process at time $\tau$ by setting $\widehat{z}_m(\tau)
= \alpha_m({\bf0}) \tau + \sqrt{\alpha_m({\bf0}) \tau}
\: g^{(0)}_m$, for every $m \in \sM$, where $g^{(0)}_m$,
$m \in \sM$, are samples independently drawn from the
standard normal distribution. Then, we can approximate
the DA process at time $2\tau$ by setting $\widehat{z}_m(2\tau)
= \widehat{z}_m(\tau)+\alpha_m({\widehat{\bfz}(\tau)}) \tau
+ \sqrt{\alpha_m(\widehat{\bfz}(\tau))\tau} \: g^{(1)}_m$,
for every $m \in \sM$, where $g^{(1)}_m$, $m \in \sM$, are
new samples independently drawn from the standard normal
distribution, and so on.

Unfortunately, the previous method may result in crude
approximations of the DA and population processes~\cite{Gout:06}.
The main culprit is our difficulty in determining an
appropriate time step $\tau$ so that the two required
conditions mentioned above are \textit{simultaneously}
satisfied. For example, we may try to reduce~$\tau$ so
that the propensity functions do not change appreciably
during any time interval $[t,t+\tau)$, thus satisfying
the first condition. However, if the reaction network
contains ``slow'' reactions (a situation that happens
often in practice), these reactions will occur infrequently
during the time interval $[t,t+\tau)$, which will result
in violating the second condition. Finally, the method
may produce reaction occurrences within a time interval
$[j\tau,(j+1)\tau)$ that may result in negative species
populations [see also the discussion by~\textcite{Mely:10},
pp.\ 65-71], which may not be appropriate in certain types
of networks (e.g., in biochemical reaction networks).

It is worthwhile noticing here that, in the limit as
$\tau \rightarrow 0^+$, Eq.~(\ref{eq-IV-12}) converges
to the following \textit{Langevin equations}~\cite{Gill:96,Gill:00}:
\beq
d \widehat{Z}_m(t) = \alpha_m(\widehat{\bfZ}(t))dt
+ \sqrt {\alpha_m(\widehat{\bfZ}(t))} \: d W_m(t),
\label{eq-IV-13}
\eeq
for $t >0$, $m \in \sM$, where $\{W_m, m \in \sM\}$ are
mutually independent standard Brownian motions whose
increments $\{d W_m(t),m \in \sM\}$ at time $t$ are also
independent of $\widehat{\bfZ}(t)$, which can be used to
approximate the master equation~(\ref{eq-II-6}). Note
that Eq.~(\ref{eq-IV-12}) provides a numerical method
for solving the Langevin equations, obtained by discretizing
Eq.~(\ref{eq-IV-13}) using the well-known Euler-Maruyama
method~\cite{High:01}. For this reason, the approximation
method based on Eq.~(\ref{eq-IV-12}), is usually referred
to as the \emph{Langevin approximation} (LA) method. Note that,
when using Eq.~(\ref{eq-IV-12}), we must make sure that $\tau$ is
small enough so that we obtain a good approximation to the
time-continuous DA process $\widehat{\bfZ}(t)$ governed
by the Langevin equations.

Finally, \textcite{MelyBurrZyga:10} has recently shown that
there are many alternative ways to formulate the Langevin
equations, which result in the \emph{same} finite-dimensional
joint probability distribution for the underlying population
variables. It turns out that using one particular formulation
can considerably accelerate implementation of Monte Carlo
estimation. Despite this advantage, and for the reasons
discussed above, caution should be exercised when replacing
the master equation with the Langevin equations.

\vspace{-12pt}

\subsubsection{Poisson approximation}

\vspace{-6pt}

The DA process satisfies the following equation~\cite{Kurt:80}:
\beq
Z_m(t) = P_m \! \left [ \int_0^t \!\! \alpha_m(\bfZ(t')) \: dt'
\right ] \! , \nn
\label{eq-IV-14}
\eeq
for $t >0$, $m \in \sM$, where $P_m$, $m \in \sM$, are statistically
independent Poisson random variables with unit rate. As a consequence
of the Markovian nature of the process, we also have that
\beq
Z_m(t+\tau) = Z_m(t) + P_m \!\! \left [ \int_t^{t+\tau}
\!\!\!\!\!\!\!\! \alpha_m(\bfZ(t')) \: dt' \right ] \! ,
\label{eq-IV-15}
\eeq
for $t >0$, $m \in \sM$. This result can be used to construct
a better technique than the LA method for
approximately sampling the master equation. In particular,
we can employ a time step~$\tau$ so that occurrence of
reactions within the time interval $[t,t+\tau)$ does not
appreciably affect the propensity functions $\alpha_m(\bfz(t))$,
$m \in \sM$. This is the first condition required by the
LA method, which is commonly referred to as the
\textit{leap condition}. In this case, given that $\bfZ(t)
= \bfz(t)$, the number of occurrences of the $\mth$ reaction
within the time interval $[t,t+\tau)$ will approximately
follow a Poisson distribution with mean and variance
$\alpha_m(\bfz(t))\tau$. As a consequence, Eq.~(\ref{eq-IV-15})
becomes
\beq
\widehat{Z}_m((j+1)\tau) = \widehat{Z}_m(j\tau)
+ P_m \! \left [ \alpha_m(\widehat{\bfZ}(j \tau)) \tau \right ] \! ,
\label{eq-IV-16}
\eeq
for $j=0,1,\ldots, \quad m \in \sM$, initialized by $\widehat{Z}_m(0)=0$,
for every $m \in \sM$. The resulting method is usually referred
to as the \emph{Poisson approximation} (PA) method.

By using Eq.~(\ref{eq-IV-16}), we expect to obtain accurate samples
of the DA process, provided that we choose a time step~$\tau$ that
sufficiently satisfies the leap condition. Hence, an important
practical problem here is to determine an appropriate value for
$\tau$ so that the leaping condition is approximately satisfied.
We would like this value to be as large as possible so that the
resulting method is appreciably faster than exact sampling using
the Gillespie algorithm. Practical considerations however dictate
that $\tau$ must not be very large, otherwise the method may
produce reaction occurrences within a time interval
$[j\tau,(j+1)\tau)$ that may result in negative species
populations, which is the same problem as the one
encountered when using the LA method.

The problem of determining the largest value of $\tau$ so that
the leap condition is satisfied has been addressed by~\textcite{Gill:01},
\textcite{GillPetz:03}, and \textcite{CaoGillPetz:06}. The latest
procedure is accurate, easy to code, and results in faster
implementation than the previous methods. To avoid negative
populations, it has been suggested by~\textcite{TianBurr:04}
and \textcite{ChatMayaEdwa-etal:05,ChatVlacKats:05} to
approximate the Poisson distribution by a binomial distribution.
The main rationale behind this choice is that the maximum number
of occurrences produced by a binomial distribution is always
bounded and easily controlled by one of the two parameters used
to specify the distribution. This however is not true for the
Poisson distribution, which can produce a very large number of
occurrences within a small time interval (a Poisson random
variable takes values between $0$ and $\infty$) that can falsely
result in negative populations. Some improvements of the original
$\tau$-leaping methods can be found in~\textcite{PengZhouWang:07}
and~\textcite{PettResa:07}.

It turns out that we can still use a Poisson distribution for the
occurrence of reactions and always guarantee nonnegative populations.
This has been recognized by~\textcite{CaoGillPetz:05b}, who proposed a
sampling method that is easier to implement than the binomial
$\tau$-leaping algorithm and is more accurate in general than
the original Poisson $\tau$-leaping technique. An improved version of
this approach, which employs a post-leap check to improve sampling
accuracy, has been proposed by~\textcite{Ande:08}.

Most $\tau$-leaping sampling methods available in the literature
require specification of the mean occurrence of a reaction during a
leap step. The value used is usually not the true mean value and,
as a result, a bias is introduced that reduces the accuracy and
speed of sampling. This problem has been addressed
in~\textcite{XuCai:08} by specifying an appropriate value for
the mean occurrence rate obtained directly from the master equation.

Finally, we refer the reader to~\textcite{CaiXu:07}, \textcite{Lips:07},
\textcite{Hell:08}, \textcite{SlepThomPlim:08}, \textcite{CaiWen:09},
\textcite{MjolOrenChat-etal:09}, \textcite{RamaSegrSbal:09}, and
\textcite{WuFuCao-etal:11}, for alternative simulation algorithms
designed to accelerate exact Monte Carlo sampling of the master
equation under certain conditions, as well as to~\textcite{LuVolfTsim-etal:04},
\textcite{Ande:07}, \textcite{Cai:07}, \textcite{RamaSbal:11},
and \textcite{YiZhuaDa-etal:12} for methods dealing with
time-varying propensity functions and delays.

\vspace{-12pt}

\subsubsection{Weighted sampling}

\vspace{-6pt}

We can also use Monte Carlo sampling to estimate the probability
of an event $\sE$, where $\sE$ is the collection of all
trajectories sampled from the master equation that satisfy a specific
condition of interest (e.g., that the population $X_n(t)$ of the $\nth$
species exceeds a given threshold during a time interval $[0,t_0]$).
If $T_l=\{\bfz_l(t), t >0 \}$, $l=1,2,\ldots,L$, are the
trajectories obtained by sampling the master equation~(\ref{eq-II-6}),
then we can estimate the probability of an event $\sE$ by employing
the following Monte Carlo estimator:
\beq
\widehat{\Pr}[\sE] = \frac{1}{L} \sum_{l=1}^L \: [T_l \in \sE],
\label{eq-IV-17}
\eeq
where $[\: \cdot \:]$ is the Iverson bracket.

To produce a sufficiently accurate probability estimate when using
Eq.~(\ref{eq-IV-17}), we may need to use a prohibitively
large number of samples, especially when $\sE$ is a rare event
(i.e., when $\Pr[\sE] \ll 1$). Rare events are of particular
interest, since they may produce a catastrophic behavior in
reaction networks, such as the onset of cancer in biochemical
networks or mass population causalities in epidemiological
networks. When $\sE$ represents a rare event, most trajectories
sampled from the master equation will not be in $\sE$ and,
therefore, will not contribute to the summation in
Eq.~(\ref{eq-IV-17}). In this case, we need to appreciably
increase the value of $L$ in order to accurately estimate
the probability $\Pr[\sE]$.

We can remedy this situation by employing importance sampling~\cite{Liu:01},
a classical method for reducing the variance of a Monte Carlo estimator
and, thus, $L$. Importance sampling is based on generating samples drawn
from a probability distribution which assigns more probability mass to
trajectories that satisfy the desired condition and less probability
mass to the remaining trajectories. This approach has been recently
employed by~\textcite{KuwaMura:08} to estimate rare event probabilities
in stochastic chemical kinetics and has led to the development and
refinement of an innovative approach for sampling the master equation,
known as \textit{weighted sampling}~\cite{KuwaMura:08,GillRohPetz:09,%
RohGillPetz:10,DaigRohGill-etal:11}.

Weighted sampling is based on defining a new set $\{\alpha^\prime_m(\bfz),
m \in \sM\}$ of propensity functions, given by $\alpha^\prime_m(\bfz)
= \lambda_m \alpha_m(\bfz)$, where $\lambda_m, m \in \sM$, are
appropriately chosen positive constants so that sampling the master equation
with propensity functions $\alpha^\prime$ produces trajectories
$T^\prime_l$, $l=1,2,\ldots,L^\prime$, which are in $\sE$ with
high probability.\footnote{Choosing these values requires a great
deal of intuition about the behavior of the reaction system or
advanced algorithmic techniques, such as those discussed
in~\textcite{DaigRohGill-etal:11}.} In this case, the Monte Carlo
estimator for the probability of event $\sE$ to occur will be given by
\beq
\widehat{\Pr}[\sE] = \frac{1}{L'} \sum_{l=1}^{L^\prime} \:
w_l (T'_l) [T'_l \in \sE] , \nn
\label{eq-IV-18}
\eeq
where $w_l(T'_l)$, $l=1,2,\ldots,L'$, are weights that account for
the bias introduced by sampling the master equation with propensity
functions $\alpha^\prime$ instead of $\alpha$.

To compute the weights, note that a trajectory $T'_l$ can be specified
as $T'_l=\{\tau_1,m_1,\ldots,\tau_{K_l},m_{K_l}\}$, where $m_1,
\ldots,m_{K_l}$ are the $K_l$ reactions that occur within the time
interval of interest $[0,t_0]$ and $\tau_1,\ldots, \tau_{K_l}$ are the
time steps leading to these reactions. Then, the probability of
sampling a trajectory $T'_l=\{\bfz'_l(t), t \in [0,t_0]\}$ from the
master equation with propensity functions $\alpha'$ is given by
\beq
\Pr [T'_l] = \prod_{i=1}^{K_l} \alpha'_{m_i}(\bfz'_l(t_i))
\exp \! \left \{ -\tau_i \! \sum_{m \in \sM} \alpha'_m(\bfz'_l(t_i))
\right \} , \nn
\label{eq-IV-19}
\eeq
by virtue of Eqs.~(\ref{eq-IV-10}) and~(\ref{eq-IV-11}), where
$t_i=\sum_{k=1}^i \tau_k$. It turns out that the weight of each biased
trajectory must be equal to the ratio of the probability that the
trajectory was sampled from the master equation with propensity
functions~$\alpha$ to the probability that it was sampled from the master
equation with propensity functions~$\alpha^\prime$. As a consequence,
\beq
w_l (T'_l) = \prod_{i=1}^{K_l} \frac{1}{\lambda_{m_i}} \exp \left \{ \tau_i
\! \left[ \sum_{m \in \sM} \!\! \left ( 1 - \frac{1}{\lambda_m} \right )
\! \alpha'_m(\bfz'_l(t_i)) \right] \right \} \! . \nn
\label{eq-IV-20}
\eeq
In this way, the weighted sampling algorithm can be used to compute
intractable rare event probabilities using the previously discussed
(exact and approximate) sampling techniques, since accurate estimation
of such probabilities usually requires $L^\prime \ll L$ number of
sampled trajectories.

\vspace{-12pt}

\subsubsection{Maximum entropy approximation}

\vspace{-6pt}

As we mentioned before, estimating the probability distributions
$\pz(\bfz;t)$ and $\px(\bfx;t)$ by sampling the master equation
can be computationally demanding and in most cases intractable.
Depending on available data, the size of the reaction network
at hand, and available computational resources, it may only be
possible to accurately estimate the first few moments $\E[X^k_n(t)]$,
$k=1,2,\ldots,K$, of the population process $X_n(t)$. In this case,
by invoking the principle of \textit{maximum entropy}~(MaxEnt), we
may be able to approximately derive an analytical form for the
\textit{marginal} probability distribution $\px(x_n;t) :=
\sum_{x_1,\ldots,x_{n-1},x_{n+1},\ldots,x_N} \px(\bfx;t)$. As a
matter of fact, using MaxEnt to determine an appropriate distribution
compatible with given moment information has produced surprisingly
good results in many diverse scientific disciplines.

The principle of maximum entropy states that an appropriate
approximation of the true-but-unknown distribution of $X_n(t)$
is the probability distribution $\widehat{p}_{\ss \bfX}(x_n;t)$
that maximizes the Shannon entropy
\beq
S(\px;t) = - \! \sum_{x_n} \px(x_n;t) \ln \px(x_n;t), \nn
\label{eq-IV-21}
\eeq
subject to known information about $X_n(t)$ [e.g., knowledge of the
support of $\px(x_n;t)$ and of the moments $\E[X^k_n(t)]$, $k=1,2,
\ldots,K$]~\cite{MeadPapa:84,Kapu:90}. This approach is based on a
well-known principle of scientific objectivity that leads us to choose
the probability distribution, out of all distributions consistent
with the given information, which maximizes our uncertainly (Shannon
entropy) about the true distribution. Given the moments $\E[X^k_n(t)]$,
$k=1,2,\ldots,K$, and the fact that $x_n$ is a nonnegative
integer-valued variable, we can show that $\widehat{p}_{\ss \bfX}(x_n;t)$
is a univariate \textit{Gibbs} distribution of the form:
\beq
\widehat{p}_{\ss \bfX}(x_n;t) = \frac{1}{\zeta(t)} \exp \left \{
- \sum_{k=1}^K \lambda_k(t) x^k_n \right \}, \nn
\label{eq-IV-22}
\eeq
for $x_n \geq 0$, $t > 0$, where the partition function $\zeta(t)$
is defined by
\beq
\zeta(t) := \sum_{u_n} \exp \left \{ - \sum_{k=1}^K
\lambda_k(t) u^k_n \right \} . \nn
\label{eq-IV-23}
\eeq
The values of parameters $\lambda_k(t)$, $k=1,2,\ldots,K$, must be chosen
so that
\beq
\sum_{x_n} x^k_n  \:\! \widehat{p}_{\ss \bfX}(x_n;t)
= \hatmux^{(k)}(n;t), \nn
\label{eq-IV-24}
\eeq
for $t > 0$, $k=1,2,\ldots,K$, where $\hatmux^{(k)}(n;t)$ is the
value of the $\kth$ moment of $X_n(t)$ obtained by Monte Carlo
sampling of the master equation~(\ref{eq-II-8}) or estimated
from available data. When only an estimate $\hatmux^{(1)}(n;t)$
of the mean of the population process $X_n(t)$ is available, the
MaxEnt approximation of $p_{\ss \bfX}(x_n;t)$ is a \textit{geometric}
distribution, given by

\newpage

$~~~~~~~~~~~~~~~~~~~~~~~$

\vspace{-30pt}

\beq
\widehat{p}_{\ss \bfX}(x_n;t) = \left [ \frac{1}{1+\hatmux^{(1)}(n;t)}
\right ] \left [ \frac{\hatmux^{(1)}(n;t)}
{1+\hatmux^{(1)}(n;t)} \right ]^{x_n} \!\!\!\!\!\!\!,
\label{eq-IV-25}
\eeq
for $x_n \geq 0$, $t > 0$. On the other hand, when only estimates
of the first two moments of the population process $X_n(t)$ are
available, the MaxEnt approximation of $p_{\ss \bfX}(x_n;t)$ is
a quadratic \textit{Gibbs} distribution, given by
\beqa
\widehat{p}_{\ss \bfX}(x_n;t) &\!\!\!=\!\!\!& \left ( \sum_{u \geq 0} \exp {\Bigl \{} \!
- \lambda_1(t) u - \lambda_2(t) u^2 {\Bigr \}} \right )^{\!\! -1} \nn \\
&& \!\!\!\! \times \exp {\Bigl \{} \! - \lambda_1(t) x_n -
\lambda_2(t) x^2_n {\Bigr \}}, \nn
\label{eq-IV-26}
\eeqa
for $x_n \geq 0$, $t > 0$. In this case however we need to
specify the values of the parameters $\lambda_1(t)$ and
$\lambda_2(t)$ so that $\widehat{p}_{\ss \bfX}(x_n;t)$
satisfies the underlying constraints imposed by knowing
the first two moments. Although it is not possible to
specify these parameters analytically, a number of numerical
methods, such as the method proposed by~\textcite{BandBhatBisw-etal:05},
can be used to address this problem [see also \textcite{Djaf:91}].

We can extend MaxEnt to deal with  multivariate marginal probability
distributions, such as $\px(x_n,x_{n'};t)$. Determining the MaxEnt
distribution however becomes increasingly difficult as the dimensionality
of the probability distribution increases~\cite{Abra:10}. Another problem
associated with MaxEnt is that the method can produce a probability
distribution that falsely assigns non-negligible probability mass
over population values that are not stoichiometrically possible [i.e.,
values that do not satisfy Eq.~(\ref{eq-II-4})]. We may attempt to
address this problem by calculating an approximation
$\widehat{p}_{\ss \bfZ}(\bfz;t)$ of the joint probability distribution
$\pz(\bfz;t)$ using MaxEnt and by then estimating $\px(\bfx;t)$ from
Eq.~(\ref{eq-II-9}) by replacing $p_{\ss \bfZ}(\bfz;t)$ with
$\widehat{p}_{\ss \bfZ}(\bfz;t)$. This approach however is only
feasible in the case of small reaction networks that contain
very few reactions so that estimation of $\pz(\bfz;t)$ by
MaxEnt is possible.

\vspace{-12pt}

\subsubsection{Stiffness}

\vspace{-6pt}

In Markovian reaction networks, the firing rates of the underlying
reactions may vary widely. In this case, most computational effort
associated with the previous Monte Carlo methods will be spent on
faithfully simulating the firings of fast reactions (i.e., reactions
with large propensity values), even if simulation of such reactions
may not be important for determining a particular system behavior of
interest. This leads to \textit{stiffness}, a serious computational
problem that results in inefficiently sampling the master equation.

To address stiffness, \textcite{RathPetzCao-etal:03} have proposed a
modified version of the $\tau$-leaping method, known as \textit{implicit}
$\tau$-leaping, that allows larger $\tau$ values to be used when
applied to stiff reaction networks than the original $\tau$-leaping
method (which must use a small step-size in this case). Subsequently,
\textcite{CaoGillpetz:07} proposed an \textit{adaptive} method that
identifies stiffness at each simulation step and automatically
chooses between the standard and implicit $\tau$-leaping methods.
Moreover, it provides an appropriate value for $\tau$ to be used
during each iteration.

Although both approaches can appreciably decrease simulation time
when compared to the standard $\tau$-leaping method, they may
excessively damp fluctuations. As a consequence, these methods
may underestimate the population variances and produce stochastic
dynamics that evolve tightly around their means. This may not be
accurate, especially when dominant stochastic fluctuations are
present in the system. To ameliorate this problem,
\textcite{RathPetzCao-etal:03} have proposed a strategy that
attempts to restore overly damped fluctuations. It is not clear
however whether this strategy performs well when used in more
complex reaction networks than the simple networks considered by
the investigators.

Another problem associated with the previous techniques is their
difficulty in effectively dealing with very small populations of
species involved in very fast reactions. Moreover, these methods
may lead to non-integer and possibly negative populations, which
may not be physically meaningful in certain types of networks
(e.g., in biochemical reaction networks). Although a
stoichiometrically consistent rounding step has been proposed
by~\textcite{RathPetzCao-etal:03} to remedy the last problem,
it has been observed that rounding may seriously impair the
performance of the resulting algorithm~\cite{RathSama:07}. To
address this issue, \textcite{RathSama:07} have proposed a method
based on decomposing a Markovian reaction network into ``motifs''
and on constructing appropriate approximations for each individual
motif. This method however is cumbersome and difficult to use, since
its effectiveness relies heavily on identifying appropriate ``motifs,''
a task that may not be possible in large reaction networks.

Finally, a ``partitioned leaping'' approach has been proposed
by~\textcite{HarrClan:06} [see also~\textcite{HarrPiccMaju-etal:09}]
that is closely related to $\tau$-leaping. At each step, the
algorithm uses the expected number of firings of a reaction
within a calculated step-size~$\tau$ to classify the reaction
into four distinct categories: very slow, slow, medium, and fast.
Based on this classification, the simulation of very slow
reactions proceeds using exact sampling. On the other hand, slow
and medium reactions are simulated using the Poisson and Langevin
approximations, respectively. Finally, the number of firings of a
fast reaction is specified deterministically by multiplying the
propensity function of the reaction with $\tau$.

``Partitioned leaping'' is an attractive idea for speeding-up
Monte Carlo sampling of the master equation. Its accuracy however
depends on correctly classifying the reactions, which may not
be always possible. Although the simple examples provided
by~\textcite{HarrClan:06} and \textcite{HarrPiccMaju-etal:09}
demonstrate its effectiveness for reducing computations while
preserving accuracy, it is not clear how the method will perform
when applied on larger and more complex networks with widely
disparate reaction rates and how robust the method is to
possible misclassification of reactions.

Stiffness in Markovian reaction networks is an important problem,
directly related to the practical use of these networks.
Unfortunately, no sufficient solution has been proposed to date a
nd more research is needed to satisfactorily address this problem.
We will revisit stiffness in Section~V when we discuss multiscale
approximations to the master equation.

\vspace{-12pt}

\subsection{Moment approximation}

\vspace{-6pt}

When a reaction network contains many species and reactions, it may
not be possible to accurately estimate, in a reasonable time, the
statistical behavior of the DA and population processes by Monte
Carlo sampling. In this case, we may try an alternative technique
known as \textit{moment closure}. This method can be derived by
setting
\beq
Z_m(t) = \muz(m;t) + W_m(t),
\label{eq-IV-27}
\eeq
for $t > 0$, $m \in \sM$, where $\muz(m;t)$ is the mean of $Z_m(t)$
and $W_m(t) := Z_m(t) - \muz(m;t)$. Note that $W_m(t)$ is additive
zero-mean noise that quantifies fluctuations of the DA process
around its mean. By using the master equation~(\ref{eq-II-6}),
we can show that the means $\muz(m;t)$ and covariances $\cz(m,m';t) :=
\cov[Z_{m}(t),Z_{m'}(t)]$ $=$ $\E[W_{m}(t) W_{m'}(t)]$ of the DA
process satisfy the following system of first-order differential
equations:
\beq
\frac{d\muz(m;t)}{dt} = \E {\bigl [} \alpha_m(\bfZ(t))
{\bigr ]} ,
\label{eq-IV-28}
\eeq
for $t >0$, $m \in \sM$, and
\beqa
\frac{d\cz(m,m';t)}{dt} &\!\!=\!\!& \E {\bigl [}
\alpha_m(\bfZ(t)) {\bigr ]} \Delta(m-m')
~~~~~~~~~~~~~~~~~~~~~~ \nn \\ [4pt]
&& + \: \E {\bigl [} [Z_m(t) - \muz(m;t)] \:
\alpha_{m'}(\bfZ(t)) {\bigr ]} \nn \\ [6pt]
&& + \: \E {\bigl [} [Z_{m'}(t) - \muz(m';t)] \:
\alpha_m(\bfZ(t)) {\bigr ]} ,
\label{eq-IV-29}
\eeqa
for $t >0$, $m,m' \in \sM$, where $\Delta$ is the Kronecker delta. Note that the derivatives
$d\muz(m;t)/dt$ and $d\cz(m,m';t)/dt$ always exist at finite times,
since the master equation~(\ref{eq-II-6}) is valid only when the
probability mass function of the DA process is a continuous function
of $t$. This implies that the means and covariances will also be
continuous in $t$ and, thus, differentiable.

In general, we cannot derive an exact solution to the previous equations,
unless we employ some approximation. Since $\bfZ(t) = \bfmuz(t)
+ \bfW(t)$, we can use the following Taylor series expansion of
$\alpha_m(\bfZ(t))$ around the mean value $\bfmuz(t)$:

\newpage

$~~~~~~~~~~~~~~~~~~~~~~~$

\vspace{-30pt}

\beqa
\alpha_m (\bfZ(t))
&\!\!\simeq\!\!& \alpha_m (\bfmuz(t)) \: + \!\! \sum_{m_1 \in \sM} \!\!
h_{m,m_1}(\bfmuz(t)) \: W_{m_1}(t) \nn \\
&& \hspace{-25pt} + \frac{1}{2} \!\! \sum_{m_1 \in \sM} \sum_{m_2 \in \sM} \!\!
h_{m,m_1,m_2}(\bfmuz(t)) \: W_{m_1}(t)  W_{m_2}(t) \nn \\ [6pt]
&& \hspace{-25pt} + \frac{1}{6} \!\! \sum_{m_1 \in \sM} \sum_{m_2 \in \sM}
\sum_{m_3 \in \sM} \!\! h_{m,m_1,m_2,m_3}(\bfmuz(t)) \nn \\
&& \hspace{50pt} \times W_{m_1}(t) W_{m_2}(t)  W_{m_3}(t) ,
\label{eq-IV-30}
\eeqa
where $h_{m,m_1}(\bfmuz(t))$ is the first-order derivative of
$\alpha_m(\bfz)$ at the mean value $\bfmuz(t)$, whereas
$h_{m,m_1,m_2}(\bfmuz(t))$ and $h_{m,m_1,m_2,m_3}(\bfmuz(t))$ are the
second- and third-order derivatives, respectively.\footnote{For simplicity,
we assume here that the propensity functions are sufficiently smooth so
that the derivatives of order $\geq 4$ are all negligible. This condition
is satisfied in many cases of interest.} Then,
Eqs.~(\ref{eq-IV-28})--(\ref{eq-IV-30}) imply that
\beqa
&& \hspace{-35pt} \frac{d\muz(m;t)}{dt} \simeq \alpha_m(\bfmuz(t)) \nn \\
&& \hspace{-12pt} +  \frac{1}{2}
\!\! \sum_{m_1 \in \sM} \sum_{m_2 \in \sM} \!\! h_{m,m_1,m_2}(\bfmuz(t)) \:
\cz(m_1,m_2;t) \nn \\ [6pt]
&& \hspace{-12pt} + \frac{1}{6} \!\! \sum_{m_1 \in \sM} \sum_{m_2 \in \sM}
\sum_{m_3 \in \sM} h_{m,m_1,m_2,m_3}(\bfmuz(t)) \nn \\
&& \hspace{80pt} \times \: \cz(m_1,m_2,m_3;t),
\label{eq-IV-31}
\eeqa
for $t >0$, $m \in \sM$, and
\beqa
&& \hspace{-3mm}
\frac{d\cz(m,m';t)}{dt} \simeq \alpha_m (\bfmuz(t)) \Delta(m-m') \nn \\ [6pt]
&& + \!\!\! \sum_{m_1 \in \sM} {\Bigl [} h_{m',m_1}(\bfmuz(t)) \: \cz(m,m_1;t) \nn \\ [-9pt]
&& \hspace{20mm} + \: h_{m,m_1} (\bfmuz(t)) \: \cz(m',m_1;t) {\Bigr ]} \nn \\
&& + \frac{1}{2} {\Bigl [} \!\! \sum_{m_1 \in \sM} \sum_{m_2 \in \sM}
h_{m,m_1,m_2}(\bfmuz(t)) \: \cz(m_1,m_2;t) {\Bigr ]} \nn \\ [-6pt]
&& \hspace{30mm} \times \: \Delta(m-m') \nn \\ [3pt]
&& + \frac{1}{2} \!\!\! \sum_{m_1 \in \sM} \sum_{m_2 \in \sM} {\Bigl [}
h_{m',m_1,m_2}(\bfmuz(t)) \: \cz(m,m_1,m_2;t) \nn \\ [-9pt]
&& \hspace{25mm} + \: h_{m,m_1,m_2} (\bfmuz(t)) \: \cz(m',m_1,m_2;t) {\Bigr ]} \nn \\ [6pt]
&& + \frac{1}{6} {\Bigl [} \!\!\! \sum_{m_1 \in \sM}
\sum_{m_2 \in \sM} \sum_{m_3 \in \sM} \!\!\! h_{m,m_1,m_2,m_3}(\bfmuz(t))  \nn \\ [-6pt]
&& \hspace{30mm} \times \: \cz(m_1,m_2,m_3;t) {\Bigr ]} \Delta(m-m') \nn \\ [3pt]
&& + \frac{1}{6} \! \sum_{m_1 \in \sM} \sum_{m_2 \in \sM} \sum_{m_3 \in \sM}
{\Bigl [} h_{m',m_1,m_2,m_3}(\bfmuz(t)) \nn \\ [-6pt]
&& \hspace{40mm} \times \: \cz(m,m_1,m_2,m_3;t) \nn \\
&& \hspace{6mm} + \: h_{m,m_1,m_2,m_3}(\bfmuz(t)) \: \cz(m',m_1,m_2,m_3;t) {\Bigr ]},
\label{eq-IV-32}
\eeqa
for $t >0$, $m,m' \in \sM$, where $\cz(m_1,m_2,m_3;t)$ and
$\cz(m_1,m_2,m_3,m_4;t)$ are the third- and fourth-order central
moments of $\bfZ(t)$, respectively. Eqs.~(\ref{eq-IV-31})
and~(\ref{eq-IV-32}) show that the mean and covariance dynamics of
the DA process $\bfZ(t)$ are in general governed by a system of
\textit{coupled} first-order differential equations driven by the
third- and fourth-order central moments. Moreover, the dependency
of these equations on propensity function derivatives tells us how
the mean and fluctuation dynamics are affected by the presence of
nonlinearities in the propensity functions.

Knowledge of the mean and covariance dynamics $\{\bfmuz(t)$,
$\C_{\ss \bfZ}(t)$, $t >0\}$ of the DA process allows us to
directly calculate the mean and covariance dynamics of the
population process by $\bfmux(t) = \bfx_0 + \S \bfmuz(t)$
and $\C_{\ss \bfX}(t) = \S \C_{\ss \bfZ}(t) \S^{\ss T}$,
respectively. These relationships can be used to derive
differential equations similar to Eqs.~(\ref{eq-IV-31})
and~(\ref{eq-IV-32}) that govern the mean and covariance
dynamics associated with the population process.

If the propensity functions are \textit{all} linear, then the means
can be calculated independently from the covariances and, in general,
the equations for the $\kth$ order moments will decouple from all
moments of order greater than $k$~\cite{Engb:06}. In all other cases
however, evaluation of the mean and covariance dynamics using the
previous differential equations requires calculating the dynamics
of at least third-order central moments. In theory, these dynamics
can be evaluated by differential equations similar to the ones above,
which require evaluation of higher-order central moments. However,
calculating high-order moment dynamics is a formidable task,
especially when dealing with large reaction networks. Note that a
reaction network comprised of $M$ reactions requires $M$ equations
for the DA means, $M(M+1)/2$ equations for the DA covariances and,
in general, $\sO(M^{\! k})$ equations for the $\kth$-order DA moments.
As a consequence, including differential equations governing the
dynamics of third- and higher-order central moments of the DA process
may make sense only when dealing with very small reaction networks.
Therefore, a practical treatment of reaction networks by calculating
moments is most often limited to evaluating only the mean and
covariance dynamics.

When the propensity functions are nonlinear, the moment equations
always form an \textit{infinite} hierarchy, with lower order moments
depending on higher order moments, indicating that exact solutions
cannot be obtained in practice. To address this problem, we can
replace the moments at some stage of the hierarchy with appropriately
chosen functions of lower-order moments~\cite{Whit:57,%
Keel:00,MurrDiecLaw:04,KrisCookMari-etal:05,Gill:09}. This
approach results in a \textit{moment closure} scheme that
produces a self-contained system of differential equations
whose solution provides approximate values for the moments
of the DA and populations processes. The resulting method
is usually referred to as the \textit{moment approximation}~(MA)
method.

\newpage

A method to construct appropriate functions for moment closure is
based on making an \textit{ansatz} for the joint probability
distributions of the DA or population process. For example, we
may assume a joint probability distribution for the DA process
that can be uniquely specified from the means and covariances.
This distribution will then impose functional relationships between
the third-order central moments and the means and covariances of
$\bfZ(t)$, which can be used to close the system of Eqs.~(\ref{eq-IV-31})
and~(\ref{eq-IV-32}). Evaluation of the mean and covariance
dynamics will now require solving a \textit{coupled} system of
first-order differential equations comprised of $M$ equations for
the means and $M(M+1)/2$ equations for the covariances.

To illustrate this method, let us consider a reaction network whose
propensity functions are at most quadratic. If we make the
\textit{ansatz} that the probability distribution of the DA process
$\bfZ(t)$ is approximately normal, then the third-order central
moment will be zero and, in this case, Eqs.~(\ref{eq-IV-31})
and~(\ref{eq-IV-32}) will be \textit{exact}, resulting in
\beqa
&& \hspace{-40pt} \frac{d\muz(m;t)}{dt} = \alpha_m(\bfmuz(t)) \nn \\
&& + \frac{1}{2} \!\! \sum_{m_1 \in \sM} \sum_{m_2 \in \sM}
\!\! h_{m,m_1,m_2} \: \cz(m_1,m_2;t) ,
\label{eq-IV-33}
\eeqa
for $t >0$, $m \in \sM$, and
\beqa
&& \hspace{-15pt}
\frac{d\cz(m,m';t)}{dt} = {\Bigl [} \alpha_m (\bfmuz(t)) \nn \\ [6pt]
&& + \frac{1}{2} \!\! \sum_{m_1 \in \sM} \sum_{m_2 \in \sM} h_{m,m_1,m_2}
\: \cz(m_1,m_2;t) {\Bigr ]} \! \Delta(m-m') \nn \\ [6pt]
&& + \!\!\! \sum_{m_1 \in \sM} {\Bigl [} h_{m',m_1}(\bfmuz(t))
\: \cz(m,m_1;t) \nn \\
&& \hspace{40pt} + \: h_{m,m_1} (\bfmuz(t)) \: \cz(m',m_1;t) {\Bigr ]},
\label{eq-IV-34}
\eeqa
for $t >0$, $m,m' \in \sM$,
where the second-order derivatives $h_{m,m_1,m_2}$ of the propensity
functions do not depend on the means. More details on this \textit{normal}
MA scheme can be found in~\textcite{Gout:07}, \textcite{UribVerg:07},
\textcite{FermLotsHell:08}, \textcite{BousDest:09}, \textcite{LeeKimKim:09},
\textcite{UllaWolk:09}, \textcite{LafuTora:10}, and \textcite{MilnGillWilk:11}.

In addition to the normal distribution, a number of alternative
approximating distributions have been suggested in the literature,
such as log-normal~\cite{KrisCookMari-etal:05,Nase:03b,Keel:00},
Poisson~\cite{Nase:03b} and near-Poisson~\cite{BuicCowaChow:10},
binomial~\cite{Nase:03b,KissSimo:12}, beta
binomial~\cite{KrisCookMari-etal:05}, and mixtures of
distributions~\cite{KrisCookMari-etal:05,KrisMariGibs:07}. Using
the log-normal distribution has some advantages over using the normal
distribution~\cite{Keel:00,Nase:03b,KrisCookMari-etal:05}. In
particular, the log-normal distribution has nonnegative support and
exhibits nonzero skewness, two properties that are important in the
context of certain nonlinear reaction networks, such as biochemical
reaction networks. On the other hand, the beta binomial distribution
is a discrete distribution with a flexible shape that, in some cases,
can capture the dynamic evolution of the true probability distribution
of the population process better than other
distributions~\cite{KrisCookMari-etal:05}. Its use however is limited
to \textit{closed} reaction networks (i.e., reaction networks with
fixed total population) that contain only two species. Finally,
using mixture distributions for deriving a moment closure scheme
shows great promise but has only been employed in few limited
cases~\cite{KrisCookMari-etal:05,KrisMariGibs:07}.

We should note here that it may be difficult to specify an
appropriate probability distribution for the population process,
since this distribution must assign zero probability to
stoichiometrically impossible populations [i.e., populations that
do not satisfy Eq.~(\ref{eq-II-4})]. On the other hand, it may be
easier to specify a probability distribution for the DA process,
since this process is usually confined within a well-defined subset
of the positive orthant of the multidimensional DA state-space.
Note also that approximating the moments of the DA process by
employing continuous distributions, such as log-normal, may be
difficult to justify due to the discrete nature of this process
(see however the following subsection for a case in which this
may be possible). But more importantly, assuming a specific form
for the probability distributions of the DA process may lead to
serious problems, since the differential equations derived from
the master equation will not be consistent with the moment
structure imposed by the assumed distribution, unless the
solution to the master equation coincides with that distribution.
To ameliorate this problem, note that we do not necessarily need
to specify the exact probability distribution for the DA process
in order to close the system of moment equations. For example,
we can use Eqs.~(\ref{eq-IV-31}) and~(\ref{eq-IV-32}) and assume
that, for every $t >0$, the third- and fourth-order central moments
are related to the first- and second-order central moments by the
same formulas as the ones associated with a multivariate normal or
log-normal distribution~\cite{Isse:18,Skou:08,LafuTora:10}. This
is a weaker \textit{ansatz} than assuming that the probability
distribution of the DA process is multivariate normal or log-normal,
which may work well in certain circumstances. As a matter of fact,
it has been shown by~\textcite{SingHesp:07,SingHesp:11} that,
under certain conditions (at most quadratic propensity functions
and a moment closure formula that has a particular separable form),
this assumption is a consequence of matching time derivatives of
the exact (not closed) moment equations at the initial time $t=0$
with that of the approximate (closed) moment equations.

Although, in some problems, the previous strategy may lead to a
sufficiently accurate estimation of the low-order moments of the
DA and population processes, it cannot provide an analytical
expression for the probability distributions of these processes.
We can however address this problem by using the MaxEnt approach
discussed in Section~IV-C-5. For example, if the propensity
functions of the reaction network at hand are at most quadratic
and if we employ a log-normal-based moment closure scheme, then
the MaxEnt approximation of the probability distribution
$p_{\ss \bfZ}(z_m;t)$ of the DA process $Z_m(t)$ associated with
the $\mth$ reaction will be given by the following Gibbs distribution:
\beqa
&& \hspace{-15pt} \widehat{p}_{\ss \bfZ}(z_m;t) \! = \! {\Bigl (}
\sum_{u \geq 0} \exp {\Bigl \{}
\! - \lambda_1(t) u - \lambda_2(t) u^2 - \lambda_3(t) u^3 {\Bigr \}}
{\Bigr )}^{\!\! -1} \nn \\
&& \hspace{40pt} \times \exp {\Bigl \{}
\! - \lambda_1(t) z_m - \lambda_2(t) z_m^2 - \lambda_3(t)
z_m^3 {\Bigr \}}, \nn
\label{eq-IV-35}
\eeqa
where the coefficients $\lambda_1(t)$, $\lambda_2(t)$, and $\lambda_3(t)$
must be determined so that
\beqa
&& \sum_{z_m \geq 0} \! z_m \widehat{p}_{\ss \bfZ}(z_m;t)
= \muz(m;t) \nn \\ [6pt]
&& \sum_{z_m \geq 0} \! z^2_m \widehat{p}_{\ss \bfZ}(z_m;t)
= \cz(m,m;t) + \muz^2(m;t) \nn \\
&& \sum_{z_m \geq 0} \! z^3_m \widehat{p}_{\ss \bfZ}(z_m;t)
= \left [ \frac{\cz(m,m;t) + \muz^2(m;t)}{\muz(m;t)} \right ]^3 \!\!\!, \nn
\label{eq-IV-36}
\eeqa
where the last constraint is due to the fact that
\beqa
&& \hspace{-30pt} \E[Z_1(t)Z_2(t)Z_3(t)] \nn \\ [6pt]
&& = \frac{\E[Z_1(t)Z_2(t)]\E[Z_1(t)Z_3(t)]\E[Z_2(t)Z_3(t)]}
{\E[Z_1(t)]\E[Z_2(t)]\E[Z_3(t)]}
\label{eq-IV-37}
\eeqa
for the adopted log-normal-based moment closure scheme. In this
case, the mean and covariance dynamics can be calculated from
Eqs.~(\ref{eq-IV-31}) and~(\ref{eq-IV-32}), by setting the
fourth-order derivatives $h_{m_1,m_2,m_3,m_4}$ equal to zero
and by using the relation in Eq.~(\ref{eq-IV-37}) for the
third-order moments.

Another approach to deal with the problem of moment closure is to
replace the true (but unknown) values of the higher-order moments
required by the MA method [such as the third- and fourth-order
central moments in Eqs.~(\ref{eq-IV-31}) and~(\ref{eq-IV-32})]
with estimated values derived from available data or from sampling
the master equation using Monte Carlo~\cite{ChevSama:11,RuesArgeSUmm-etal:11}.
A technique proposed by~\textcite{RuesArgeSUmm-etal:11} employs
a small number of Monte Carlo samples to obtain crude estimates of
the higher-order moments. The resulting estimates are interpreted
as noisy measurements of the true moment values and an extended
Kalman filtering approach is then used to obtain more accurate
estimates of these values. On the other hand, a strategy proposed
by~\textcite{ChevSama:11} replaces the unknown moment values with
estimated values obtained from available data. Both approaches may
work well when the estimation error is small. However, large errors
may lead to erroneous calculations due to potential amplification
of these errors when solving the differential equations that
govern the dynamic evolution of the lower-order moments [such
as Eqs.~(\ref{eq-IV-31}) and~(\ref{eq-IV-32})]. To address this
problem, a large number of Monte Carlo samples may be needed
when estimating the higher-order moments, which can appreciably
decrease the computational efficiency of the first method.
On the other hand, the second approach requires a large amount
of data to be available for reliable estimation of moments
and, therefore, it is limited to a small number of problems
in which this may be possible.

\vspace{-12pt}

\subsection{Linear noise approximation}

\vspace{-6pt}

In certain circumstances, the joint probability distributions of the DA
and population processes can be well approximated by multivariate
normal distributions. To see why this is true, we will assume the
existence of a system parameter~$\Omega$ that measures the relative
size of stochastic fluctuations in a Markovian reaction network,
such that fluctuations are small for large $\Omega$. This is
motivated by the fact that, in chemical reaction systems, stochastic
fluctuations gradually diminish as the system approaches the
\textit{thermodynamic limit} at which the population of each species
and the system volume approach infinity in a way that the concentrations
remain the same.\footnote{We simply denote the thermodynamic limit as
$\Omega \rightarrow \infty$.}

It is intuitive to expect that the probability of a reaction to
occur within the infinitesimally small time interval $[t,t+dt)$
depends on the ``density'' $\bfx(t)/\Omega$ of the population
process at time $t$ and that this probability does not change
when~$\Omega$ varies as long as the population densities remain
fixed~\cite{Kamp:07}. This implies that the propensity functions
$\pi_m$ must satisfy $\pi_m(\bfx;\Omega)$ $=$
$\widetilde{\pi}_m(\bfx/\Omega)$, where $\widetilde{\pi}_m$
does not depend on $\Omega$.\footnote{When necessary, we
explicitly denote the dependance of various quantities on
$\Omega$.} To be more general, we may also add a term
$\Omega^{-1} \widetilde{\pi}'_m(\bfx/\Omega)$, in which
case we would like $\pi_m(\bfx;\Omega) = \widetilde{\pi}_m(\bfx/\Omega) +
\Omega^{-1} \widetilde{\pi}'_m(\bfx/\Omega)$.\footnote{As a matter
of fact, we can also add terms of $\sO(\Omega^{-2})$,
$\sO(\Omega^{-3})$, etc., if necessary. These terms can be easily
accommodated in the formulation. However, it is not necessary to do
that here.} Moreover, we can assume that $\widetilde{\pi}_m(\:\cdot\:)$
and $\widetilde{\pi}'_m(\: \cdot \:)$ are analytic. Finally, we
may allow an arbitrary positive factor $f(\Omega)$, such that
\beq
\pi_m(\bfx;\Omega) = f(\Omega) \left [ \widetilde{\pi}_m(\bfx/\Omega)+
\Omega^{-1} \widetilde{\pi}'_m(\bfx/\Omega) \right ].
\label{eq-IV-38}
\eeq
As a consequence, we can assume the following scaling law for the
propensity functions of the DA process:
\beq
\alpha_m(\bfz;\Omega) = f(\Omega) \! \left [ \widetilde{\alpha}_m(\bfz/\Omega)
+ \Omega^{-1} \widetilde{\alpha}'_m(\bfz/\Omega) \right ],
\label{eq-IV-39}
\eeq
for $m \in \sM$, where $\widetilde{\alpha}_m(\bfz/\Omega)$ $:=$
$\widetilde{\pi}_m(\bfx_0/\Omega+\S\bfz/\Omega)$ and
$\widetilde{\alpha}'_m(\bfz/\Omega)$ $:=$
$\widetilde{\pi}'_m(\bfx_0/\Omega+\S\bfz/\Omega)$.

To proceed, we can make the following \textit{ansatz}:
\beq
\widetilde{Z}_m(t;\Omega) = \zeta_m(t) + \frac{1}{\sqrt{\Omega}}\:
\Xi_m(t),
\label{eq-IV-40}
\eeq
for $t >0$, $m \in \sM$, where $\widetilde{Z}_m(t;\Omega)$ is
the ``density'' $Z_m(t;\Omega)/\Omega$
of the DA process, $\Xi_m(t)$ is a noise component that quantifies the
fluctuations associated with the DA process, and $\zeta_m(t)$ is a
deterministic process that satisfies:
\beq
\frac{d\zeta_m(t)}{dt} = \widetilde{\alpha}_m(\bfzeta(t)),
\label{eq-IV-41}
\eeq
for $t >0$ and $m \in \sM$, initialized with $\zeta_m(0)=0$.
For each~$\Omega$, Eq.~(\ref{eq-IV-40}) decomposes the random DA
density $\widetilde{Z}_m(t;\Omega)$ into a deterministic component
$\zeta_m(t)$ and an additive noise component $\Xi_m(t)$. Clearly,
this equation is based on the premise that the fluctuations diminish
to zero as fast as $\Omega^{-1/2}$. In contrast to Eq.~(\ref{eq-IV-27}),
which is exact, Eq.~(\ref{eq-IV-40}) must be justified. This can be
done by a central limit theorem for the behavior of the probability
density function of the DA density process $\widetilde{\bfZ}(t;\Omega)$,
as $\Omega \rightarrow \infty$, similar to that shown
by~\textcite{Kurt:71,Kurt:72} for the case of biochemical reaction
networks.

By using Eqs.~(\ref{eq-IV-39})--(\ref{eq-IV-41}) and the $\Omega$-expansion
method of van Kampen, it can be shown that, for a sufficiently large~$\Omega$,
the dynamic evolution of the probability density function $\pxi(\bfxi;t)$
of the noise vector $\bfXi(t)$ is approximately governed by the following
\textit{linear} Fokker-Planck equation~\cite{Kamp:61,Kamp:76,Kamp:07}:
\beqa
\frac{\partial \pxi(\bfxi;t)}{\partial t} &\!\!=\!\!& \frac{1}{2} \sum_{m \in \sM}
\widetilde{\alpha}_m(\bfzeta(t))
\frac{\partial^2 \pxi(\bfxi;t)}{\partial \xi^2_m} \nn \\
&& - \sum_{m \in \sM}
\sum_{m' \in \sM} \frac{\partial \widetilde{\alpha}_m(\bfzeta(t))}
{\partial \zeta_{m'}} \frac{\partial [ \xi_{m'}
\pxi(\bfxi;t)]}{\partial \xi_m} , \nn
\label{eq-IV-42}
\eeqa
for $t >0$, initialized with $\pxi(\bfxi;0) = \delta(\bfxi)$,
where $\delta ( \cdot )$ is the Dirac delta function. In this case,
${\bf \Xi}(t)$ will approximately be a normal random vector with
zero mean and correlation matrix~$\C_{\ss \bf \Xi}(t)$ that satisfies
the following Lyapunov matrix differential equation:
\beq
\frac{d\C_{\ss \bf \Xi}(t)}{dt}=\A(t)+\G(t)\C_{\ss \bf \Xi}(t)  +
\C_{\ss \bf \Xi}(t) \G^T(t),
\label{eq-IV-43}
\eeq
for $t >0$, initialized with $\C_{\ss \bf \Xi}(0)=0$. In this equation,
$\A(t)$ and $\G(t)$ are two $M \times M$ matrices with elements
\beqa
a_{m,m'}(t) &\!\!=\!\!& \widetilde{\alpha}_m(\bfzeta(t)) \: \Delta(m-m') \nn \\ [6pt]
g_{m,m'}(t) &\!\!=\!\!& \frac{\partial
\widetilde{\alpha}_m(\bfzeta(t))} {\partial \zeta_{m'}} \:, \nn
\label{eq-IV-44}
\eeqa
respectively, where $\Delta$ is the Kronecker delta function. As a
consequence, and for a sufficiently large $\Omega$, we can
approximate the probability distribution $p_{\ss {\widetilde{\bfZ}}}
(\widetilde{\bfz};t)$ of the DA density process by a multivariate
normal probability density function with mean $\bfzeta(t)$ and
covariance matrix $\C_{\ss \bf \Xi}(t)/\Omega$. Due to Eq.~(\ref{eq-II-4}),
this also allows us to approximate the probability distribution
$p_{\ss {\widetilde{\bfX}}}(\widetilde{\bfx};t)$ of the population
density process $\widetilde{\bfX}(t;\Omega) := \bfX(t)/\Omega$
by a multivariate normal probability density function with mean
$\bfx_0/\Omega + \S \bfzeta(t)$ and covariance matrix
$\S \C_{\ss \bf \Xi}(t) \S^T$. As a consequence, and since $\bfZ(t)
= \Omega \widetilde{\bfZ}(t)$, we can also approximate the probability
distribution $\pz(\bfz;t)$ of the DA process with a multivariate
normal distribution, with mean $\Omega \bfzeta(t)$ and
covariance matrix $\Omega \C_{\ss \bf \Xi}(t)$, whereas, we can
approximate the probability distribution $\px(\bfx;t)$ of the
population process with a multivariate normal distribution
with mean  $\bfx_0+\Omega \S \bfzeta(t)$ and covariance matrix
$\Omega \S \C_{\ss \bf \Xi}(t) \S^T$.

Because the fluctuations in the reaction network are governed by
the linear ``signal-plus-noise'' \textit{ansatz} given by
Eq.~(\ref{eq-IV-40}), the previous method is known as \textit{linear
noise approximation}~(LNA). Its use requires specification of an
appropriate fluctuation size parameter~$\Omega$, such that
Eq.~(\ref{eq-IV-40}) is satisfied, and a sufficiently large
value for this parameter so that the method produces a reasonable
approximation of the two probability distributions
$p_{\ss \widetilde{\bfZ}}(\widetilde{\bfz};t)$ and
$p_{\ss \widetilde{\bfX}}(\widetilde{\bfx};t)$. Implementation
of the method requires one to separately solve the system of~$M$
first-order differential equations~(\ref{eq-IV-41}) and the
system of $M(M+1)/2$ first-order differential equations~(\ref{eq-IV-43}).
In sharp contrast to the MA method, the LNA method decouples the
computation of the means from the computation of the covariances.
It turns out that the LNA method is substantially faster than Monte
Carlo estimation and can be used to provide a rapid assessment of
the statistical behavior of some Markovian reaction
networks~\cite{Gout:06}. This method has already been used to study
biochemical reaction networks~\cite{ElfEhre:03,TomiKimuKoba-etal:04,%
HayoJaya:04,TaoJiaDewe:05,ScotIngaKaer:06}, epidemiological
networks~\cite{ChenBokk:05}, ecological networks~\cite{DattDeliLaw:10},
social networks~\cite{LamaSzenIgle-etal:06}, and neural
networks~\cite{Bres:09,BenaCowaDron-etal:10}.

\vspace{-12pt}

\subsection{Macroscopic solution}

\vspace{-6pt}

For large nonlinear reaction networks, the MA and LNA methods can
become computationally intractable, since evaluation of the
covariances requires solving a system of $\sO(M^2)$ differential
equations. If that turns out to be the case, then the only option
left to characterize the dynamic behavior of the reaction network
is in terms of DA or population densities by using, for example,
the macroscopic (fluctuation-free) system of $M$ differential
equations given by Eq.~(\ref{eq-IV-41}). As a matter of fact,
Eq.~(\ref{eq-IV-40}) implies that, for any $t >0$, the DA density
process $\widetilde{Z}_m(t;\Omega)$ converges in distribution to
$\zeta_m(t)$ as $\Omega \rightarrow \infty$. On the other hand,
the difference between the DA density dynamics predicted by the
macroscopic system and the MA method grows as $\Omega$ decreases.
Indeed, let $\delta \zeta_m(t;\Omega) := \muz(m;t)/\Omega
- \zeta_m(t)$, where $\muz(m;t)$ is the mean value of the DA process
$Z_m(t)$ predicted by the MA method. Then, from Eqs.~(\ref{eq-IV-27})
and (\ref{eq-IV-40}) we have $\Omega^{-1} W_m(t) = \Omega^{-1/2}
\Xi_m(t) - \delta \zeta_m(t;\Omega)$ and, since $W_m(t)$ is zero mean,
we obtain
\beq
\delta \zeta_m(t;\Omega) = \frac{1}{\sqrt{\Omega}} \: \E[\Xi_m(t)]
= \sO(\Omega^{-1/2}),  \nn
\label{eq-IV-45}
\eeq

\newpage

\noindent
for $t > 0$ and $m \in \sM$.
Clearly, for sufficiently large~$\Omega$, $\delta \zeta_m(t;\Omega)
\simeq 0$, in which case the macroscopic density dynamics obtained by
Eq.~(\ref{eq-IV-41}) and the mean density dynamics obtained by the MA
method will approximately coincide. However, for small~$\Omega$,
$\delta \zeta_m(t;\Omega) \not\simeq 0$ and Eq.~(\ref{eq-IV-41}) may
fail to correctly predict the mean density dynamics of the DA process.
As a matter of fact, it has been demonstrated in the literature that,
for reaction networks with small species populations and appreciable
stochastic fluctuations (a situation that occurs at small $\Omega$
values), the MA method may reveal behavior that cannot be predicted
by the macroscopic equation~(\ref{eq-IV-41})~\cite{McQuJachRuss:64,%
ThakRescLisi:78,LeonReic:90,ZhenRoss:91,SrivYouSumm-etal:02,RaoArki:03,Gout:07,UribVerg:07,%
VellQian:07}.

Similarly to the DA density process, the population density process
$\widetilde{\bfX}(t;\Omega)$ converges in distribution, as
$\Omega \rightarrow \infty$, to the deterministic process $\bfchi(t)$
that satisfies the following macroscopic equations:
\beq
\frac{d \chi_n(t)}{dt} = \!\! \sum_{m \in \sM} \!\! s_{nm}
\widetilde{\pi}_m(\bfchi(t)),
\label{eq-IV-46}
\eeq
for $t > 0$, $n \in \sN$, where $\widetilde{\pi}_m(\widetilde{\bfx})
:= \Omega^{-1}\pi_m(\Omega \widetilde{\bfx})$, provided that these
equations are initialized with the same condition as the master
equation~(\ref{eq-II-8}). This is clearly true at finite times.
It is also true in the limit as $t \rightarrow \infty$, provided
that the macroscopic equations~(\ref{eq-IV-46}) have a
\textit{unique} asymptotically stable stationary solution that is
independent of the initial state~\cite{MansBroeNico-etal:81,Kamp:07}.

\vspace{-12pt}

\subsection{Remarks}

\vspace{-6pt}

\noindent
\textbf{1.}~It has been shown by~\textcite{GrimThomStra:11} that,
for monostable Markovian reaction networks, the LA method produces
means and covariances that are accurate to $\sO(\Omega^{-3/2})$
for systems of size $\Omega$ which are away from thermodynamic
equilibrium (i.e., systems that do not obey detailed balance --
see~Section~VIII) and at least accurate to $\sO(\Omega^{-2})$ for
systems that are at thermodynamic equilibrium (i.e., obey detailed
balance). As a consequence, the LA method will in general result
in more accurate means and covariances than the LNA method, which
produces means accurate to $\sO(\Omega^{-1/2})$ and covariances
accurate to $\sO(\Omega^{-3/2})$. Therefore, Monte Carlo estimation
of the means and covariances of the DA and population processes based
on the LA method will result in excellent estimation of the exact
values at sufficiently large system sizes $\Omega$, provided
that we use a small time step $\tau$ and a large number $L$ of
Monte Carlo samples. However, the LA method will produce
continuous-valued DA trajectories $\{\bfz(t)$, $t >0\}$,
which disagrees with the fact that these trajectories are
integer-valued.

\vspace{6pt}
\noindent
\textbf{2}.~It can be seen from Eqs.~(\ref{eq-IV-15}) and~(\ref{eq-IV-16}) that,
in the limit as $\tau \rightarrow 0^+$, the means and covariances of
the approximating DA process $\widehat{\bfZ}(t)$ associated with
the PA method will approach the means and covariances of the actual
DA process $\bfZ(t)$ governed by the master equation. Therefore, Monte
Carlo estimation of the means and covariances of the DA process
based on the PA method will result in excellent estimates of
the exact values, provided that we use a sufficiently small time step
$\tau$ and a sufficiently large number~$L$ of Monte Carlo samples.
Note also that, in sharp contrast to the Langevin approximation,
the Poisson approximation produces DA trajectories $\{\bfz(t)$,
$t >0\}$ that are integer-valued.

\vspace{6pt}
\noindent
\textbf{3}.~When the propensity functions of a Markovian reaction network satisfy
Eq.~(\ref{eq-IV-38}), the macroscopic solution, the LNA method, and
the MA method provide a hierarchy of approximations to the master
equation~\cite{FermLotsHell:08}. At large values of $\Omega$,
close to the thermodynamic limit, the macroscopic equations may
provide a sufficiently accurate description of the reaction network.
For smaller values of $\Omega$, the LNA method will be more preferable,
whereas, for even smaller values of $\Omega$, the MA method must be
employed. Unfortunately, there is no way to determine the range of
$\Omega$ values for which each approach is valid. Moreover, for very
small values of~$\Omega$, these approximations may not be accurate and
Monte Carlo simulation methods should be employed instead. It is
therefore very difficult to determine a priori which method provides
accurate estimation of the network dynamics for a given value of
$\Omega$.

\vspace{6pt}
\noindent
\textbf{4}.~For networks governed by the mass-action law, $\Omega$ usually
represents the system volume. In this case, the specific probability
rate constant $\k$ of a reaction that involves two species will be
proportional to $\Omega^{-1}$ with proportionality factor $k$; i.e.,
$\kappa=k\Omega^{-1}$~\cite{Gill:92}. This naturally implies that
the frequency of the reaction to occur within the infinitesimally
small time interval $[t,t+dt)$ will be reduced at a rate that is
inversely proportional to system volume. As a consequence, the
propensity function $\pi(\bfx;\Omega)$ of a reaction $X_1+X_2
\rightarrow X_3$ will satisfy Eq.~(\ref{eq-IV-38}) with
$f(\Omega)=\Omega$, $\widetilde{\pi}(\bfx/\Omega) = k (x_1/\Omega)
(x_2/\Omega)$, and $\widetilde{\pi}'(\bfx/\Omega)=0$. On the other
hand, the propensity function of a reaction $2X_1 \rightarrow X_2$
will satisfy Eq.~(\ref{eq-IV-38}) with $f(\Omega)=\Omega$,
$\widetilde{\pi}(\bfx/\Omega) = (k/2)(x_1/\Omega)^2$, and
$\widetilde{\pi}'(\bfx/\Omega) = -(k/2)(x_1/\Omega)$.

\vspace{6pt}
\noindent
\textbf{5}.~An important issue associated with the \textit{ansatz} given by
Eq.~(\ref{eq-IV-40}) is that, for a given value of $\Omega$, the portion
of the left tail of the Gaussian distribution of the noise component
$\Xi_m(t)$ that extends below zero may have appreciable mass when the
standard deviation is large, falsely producing negative populations
with non-zero probability when such populations are not possible.
This problem can be taken care of by sufficiently increasing $\Omega$,
provided that the standard deviation of $\Xi_m(t)$ is finite. As a
consequence, justifying the \textit{ansatz} given by Eq.~(\ref{eq-IV-40}),
and thus the applicability of the LNA method, requires that the
covariance matrix $\C_{\ss {\bf \Xi}}(t)$ is finite. This however
may not always be true. To see why, note that the unique solution
of the Lyapunov matrix equation~(\ref{eq-IV-42}) is given
by~\cite{KandFreiIone-etal:03}
\beq
\C_{\ss {\bf \Xi}}(t) = \int_0^t \!\! \bfPhi_{\ss G}(t;\tau)
\A(\tau) \bfPhi^T_{\ss G}(t;\tau) d \tau,
\label{eq-IV-60}
\eeq
where $\bfPhi_{\ss G}(t;\tau)$ is an $M \times M$ matrix, where
$M$ is the number of reactions. This matrix satisfies
\beq
\frac{\partial \bfPhi_{\ss G}(t;\tau)}{\partial t } = \G(t)
\bfPhi_{\ss G}(t;\tau), \quad t \geq \tau, \quad \bfPhi_{\ss G}
(\tau;\tau) = \I_M, \nn
\label{eq-IV-61}
\eeq
with $\I_M$ being the $M \times M$ identity matrix. Since
$\A(t)$ is a diagonal (and thus symmetric) matrix with nonnegative
elements, $\C_{\ss {\bf \Xi}}(t)$ will be a symmetric positive
semidefinite matrix. These properties are required so that
$\C_{\ss {\bf \Xi}}(t)$ is a correlation matrix. If the real
parts of the eigenvalues of the Jacobian matrix $\G(t)$ are
\textit{all} negative, for every $t >0$, then, for any fixed~$\tau$,
$\bfPhi_{\ss G}(t;\tau)$ will be finite, for every $t> \tau$, and
$\lim_{t \rightarrow \infty} \bfPhi_{\ss G}(t,\tau) = 0$, in which
case $\C_{\ss {\bf \Xi}}(t) < \infty$, for every $t>0$.\footnote{Note
that $\A(t)$ is bounded for every $t>0$ due to the assumption that
the propensity functions are analytic.} This condition is equivalent
to saying that the solution of Eq.~(\ref{eq-IV-41}) must be
asymptotically stable. As a consequence, lack of asymptotic
stability of the dynamic evolution of the means of the DA density
process may result in infinite fluctuations, thus violating the
\textit{ansatz} given by Eq.~(\ref{eq-IV-40}) and rendering the
LNA method invalid. Examples on what can happen in this case can
be found in~\textcite{Kamp:76}.

\vspace{6pt}
\noindent
\textbf{6}.~It is interesting to note that $\A(t)$ in Eq.~(\ref{eq-IV-60}) is
a diffusion matrix that tells us how the stochastic properties of
a reaction network, determined by the propensity functions, change
at each time point along the mean DA trajectory. It turns out that
$\A(t)$ represents the growth of stochastic fluctuations about the
mean DA trajectory as time progresses. On the other hand, the
Jacobian matrix $\G(t)$ produces the dissipation matrix
$\bfPhi_{\ss G}(t;\tau)$ which locally damps the stochastic
fluctuations along the mean trajectory and squeezes this growth.

\vspace{6pt}
\noindent
\textbf{7}.~The LNA method is not appropriate when the probability distributions
of the DA and population processes are not unimodal, a situation that
arises in multistable reaction networks~\cite{Kamp:07,Qian:11}. In
such cases, the approximation can still be applied but only during
sufficiently short timescales with an initial condition that is
inside the domain of attraction of an equilibrium point of
Eq.~(\ref{eq-IV-41}). Moreover, the Gaussian nature of the LNA
method implies that both processes must be continuous-valued.
Although this may be approximately true for large values of $\Omega$,
it is not necessarily true for smaller values. This is due to the
fact that, for small $\Omega$, the reaction system may contain a
small number of species interacting through infrequently occurring
reactions, in which case $\widetilde{\bfZ}(t)$ may take only a
small number of possible values, at least during an appreciably
long initial time interval. Note also that, for every value of
$\Omega$, the left tail of the Gaussian distribution extends below
zero. For large $\Omega$, the total probability mass over negative
DA values is negligible and poses no practical problem. Since
however the covariance matrix of $\widetilde{\bfZ}(t)$ is inversely
proportional to~$\Omega$ while the mean does not depend on~$\Omega$,
the approximation may falsely produce negative DA values with
appreciable probability when~$\Omega$ is small, which may
erroneously predict negative populations (which is also true
in the case of the LA and PA methods), when such populations
are not possible~\cite{BostKessShne-etal:12}.

\vspace{6pt}
\noindent
\textbf{8}.~As a consequence of the previous remarks, the LNA method provides
a reasonable approximation of a Markovian reaction network exhibiting
a single and asymptotically stable behavior subject to relatively small
stochastic fluctuations. By appropriately modifying this method,
we can also use it to approximate fluctuations around mean trajectories
$\bfzeta(t)$ that produce Jacobian matrices $\G(t)$ with purely
imaginary eigenvalues. In this case however the fluctuations will
only be driven by the diffusion matrix~$\A(t)$ and, thus, will grow
as a function of time. As a consequence, the LNA method can only be
used over an initial time interval during which the fluctuations will
be sufficiently small so that the \textit{ansatz} given by
Eq.~(\ref{eq-IV-40}) can be justified. An appropriate modification of
the $\Omega$-expansion method can lead in this case to approximating
fluctuations using the nonlinear Fokker-Planck equation or other
nonlinear partial differential equations~\cite{Kamp:76,Kamp:07,Grim:10,%
Grim:11,GrimThomStra:11}. Finally, it may also be possible under
certain circumstances to modify the LNA method to deal with cases
in which the eigenvalues of the Jacobian matrix $\G(t)$ have positive
real parts [i.e., when the mean trajectories $\bfzeta(t)$ are
unstable [see~\textcite{Kamp:76,Kamp:07} and ~\textcite{TomiOhtaTomi:74}].
For a recent example, see~\textcite{ScotIngaKaer:06}.

\vspace{6pt}
\noindent
\textbf{9}.~A number of moment closure schemes have been proposed in the literature
based on truncating high-order central moments or cumulants~\cite{Nase:03a,%
Nase:03b,Engb:06,MatiKiff:99,MatiKiff:02,LeeKimKim:09}. The typical
assumption behind these methods is that the solution of the master
equation has negligible high-order central moments or cumulants,
which can be set equal to zero without affecting the mean and
covariance dynamics. This assumption is not true in general, with the
exception of normal random variables whose central moments of odd
order and cumulants of order $\geq 3$ are zero. For non-normal random
variables, there is an infinite number of non-vanishing moments or
cumulants in general~\cite{Gard:10}. For example, all cumulants of
a Poisson random variable are equal to the mean value. As a
consequence, truncation of high-order moments or cumulants cannot
be easily justified and may lead to a non-valid probability
distribution~\cite{HangTalk:80,Keel:00,Hieb:06}. It is worthwhile
noticing however that low-order cumulants can be used to naturally
construct univariate and bivariate approximations to probability
distributions of certain nonlinear Markovian processes~\cite{Rens:98,%
Rens:00}. Moreover, it has been suggested by~\textcite{BuicCowaChow:10}
that an appropriately defined change of variables that measures
the deviation of each cumulant from its value under a Poisson
assumption (i.e., from the mean) produces a moment hierarchy
that can be naturally and justifiably truncated in the case of
neural networks since, in this case, the solution of the master
equation is expected to be near Poisson. Finally, it has been
shown by~\textcite{Grim:12} that, for a monostable Markovian
reaction network at sufficiently large size $\Omega$ with at
most quadratic propensity functions, a normal approximation
of the mean of the population density process $\bfX(t)/\Omega$
(obtained by setting its third- and higher-order cumulants equal
to zero) is at least as accurate as the approximation produced
by the macroscopic equations obtained by the $\Omega$-expansion method
in the thermodynamic limit of $\Omega \rightarrow \infty$. This
approximation however may lead to inaccurate covariance dynamics.
On the other hand, a moment approximation scheme constructed by
setting the fourth- and higher-order cumulants of the population
density process equal to zero (thus including third-order central
moments in the formulation) will produce more accurate mean and
covariance dynamics than the normal approximation, provided that
the system size $\Omega$ is sufficiently large.

\begin{figure*}
\includegraphics[width=7in]{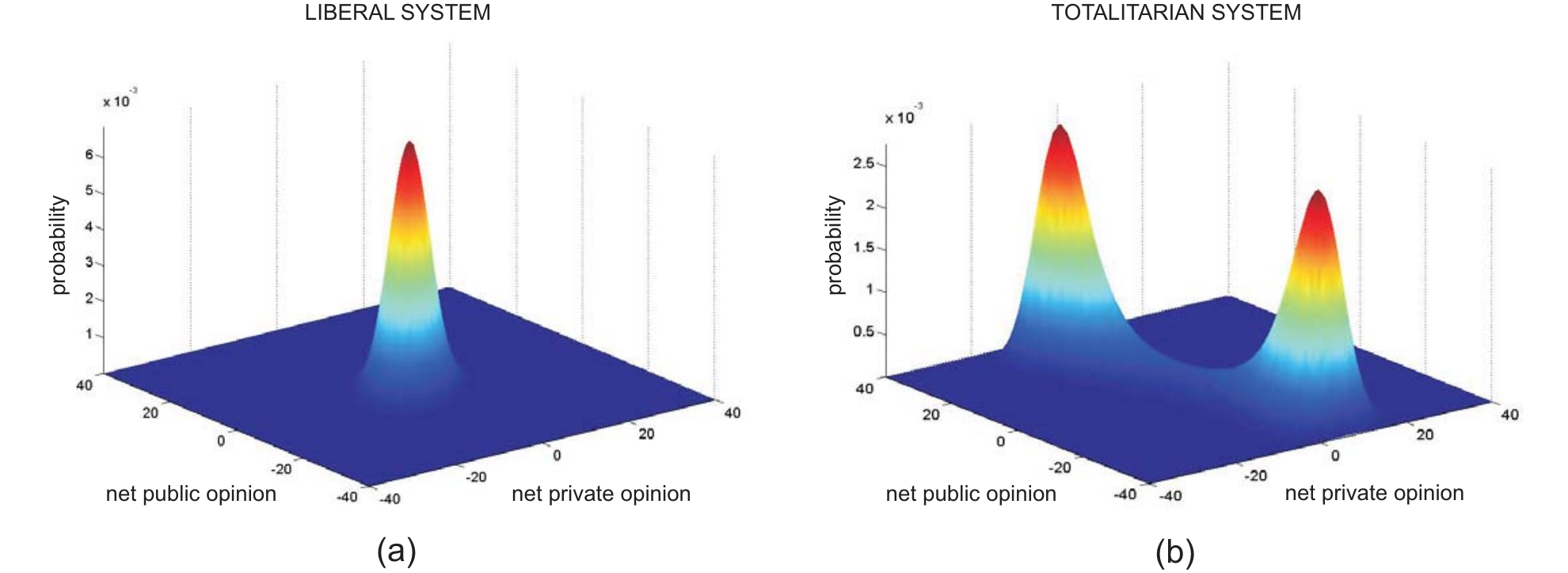}
\caption{The joint probability distributions of the net public and
private opinions in the liberal and totalitarian systems of opinion
formation example of Section~IV-I at steady-state computed by the
KSA method. (a)~The liberal system is characterized by a \textit{unimodal}
stationary distribution with its mode located at the point of zero
net public and private opinions. (b)~The totalitarian system is
characterized by a \textit{bimodal} stationary distribution with
the modes corresponding to two different totalitarian states: one
in which a large number of individuals publicly support the ideology
of the ruling party, while a small number of individuals are privately
against this ideology, and one in which a large number of individuals
are publicly opposing the current ruling ideology, while a small number
of individuals privately support it. In the latter case, the ruling
party effectively switches its current ideology to fit public opinion,
thus maintaining itself in power.}
\label{fig-3}
\vspace{-12pt}
\end{figure*}

\vspace{6pt}
\noindent
\textbf{10}.~The MA method is based on replacing the mean value $\E[\alpha_m(\bfZ(t))]$
by $\alpha_m(\bfmuz(t))+T_m(\bfmuz(t))$, where $T_m$ is a term calculated
by solving differential equations for higher-order moments [see
Eqs.~(\ref{eq-IV-31})--(\ref{eq-IV-34})]. When the propensity function
$\alpha_m(\bfz)$ is \textit{convex}, then Jensen's inequality implies that
$\E[\alpha_m(\bfZ(t))] \geq \alpha_m(\E[\bfZ(t)]) = \alpha_m(\bfmuz(t))$.
As a consequence, we must have $T_m(\bfmuz(t)) \geq 0$, since
$\E[\alpha_m(\bfZ(t))] = \alpha_m(\bfmuz(t)) +T_m(\bfmuz(t)) \geq
\alpha_m(\bfmuz(t))$. This however may not necessarily be the true,
in which case the method may result in additional errors that can
lead to instabilities. To address this problem, we can replace
$\E[\alpha_m(\bfZ(t))]$ with $\alpha_m(\bfmuz(t)) +
\max\{0,T_m(\bfmuz(t))\}$. In many instances, this simple modification
results in a more accurate and more stable implementation of the
MA method.

\vspace{6pt}
\noindent
\textbf{11}.~A moment closure scheme used in a particular application must produce
moments that satisfy a number of necessary conditions. For example, all
moments must be nonnegative and invariant under permutations. In addition,
if $Z_{m_1}(t)$, $Z_{m_2}(t)$, and $Z_{m_3}(t)$ are mutually uncorrelated,
then we must have $\E[Z_{m_1}(t)Z_{m_2}(t)Z_{m_3}(t)]$ $=$ $\E[Z_{m_1}(t)]
\E[Z_{m_2}(t)]\E[Z_{m_3}(t)]$, whereas, if $Z_{m_3}(t)$ is uncorrelated from
$Z_{m_1}(t)$ and $Z_{m_2}(t)$, we must have $\E[Z_{m_1}(t)Z_{m_2}(t)Z_{m_3}(t)]$
$=$ $\E[Z_{m_1}(t)] \E[Z_{m_2}(t)Z_{m_3}(t)]$. Moreover, the resulting
covariances must define a symmetric positive semi-definite matrix.

\vspace{6pt}
\noindent
\textbf{12}.~The assumptions underlying a given moment closure scheme may be
inconsistent with the statistics of a particular reaction network
at hand. In this case, the moment equations may have no solution, may
produce a number of unrealistic solutions, or result in unstable and
unbounded solutions~\cite{Nase:03a,Nase:03b}. Just like the LNA method,
normal MA techniques cannot characterize bimodality or highly skewed
probability distributions and may result in undesirable negative DA
and population values~\cite{KrisCookMari-etal:05}. Finally, much work
must be done to determine conditions for the stability of the differential
equations obtained by MA methods.

\vspace{-12pt}

\subsection{Example: Opinion formation}

\vspace{-6pt}

To illustrate and compare the previous methods for solving
the master equation, we focus on the opinion formation model
discussed in Section~III-E. The simplicity of this model
permits us to solve the underlying master equation using
a numerical approach. On the other hand, the complexity
introduced by the nonlinear nature of the propensity
functions given by Eq.~(\ref{eq-III-18}), allows us to
illustrate some intricate behavior. We consider two
parameterizations of the model corresponding to a liberal
democracy and a totalitarian regime~\cite{Weid:06}. In
particular, we assume a social system of $80$ individuals
(in which case $L=40$) and set the values of the specific
probability rate constants associated with the reactions in
Eq.~(\ref{eq-III-17}) to $\kappa_1 = 1/2~\mbox{day}^{-1}
\mbox{individual}^{-1}$ and $\kappa_2 = 1~\mbox{day}^{-1}
\mbox{individual}^{-1}$.

The main difference between the two social systems under
consideration is that, in the liberal system, there is no
pressure inflicted on public opinion with individuals
having an affirmative private bias towards the ideology
of the ruling party. On the other hand, there is heavy
pressure in the totalitarian system inflicted on public
opinion with individuals having a weakly dissident private
opinion bias against the ruling ideology. We quantify these
differences by setting $a_1=0~\mbox{individual}^{-1}$ and
$a_3=1/80~\mbox{individual}^{-1}$ in the liberal system, and
$a_1=3/80~\mbox{individual}^{-1}$, $a_3=-1/320~\mbox{individual}^{-1}$
in the totalitarian system. In addition, we assume that the two
systems differ on how strongly  privately held beliefs influence
publicly stated opinions and set $a_2 = 1/80~\mbox{individual}^{-1}$
in the liberal system and $a_2 = 1/40~\mbox{individual}^{-1}$ in
the totalitarian system. Finally, we consider the condition
$X_1(0)=X_2(0)=0$, which represents completely neutral initial
net publicly and privately held opinions.

We can numerically solve the master equation associated with the
opinion formation model by employing the KSA method. This method
is more appropriate than the IE method, since the DAs of the
underlying reactions can grow rapidly in this case, whereas the
populations $X_1(t)$ (net public opinion) and $X_2(t)$ (net private
opinion) are bounded, taking values between $-40$ and $40$. To
implement the KSA method, we used a Krylov subspace of dimension
$K_0 = 40$. The joint probability distributions of the net public
and private opinions in the liberal and totalitarian systems at
steady-state are depicted in Fig.~\ref{fig-3}, whereas, movies
encapsulating the entire dynamic evolutions of these distributions
for a period of $15$~days can be found in the accompanying Matlab
software. Evaluation of each solution took about $15$~seconds of
CPU time on a 2.20 GHz Intel Core 2 Duo processor running Windows~7.

In the liberal system, the stationary joint probability
distribution of public and private opinions is \textit{unimodal}
and almost identical to a sampled normal distribution;
see Fig.~\ref{fig-3}(a). This distribution characterizes the
fact that there is no need for an individual to hide her private
opinion (thus the high correlation between the public and private
opinions). As a consequence, the net opinions nearly balance
around the origin, which represents an equal number of privately
held and publicly pronounced opinions for or against the ruling
ideology. On the other hand, the stationary joint probability
distribution of public and private opinions in the totalitarian
system is \text{bimodal}; see Fig.~\ref{fig-3}(b). Here, weakly
dissident individuals tend to privately disapprove the ruling
ideology, but pressure on public opinion ensures that most
individuals publicly approve this ideology. As a consequence,
dissidence is not strong enough to destabilize the totalitarian
state and the system will operate around the peak located at
point $(30,-6)$ with strong net public opinion in support of
the ruling ideology and a rather weak private opinion against
this ideology.

\begin{figure}
\includegraphics[width=3.4in]{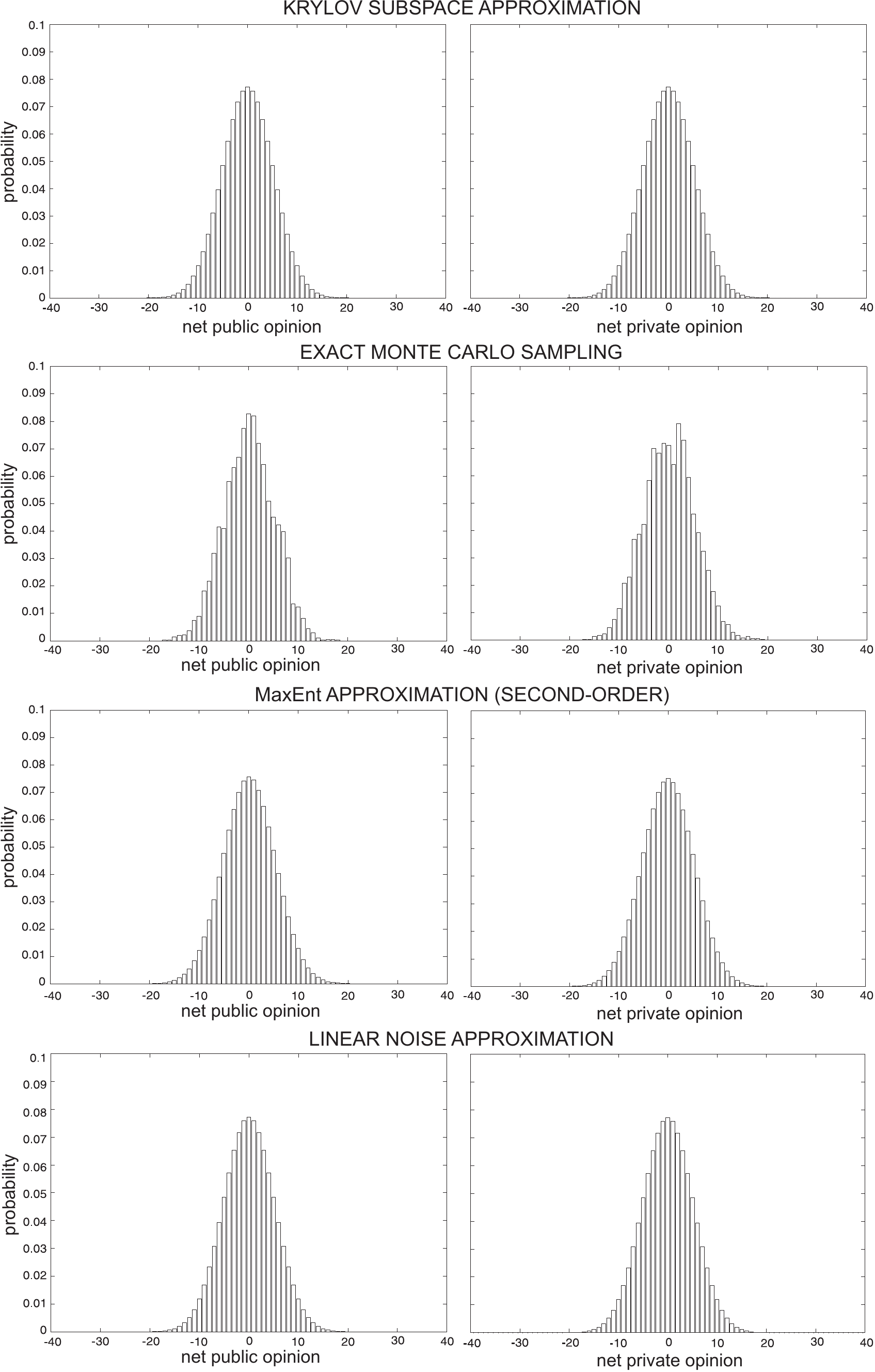}
\caption{The stationary marginal probability distributions of the net
public and private opinions in the liberal system of opinion formation
example of Section~IV-I computed by the Krylov subspace approximation
method, exact Monte Carlo sampling, second-order MaxEnt approximation,
and the lineal noise approximation method.}
\label{fig-4}
\vspace{-18pt}
\end{figure}

A totalitarian society may move to another peak located at
point $(-30,6)$ with strong net public opinion against
the ruling ideology and a rather weak private opinion in support
of this ideology. This situation can be easily remedied by the
ruling party, which can effectively change its ideology to fit
the prevailing public opinion, thus preventing public upheaval
and subsequent removal of the totalitarian regime from power.

\begin{figure}
\includegraphics[width=3.4in]{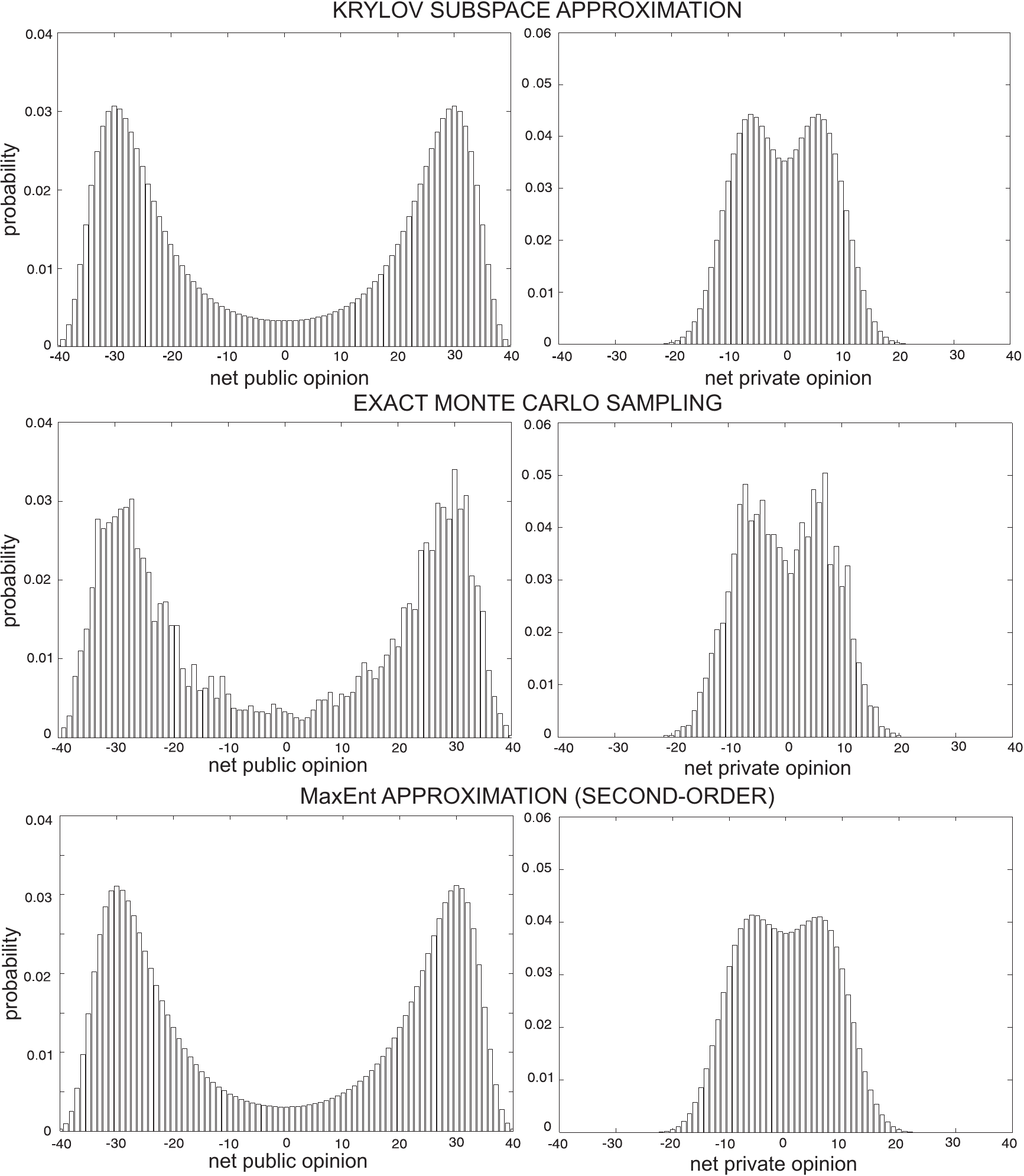}
\caption{The stationary marginal probability distributions of the net
public and private opinions in the totalitarian system of opinion formation
example of Section~IV-I, computed by the Krylov subspace approximation
method, exact Monte Carlo sampling, and eighth-order MaxEnt approximation.}
\label{fig-5}
\vspace{-18pt}
\end{figure}

Finally, the dynamic evolutions of the joint probability
distributions provided in the supplemental files demonstrate a
rapid convergence of the liberal system to its stable stationary
mode after $3$ days and a slower convergence of the totalitarian
system to one of its two modes after $14$ days.

Instead of using the KSA method, we can estimate the probability
distributions of the liberal and totalitarian systems by Monte Carlo
sampling. To do so, we employ $4,\!000$ trajectories obtained from
the master equation using the exact sampling algorithm of Gillespie
at a cost of about $230$~seconds of CPU time. We depict the estimated
stationary marginal probability distributions of the net public and
private opinions in Fig.~\ref{fig-4} for the liberal system, and
in Fig.~\ref{fig-5} for the totalitarian system. For comparison,
we also depict the stationary marginal probability distributions
obtained by the KSA method.

These results demonstrate the power and
weakness of Monte Carlo estimation. Monte Carlo sampling is a
simple and robust approach for solving the master equation which,
in principle, allows us to compute almost any statistical summary
of interest with arbitrary precision. The probability of a particular
event can be estimated by counting the number of occurrences of
the event and dividing by the total number of sampled trajectories.
However, this simple and elegant procedure comes with a large
computational cost, since the number of trajectories required
to obtain a sufficiently accurate estimate is usually very large.
Even for estimating the univariate marginal probability distributions
of the net public and private opinions, $4,\!000$ iterations does
not seem to be sufficient, since the symmetric and bimodal nature
of these distributions can be obscured by estimation errors.
As a matter of fact, it is often very difficult to know
\emph{a priori} how many samples must be used to sufficiently
estimate the qualitative and quantitative properties of a
statistical summary of interest or to verify \emph{a posteriori}
when convergence has occurred. Note however that the Monte Carlo
method scales far better than the KSA method, since trajectories
can be sampled from the master equation with a relative ease, even
in the case of large Markovian reaction networks for which numerical
methods, such as the KSA or IE methods, cannot be used. For large
systems, the number of trajectories that can be sampled from the
master equation in a reasonable time will certainly not be adequate
for accurately computing complex population statistics, such as
probability distributions, but they may be sufficient for
estimating certain moments (e.g., the means and covariances) of
the DA and population processes with a desired precision.

To ameliorate the computational burden of exact Monte Carlo sampling,
we can employ a Langevin or Poisson approximation. Sampling $4,\!000$
trajectories from the master equation governing the opinion formation
model using the LA method took $30$~seconds of CPU time,
whereas, drawing the same number of samples using the PA
method required $75$~seconds of CPU time. The estimation results
are very similar to the ones obtained by exact Monte Carlo sampling
(data not shown). The LA method however does not retain
the integer-valued nature of the net opinion trajectories, which can
be an issue when accurate simulation of these trajectories is desired.

By using the PA method, we can draw $4,\!000$ samples from the master
equation and use these samples to estimate the first $K$ moments of
the opinion trajectories under consideration. We can then use the
MaxEnt method discussed in Section~IV-C-5 to derive an analytical
approximation of the marginal probability distributions of the net
public and private opinions. The resulting sampled probability
density functions depicted in Figs.~\ref{fig-4} and~\ref{fig-5}
clearly show the potential of the MaxEnt method for correctly
estimating the stationary marginal distributions in the opinion
formation model. It took about $75$~seconds of CPU time to compute
these results, which have been obtained by using the Matlab code
developed by~\textcite{Djaf:91}. For the democratic system, we only
need to estimate the first- and second-order moments by Monte
Carlo. This is due to the fact that the true stationary
distributions are almost sampled normal. On the other hand, to
obtain sufficiently accurate approximations of the stationary
marginal distributions in the totalitarian system, we need to
estimate the first eight moments by Monte Carlo. The main advantage
of MaxEnt is its ability to provide a relatively accurate
approximation of marginal probability distributions by using
appreciably fewer sample trajectories than the ones required by
Monte Carlo in order to achieve a similar level of estimation
accuracy.

Application of the LNA method for solving the master equation
associated with the totalitarian system is not possible due to
the bimodal nature of the stationary joint probability distribution
(see Remark~IV-G-7). This method however can be used in the case of
the liberal system by choosing the system size parameter $\Omega$
to be equal to the ``size'' $L$ of the net public or private
opinions. Evaluation of the solution obtained by the LNA method
took only $0.35$~seconds of CPU time. The resulting Gaussian
probability density function approximates well the stationary
solution found by the KSA method. The sampled computed stationary
marginal probability distributions of the net public and private
opinions are depicted in Fig.~\ref{fig-4}. If there were more
individuals in the system (i.e., for larger values of $\Omega$),
then the LNA method could produce a more accurate result. On the
other hand, a significantly smaller number of individuals may
dramatically reduce the accuracy of the method, since the
statistical properties of the system may appreciably deviate
from normality. Despite its clear computational advantage,
use of the LNA method is hampered by the absence of a strategy
to effectively determine for which values of $\Omega$ the
resulting normal approximation is accurate.

\begin{figure}
\includegraphics[width=3.4in]{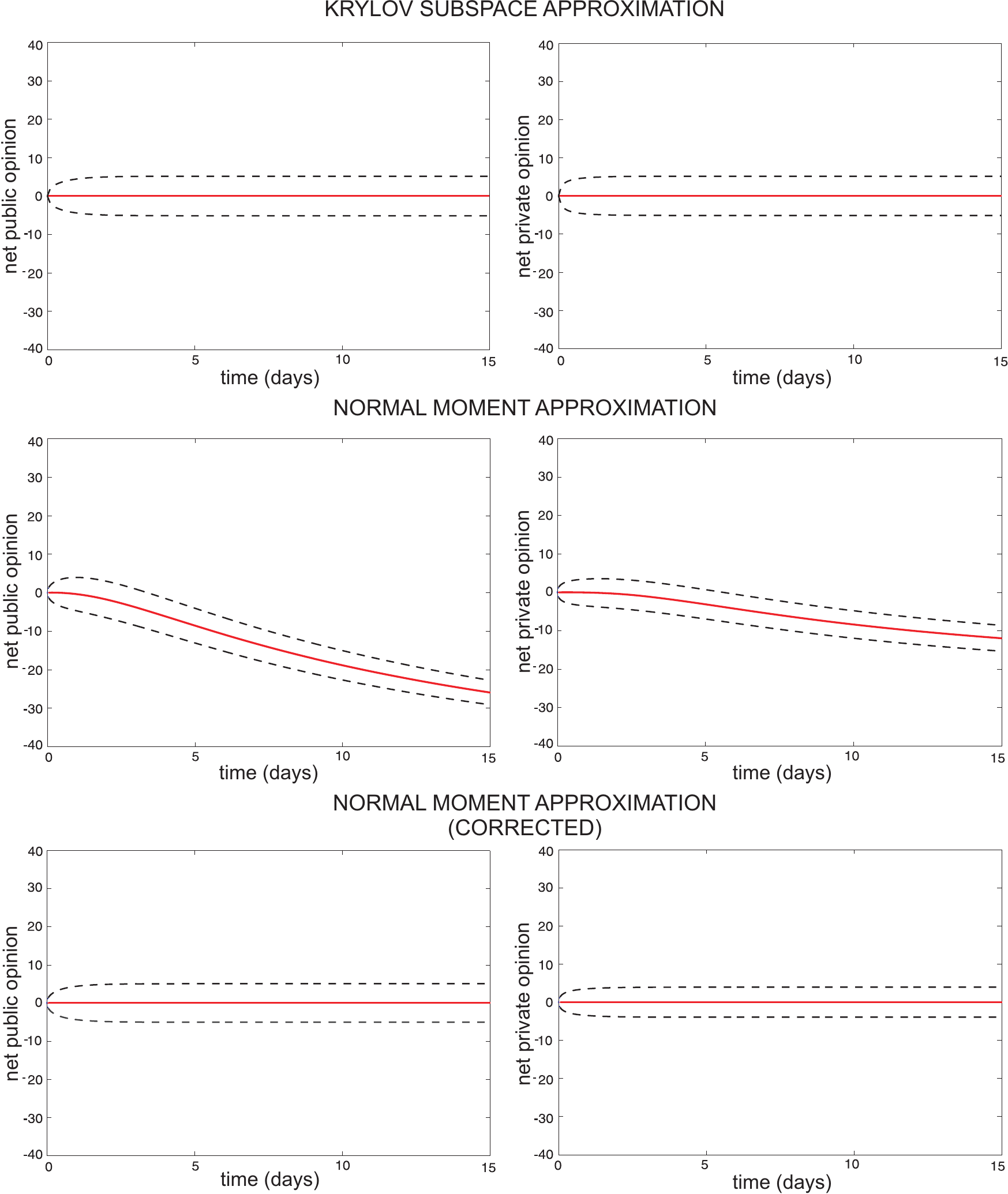}
\caption{The means (solid red lines) and the $\pm 1$ standard deviations
(dashed lines) of the net public and private opinions in the liberal
system of opinion formation example of Section~IV-I, obtained by the
Krylov subspace and normal moment approximation methods (with and
without correction using Jensen's inequality).}
\label{fig-6}
\vspace{-12pt}
\end{figure}

Finally, use of the MA method is much easier for the liberal system
than the totalitarian system, since the totalitarian system requires
at least eighth-order moments to sufficiently characterize its bimodal
stationary distribution. For simplicity, we will therefore focus
on the liberal system. Note  that, due to the exponential
nature of the propensity functions, given by Eq.~(\ref{eq-III-18}),
their effect on the moment equations can persist through infinitely
many derivatives, which can make the task of finding an appropriate
moment closure scheme very difficult. However, since the stationary
joint probability distributions of the net public and private opinions
approximate well a sampled Gaussian distribution (see Fig.~\ref{fig-4}),
we may be able to approximate the means and covariances by using the
\textit{normal} MA scheme given by Eqs.~(\ref{eq-IV-33})
and~(\ref{eq-IV-34}). In Fig.~\ref{fig-6}, we depict the means
(solid red lines) and the $\pm 1$ standard deviations (dashed lines)
of the net public and private opinions, obtained with the normal MA
method. Implementation of this method took a mere $0.6$~seconds of
CPU time. For comparison, we also depict the corresponding moments
and standard deviations obtained with the KSA method. Clearly, the
normal MA method produces unsatisfactory results.

The main culprit here is the fact that Eqs.~(\ref{eq-IV-33})
and~(\ref{eq-IV-34}) were derived for quadratic propensity
functions whose higher-order derivatives vanish, which along
with the Gaussian assumption results in a decoupling of the
means and covariances from higher-order central moments.
In the democratic system, the Gaussianity assumption is
approximately valid but errors accumulate, since Eqs.~(\ref{eq-IV-33})
and~(\ref{eq-IV-34}) neglect to account for non-vanishing
higher-order derivatives of the propensity functions. We can
however improve the closure scheme by using Jensen's inequality
(see Remark~IV-G-10), due to the convexity of the propensity
functions. The results depicted in Fig.~\ref{fig-6} clearly
demonstrate the effectiveness of this correction.

%
%

\vspace{-9pt}
\section{Multiscale methods}

\vspace{-9pt}

As we discussed earlier in this review (see Section~IV-C-6), the
reactions in a Markovian reaction network may occur at different
time scales, with slow reactions occurring infrequently and fast
reactions firing numerous times between successive occurrences of
slow reactions. This may appreciably increase the computational
effort required to sample the master equation by Monte Carlo, which
can make analysis of Markovian reaction networks very difficult to
perform in practice. In this section, we discuss methods available
to address this problem. The main idea is to eliminate the fast
reactions by approximating the master equation with one that
consists of only slow reactions.

\newpage

\vspace{-12pt}

\subsection{Partitioning approximation}

\vspace{-6pt}

In many cases of interest, it is not important to know the detailed
activities of fast reactions, since the dynamic evolution of the
state of a Markovian reaction network may be mostly determined
by the slow reactions. If that is true, we may be able to approximate
the master equation by one that consists only of slow reactions.
If a sufficiently accurate approximation of the master equation
can be found in terms of slow reactions, then it can be used to
appreciably reduce the computational complexity associated with
Monte Carlo sampling. This is due to the fact that sampling slow
reactions is appreciably more efficient than sampling fast
reactions. This idea has led to the development of techniques
for eliminating fast reactions, known as \textit{multiscale} or
\textit{partitioning approximation} methods~\cite{HaseRawl:02,%
RaoArki:03,PuchKier:04,CaoGillPetz:05a,ELiuEijn:05,Gout:05,%
HaseRawl:05,SaliKazn:05,SamaVlac:05,BallKurtPopo-etal:06,%
ELiuEijn:07,MastHaseRawl:07,UribVergTzaf:08,ChevSama:09,%
CrudDebuRadu:09,ZeroSant:10,CottZygaKevr-etal:11,PahlAtzbKham:11,%
BostKessShne-etal:12,KangKurt:12,ThomStraGrim:12}.

To illustrate the main steps underlying most multiscale approximation
schemes, let us assume that the first~$M_s$ reactions of a Markovian
reaction network are slow, whereas, the remaining $M_{\! f}=M-M_s$
reactions are fast. We can then decompose the DA process $\bfZ(t)$
into two components, $\bfZ_{\! s}(t)$ and $\bfZ_{\! f}(t)$, with the
first component corresponding to the slow reactions and the second
corresponding to the fast reactions. In this case, Eq.~(\ref{eq-II-6})
leads to the following master equation~\cite{Gout:05}:
\beqa
&& \hspace{-10mm} \frac{\partial p_{\bfz_{\! s}}(\bfz_{\! s};t)}{\partial t} ~=\!\!
\sum_{m\in\sM_{\! s}} \!\! {\Bigl \{} \alpha_m (\bfz_{\! s}
- \overline{{\bf e}}_m;t) p_{\bfz_{\! s}}(\bfz_{\! s}
- \overline{{\bf e}}_m;t) \nn \\ [-6pt]
&& \hspace{22mm} - \: \alpha_m(\bfz_{\! s};t)
p_{\bfz_{\! s}}(\bfz_{\! s}; t) {\Bigr \}},
\label{eq-V-1}
\eeqa
for $t >0$, with
\beq
\alpha_m (\bfz_{\! s};t) \: := \: \sum_{\bfz_{\! f}}
\alpha_m(\bfz_{\! s},\bfz_{\! f}) p_{\bfz_{\! f} \mid
\bfz_{\! s}}(\bfz_{\! s},\bfz_{\! f}; t) ,
\label{eq-V-2}
\eeq
for $m \in \sM_{\! s}$,
where $p_{\bfz_{\! s}}(\bfz_{\! s};t)$ is the marginal probability
distribution over the DAs of the slow reactions, $p_{\bfz_{\! f}
\mid \bfz_{\! s}}(\bfz_{\! s},\bfz_{\! f}; t)$ is the conditional
probability of the fast DAs at time $t$ given the DAs of the slow
reactions, $\sM_{\! s} := \{1,2,\ldots,M_s\}$, and $\overline{{\bf e}}_m$
is a vector comprised of the first $M_s$ rows of $\bfe_m$ (the $\mth$
column of the $M \times M$ identity matrix).

According to Eq.~(\ref{eq-V-1}), the DAs of the slow reactions follow
a master equation that is similar to the one governing the entire
Markovian reaction network, albeit with time-varying propensities.
The fast reactions exert their influence on the slow reactions
through the propensity functions given by Eq.~(\ref{eq-V-2}), which
are computed as the conditional means of the original propensity
functions given the DAs of the slow reactions. As a consequence, if
we can evaluate the mean propensity functions $\alpha_m (\bfz_{\! s};t)$,
$m \in \sM_{\! s}$, then we can efficiently simulate the stochastic
evolutions of the DAs of the slow reactions by using the exact
Gillespie algorithm, or any other appropriate technique, modified
to account for the fact that the propensity functions are now
time-dependent~\cite{RaoArki:03,HaseRawl:02}. Moreover, given that
$\bfZ_{\! s}(t)=\bfz_{\! s}$, we can estimate the population process
$X_n(t)$ by using the \textit{optimum} mean-square estimate
$\widehat{x}_n(t)$, given by [recall Eq.~(\ref{eq-II-4})]:
\beqa
&& \hspace{-10mm} \widehat{x}_n(t) = \E[X_n(t) \mid
\bfZ_{\! s}(t)=\bfz_{\! s}] \nn \\ [6pt]
&& \hspace{-1mm} = x_{0,n} + \!\!\!
\sum_{m \in \sM_{\! s}} \!\! s_{nm} z_m(t) +\!\!\!\! \sum_{m \in \sM_{\! f}}
\!\! s_{nm} \muz(m;t,\bfz_{\! s}),
\label{eq-V-3}
\eeqa
for $n \in \sN$,
where $\muz(m;t,\bfz_{\! s}):= \E[Z_m(t) \mid \bfZ_{\! s}(t)
= \bfz_{\! s}]$, for $m \in \sM_{\! f}$ $:=$ $\{M_{\! s}+1,M_{\! s}
+2,\ldots,M\}$, is the mean DA of the $\mth$ fast reaction at time~$t$,
given the state~$\bfz_{\!s}(t)$ of the slow reactions at $t$. Hence,
we can replace the Markovian reaction network by a ``slow'' Markovian
reaction subnetwork whose DA process is governed by Eqs.~(\ref{eq-V-1}),
(\ref{eq-V-2}) and whose population process is governed by
Eq.~(\ref{eq-V-3}).

Calculating the propensity functions $\alpha_m (\bfz_{\! s};t)$,
$m \in \sM_{\! s}$, and the means $\muz(m;t,\bfz_{\! s})$, $m \in
\sM_{\! f}$, requires knowledge of the conditional probability
$p_{\bfz_f \mid \bfz_s}(\bfz_s,\bfz_f;t)$. It can be shown that,
within the coarse time scale, the dynamic evolution of this
probability is approximately governed by the following master
equation~\cite{Gout:05}:
\beqa
&& \hspace{-12mm}
\frac{\partial p_{\bfz_{\! f} \mid \bfz_s}(\bfz_s,\bfz_{\! f};t)}{\partial t} \nn \\ [3pt]
&& \hspace{-8mm}
=\!\! \sum_{m\in\sM_{\! f}} \!\! {\Bigl \{} \alpha_m (\bfz_{\! s},
\bfz_{\! f} - \underline{{\bf e}}_m) p_{\bfz_{\! f} \mid \bfz_{\! s}}
(\bfz_{\! s},\bfz_{\! f}- \underline{{\bf e}}_m ;t) \nn \\ [-6pt]
&& \hspace{10mm}
- \alpha_m(\bfz_{\! s},
\bfz_{\! f}) p_{\bfz_f \mid \bfz_{\! s}}(\bfz_{\! s},\bfz_{\! f};t)
{\Bigr \}},
\label{eq-V-4}
\eeqa
for $t >0$,
where $\underline{{\bf e}}_m$ is a vector comprised of the last
$\sM_{\! f}$ rows of $\bfe_m$, which includes only the fast reactions.
This equation is derived by assuming that, within successive firings
of slow reactions, the ``slow'' Markovian reaction subnetwork behaves
like one with propensity functions that are not appreciably larger
than zero, since it is very unlikely that a slow reaction will occur
within that time scale.

It turns out that most multiscale approximation schemes currently
available in the literature follow a similar approach by decomposing
a Markovian reaction network into ``slow'' and ``fast'' Markovian
subnetworks based on appropriately chosen variables. Although the
method discussed above uses the DAs of individual
reactions~\cite{HaseRawl:02,Gout:05,SaliKazn:05}, similar methods
can be constructed using the \textit{net} DAs of reversible\footnote{A
reaction is called \textit{reversible} if it can occur in both
directions, arbitrarily labeled as ``forward'' and ``backward,''
with nonzero probability; otherwise, the reaction is called
\textit{irreversible}.} reactions~\cite{HaseRawl:05}, or the
populations of the underlying species~\cite{RaoArki:03,%
CaoGillPetz:05a,SamaVlac:05,UribVergTzaf:08,ChevSama:09,ZeroSant:10}.

Unfortunately, solving the master equation of the ``fast'' reaction
subnetwork [e.g., Eq.~(\ref{eq-V-4})] is as difficult as solving the
master equation of the entire network. Moreover, evaluating the
propensity functions of the ``slow'' subnetwork requires Monte Carlo
estimation in general, which adds to computational complexity. To
address these issues, a number of different approaches have been
proposed in the literature, based on the techniques discussed in
Section~IV. For example, it has been assumed that, within successive
firings of slow reactions, the fast reactions rapidly reach a
stationary state whose probability does not depend on time
$t$~\cite{RaoArki:03,CaoGillPetz:05a,ELiuEijn:05,HaseRawl:05,%
SamaVlac:05,ELiuEijn:07,UribVergTzaf:08,ZeroSant:10}. In this
case, we can set the right-hand-side of the master equation of
the ``fast'' reaction subnetwork equal to zero and use a numerical
technique (see Sec.~IV-B) to calculate the desired stationary
conditional probability of the ``fast'' variables given the
``slow'' variables. We can then evaluate the propensity functions
of the ``slow'' reaction subnetwork either by direct
summation~\cite{HaseRawl:05,CaoGillPetz:05a}, if computationally
feasible, or by Monte Carlo estimation~\cite{HaseRawl:05}.

Numerically solving the master equation of the ``fast'' reaction
subnetwork may not be easy, especially for large subnetworks.
Moreover, evaluating expectations by direct summation or Monte
Carlo estimation can be computationally demanding. The main
difficulty however with the previous approach is to verify that
the ``fast'' reaction subnetwork reaches a stationary state,
since there might be only a short induction time between
successive firings of slow reactions during which convergence
to steady-state may not occur.

We can avoid the previous problems by sampling the master equation
of the ``fast'' reaction subnetwork using the exact Gillespie
algorithm and evaluate the propensity functions of the ``slow''
reaction subnetwork by Monte Carlo estimation~\cite{ELiuEijn:05,%
ELiuEijn:07,SamaVlac:05}. Although this strategy is very general,
requiring no additional assumptions other than the ones leading
to Eqs.~(\ref{eq-V-1}), (\ref{eq-V-2}), and (\ref{eq-V-4}), the
embedded estimation step may require a large number of Monte Carlo
samples, which can substantially increase computations.

Exact sampling of the master equation of the ``fast'' reaction
subnetwork can be replaced by the Langevin~\cite{HaseRawl:02,%
HaseRawl:05,SaliKazn:05} or the Poisson~\cite{PuchKier:04}
approximations. Although this can speed-up sampling of the
``fast'' variables, it will not eliminate the need for evaluating
the propensity functions of the ``slow'' reaction subnetwork
using Monte Carlo estimation. Due to its discrete nature, the
Poisson approximation may be more preferable than the Langevin
approximation. However, for both approximations to be valid and
computationally efficient, it is necessary to determine a value
for the leaping parameter $\tau$ that is as large as possible
while still ensuring that occurrences of fast reactions within
a time interval $[t,t+\tau)$ do not appreciably affect the
propensity functions. Intuitively speaking, finding such value
may be possible due to the assumed futility of the fast reactions.
In practice however this may not be easy. Note finally
that the expected number of occurrences of each fast reaction
during $[t,t+\tau)$ will be much larger than one, a condition
that is required for the Langevin approximation to be valid.

The propensity functions of the ``slow'' reaction subnetwork can
be approximated by using Eq.~(\ref{eq-V-2}) and a Taylor
series expansion, such as the one given by Eq.~(\ref{eq-IV-30}),
of the propensity functions of the entire network around the
means of the ``fast'' variables. This can be done by evaluating the
conditional moments of the ``fast'' variables, given the values
of the ``slow'' variables, thus eliminating the need for
Monte Carlo estimation. If the propensity functions of the
``slow'' reaction subnetwork depend only linearly on the ``fast''
variables, then we only need to calculate the conditional means
of these variables. On the other hand, if the ``slow'' propensity
functions depend quadratically on ``fast'' variables, then we also
need to calculate the conditional covariances. We can perform
these calculations by employing a moment approximation scheme
applied on the master equation of the ``fast'' reaction
subnetwork~\cite{HaseRawl:02,RaoArki:03,CaoGillPetz:05a,Gout:05,%
HaseRawl:05,UribVergTzaf:08,ChevSama:09}. The accuracy of this
approach however depends on the degree of nonlinearity of the
network propensity functions in terms of the ``fast'' variables
and the particular moment approximation scheme used.

Most multiscale approximation methods developed so far are based
on a clear separation between fast and slow reactions. In reality
however this may not be possible. For this reason, it may be more
appropriate to develop techniques that involve more than two
separate time scales. We refer the reader
to~\textcite{ELiuEijn:05,ELiuEijn:07} and \textcite{HarrClan:06}
for two promising techniques along this direction.

We should finally point out that a few alternative multiscale
approximation schemes have been proposed in the literature,
namely two techniques related to the finite state projection
method~\cite{PeleMunsKham:06,PigoVulp:08}, a method based on
separating species in terms of their variance~\cite{HellLots:07},
a technique based on an adiabatic approximation using a stochastic
path integral~\cite{SiniHengNeme:09}, and a rigorous and versatile
technique based primarily on stochastic equations determining the
dynamic evolution of the population process
itself~\cite{BallKurtPopo-etal:06,KangKurt:12}. Although promising,
these methods have only been applied to very small reaction
networks. At this point, it is not clear how they will perform
when dealing with larger and more realistic networks, nor have
they been sufficiently compared to the other approaches discussed
in this section.

\vspace{-12pt}

\subsection{Example: Transcription regulation}

\vspace{-6pt}

To illustrate the effectiveness of a multiscale approximation method
for solving the master equation, we consider here a biochemical
reaction network comprised of six molecular species that interact
through the following ten reactions:
\beq
\begin{array}{lrcl}
\text{reaction}~1:~~    & X_1       & \rightarrow   & X_1 + X_2     \\ [2pt]
\text{reaction}~2:~~    & 2X_2      & \rightarrow   & X_3           \\ [2pt]
\text{reaction}~3:~~    & X_3       & \rightarrow   & 2 X_2         \\ [2pt]
\text{reaction}~4:~~    & X_3 + X_4 & \rightarrow   & X_5           \\ [2pt]
\text{reaction}~5:~~    & X_5       & \rightarrow   & X_3 + X_4     \\ [2pt]
\text{reaction}~6:~~    & X_3 + X_5 & \rightarrow   & X_6           \\ [2pt]
\text{reaction}~7:~~    & X_6       & \rightarrow   & X_3 + X_5     \\ [2pt]
\text{reaction}~8:~~    & X_5       & \rightarrow   & X_1 + X_5     \\ [2pt]
\text{reaction}~9:~~    & X_1       & \rightarrow   & \emptyset     \\ [2pt]
\text{reaction}~10:~~   & X_2       & \rightarrow   & \emptyset.
\end{array} \nn
\label{eq-V-5}
\eeq
This reaction network was originally proposed by~\textcite{Gout:05}
and can serve as a model for a particular type of transcription regulation
in single cells. As a matter of fact, reaction~1 can be used to model the
translation of an mRNA molecule~$X_1$ into a protein molecule~$X_2$,
whereas reactions~2 and~3 can be used to model dimerization of~$X_2$
into~$X_3$. On the other hand, reactions~4--7 can be used to model
the binding of dimer~$X_3$ on a gene~$X_4$, assuming that the promoter
of this gene has two binding sites for $X_3$, whereas, reaction~8 can be
used to model the transcription of~$X_4$ to mRNA molecules~$X_1$,
assuming that this process occurs when the promoter of $X_4$ is only
bound by one dimer~$X_3$. Finally, reactions~9 and~10 model degradation
of the mRNA and protein molecules~$X_1$ and~$X_2$, respectively. Here,
we slightly simplify the original model by assuming that the cell's
volume remains fixed at $10^{-15}$~liters (l). In this case, the
specific probability rate constants are set to $\kappa_1 =
0.043~\text{s}^{-1}$, $\kappa_2 = 0.083~\mbox{moles}\cdot
\mbox{l}^{-1}\cdot\text{s}^{-1}$, $\kappa_3 = 0.5~\text{s}^{-1}$,
$\kappa_4=0.0199~\mbox{moles}\cdot\mbox{l}^{-1}\cdot\text{s}^{-1}$,
$\kappa_5=0.4791~\text{s}^{-1}$, $\kappa_6 = 1.9926\times10^{-4}
~\mbox{moles}\cdot\mbox{l}^{-1}\cdot\text{s}^{-1}$, $\kappa_7=8.7658
\times10^{-12}~\text{s}^{-1}$, $\kappa_8=0.0715~\text{s}^{-1}$,
$\kappa_9=0.0039~\text{s}^{-1}$, and $\kappa_{10}=0.0007~\text{s}^{-1}$,
in agreement with the values used by~\textcite{Gout:05}. Finally, we
initialize the system by setting $X_1(0)=0$, $X_2(0)=2$, $X_3(0)=4$,
$X_4(0)=2$, and $X_5(0)=X_6(0)=0$; i.e., we assume that the system
contains initially two mRNA molecules, two copies of the same gene,
and four dimers.

Despite the modest size of the previous network, simulation using
exact Monte Carlo sampling of the master equation is computationally
intensive. It took about $2$~hours and $15$~minutes of CPU time on
a 2.20~GHz Intel Core~2 Duo processor running Windows~7 to obtain
$2,\!000$ samples of the population dynamics during a period of
$35$~minutes. This serious inefficiency is due to the stiffness of
the system caused by the reversible reactions associated with the
dimerization of protein $X_2$ (i.e., reactions~2 and~3) being much
faster than the remaining reactions. As a consequence, exact Monte
Carlo sampling is forced to spend a substantial amount of time
simulating the occurrences of these two reactions. Unfortunately,
we cannot appreciably reduce computational effort by using the
Poisson approximation since, for accurately solving the master
equation, stiffness constrains the leaping parameter~$\tau$ to
take a very small value thus deeming this approximation method
computationally comparable to exact sampling. Dimerization however
is reversible and occurs on a much faster timescale than the other
reactions. As a consequence, we expect its effect to largely cancel
out. Therefore, faithful simulation of dimerization may not be necessary.

If we set $\sM_s = \{1,4,5,6,7,8,9,10\}$ and $\sM_f = \{2,3\}$, then
the ``slow'' subsystem, comprised of the reactions in~$\sM_s$, will
be characterized by the master equation~(\ref{eq-V-1}) with propensity
functions given by [recall Eq.~(\ref{eq-V-2})]
\beqa
\alpha_1(\bfz_{\! s};t) &\!\!=\!\!&
\kappa_1(z_8-z_9), \nn \\
\alpha_4(\bfz_{\! s};t) &\!\!=\!\!&
\kappa_4[4-z_4+z_5-z_6+z_7+\muz(2;t,\bfz_{\! s}) \nn \\
&& \hspace{10mm} -\muz(3;t,\bfz_{\!s})](2-z_4+z_5), \nn \\
\alpha_5(\bfz_{\! s};t) &\!\!=\!\!& \kappa_5(z_4-z_5-z_6+z_7), \nn \\
\alpha_6(\bfz_{\! s};t) &\!\!=\!\!&
\kappa_6[4-z_4+z_5-z_6+z_7+\muz(2;t,\bfz_{\! s}) \nn \\
&& \hspace{10mm} -\muz(3;t,\bfz_{\!s})] (z_4-z_5-z_6+z_7), \nn \\
\alpha_7(\bfz_{\! s};t) &\!\!=\!\!&
\kappa_7(z_6-z_7), \nn \\
\alpha_8(\bfz_{\! s};t) &\!\!=\!\!&
\kappa_8(z_4-z_5-z_6+z_7), \nn \\
\alpha_9(\bfz_{\! s};t) &\!\!=\!\!& \kappa_9(z_8-z_9), \nn \\
\alpha_{10}(\bfz_{\! s};t) &\!\!=\!\!&
\kappa_{10}[2\!+\!z_1\!-\!z_{10}\!-\!2\muz(2;t,\bfz_{\! s})\!+\!2\muz(3;t,\bfz_{\!s})] , \nn
\label{eq-V-6}
\vspace{-12pt}
\eeqa
where $\muz(2;t,\bfz_{\! s})$ and $\muz(3;t,\bfz_{\!s})$ are the mean DAs
of the two fast reactions~2 and~3, respectively. Therefore, to calculate
these  propensities, we need to compute the difference $\muz(2;t,\bfz_{\! s})
-\muz(3;t,\bfz_{\!s})$. By assuming that the ``fast'' reaction subsystem
of the two dimerization reactions rapidly reaches equilibrium within
successive occurrences of slow reactions, we can show that
[see~\textcite{Gout:05}]
\beq
\muz(2;t,\bfz_{\! s})-\muz(3;t,\bfz_{\!s}) \!=\! \frac{1}{2}
{\Big [} A(\bfz_{\! s})\!-\!\sqrt{A^2(\bfz_{\! s})-4B(\bfz_{\! s})}
{\Big ]} \!,
\label{eq-V-7}
\eeq
where
\beq
\begin{array}{ll}
A(\bfz_{\! s})=1.5+z_1-z_{10}+(\kappa_3/2\kappa_2) \\ [6pt]
B(\bfz_{\! s})=0.25(1+z_1-z_{10})(2+z_1-z_{10}) \\
~~~~~~~~~~~~~ -(\kappa_3/2\kappa_2) (4-z_4+z_5-z_6+z_7) .
\end{array}
\label{eq-V-8}
\eeq
As a consequence, we can solve the master equation~(\ref{eq-V-1}) of
the ``slow'' reaction subsystem without having to solve the conditional
master equation~(\ref{eq-V-4}) of the ``fast'' subsystem, and estimate
the population process by using Eq.~(\ref{eq-V-3}), which depends only
on the difference $\muz(2;t,\bfz_{\! s})-\muz(3;t,\bfz_{\!s})$, and
Eqs.~(\ref{eq-V-7}), (\ref{eq-V-8}).

\begin{figure}
\includegraphics[width=3.4in]{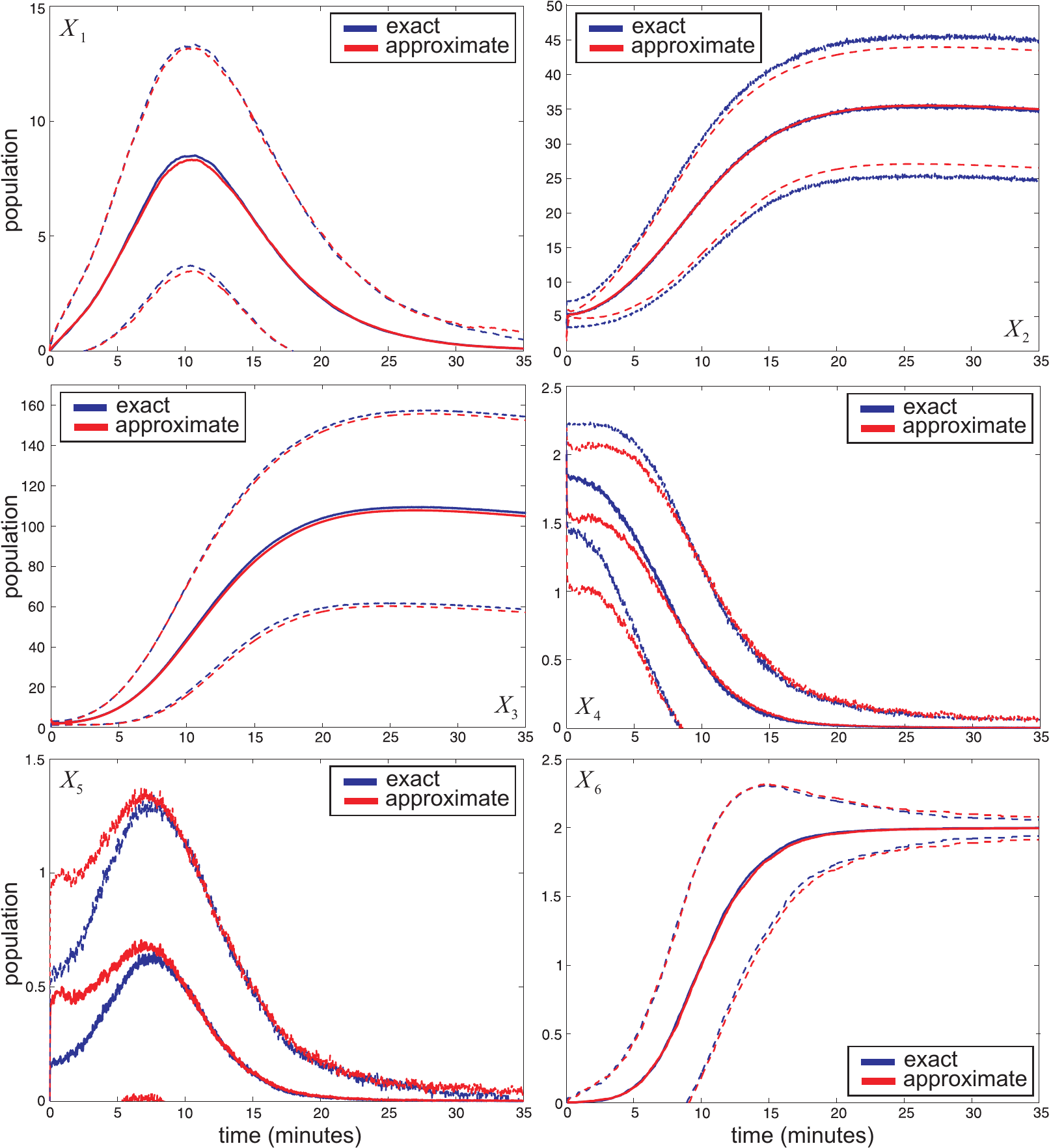}
\caption{Means (solid lines) and $\pm 1$ standard deviations (dashed lines)
of the population processes in the transcription regulation network example
of Section~V-B obtained by exact Monte Carlo sampling (blue lines) and
multiscale approximation (red lines).}
\label{fig-7}
\vspace{-12pt}
\end{figure}

It took less than a minute ($52$~seconds) of CPU time to draw $2,\!000$
Monte Carlo samples from the master equation of the ``slow'' reaction
subsystem (as compared to $135$~minutes of CPU time required by exact
Monte Carlo sampling). The mean and $\pm 1$ standard deviation dynamics
of the underlying population processes obtained by the two methods are
depicted in Fig.~\ref{fig-7}. These results clearly show that an
appropriately derived multiscale approximation of a stiff Markovian
reaction network can lead to dramatic improvements in computational
efficiency while producing a relatively accurate approximation of
the population dynamics. It turns out that the relatively large
transient errors in the population dynamics of $X_4$ and $X_5$
depicted in Fig.~\ref{fig-7} are due to incorrectly computing the
\textit{net} DA $Z(4)-Z(5)$ of reactions~4 and~5 (binding and
unbinding of dimer $X_3$ on the promoter of gene $X_4$), which is
a consequence of the imposed adiabatic approximation of the DAs of
the fast reactions~2 and~3 (dimerization) through their mean values.
Although this approximation also affects the accuracy of the
remaining population dynamics, the resulting approximate dynamics
track the ones computed by exact Monte Carlo sampling sufficiently
well.

%
%

\newpage

\vspace{-12pt}

\section{Mesoscopic (probabilistic) behavior}

\vspace{-6pt}

An important goal when studying Markovian reaction networks is to
investigate the existence, uniqueness, and stability of a stationary
solution of the underlying master equation and derive mathematical
properties of the dynamic behavior of the probability distribution
of the system state. This can be done by using a \textit{mesoscopic}
description of the network in terms of the population probabilities
$\{\px(\bfx;t), \bfx \in \sX\}$, for $t \geq 0$. To avoid mathematical
subtleties, which are outside the scope of this review, we assume
here that the cardinality of the population state-space $\sX$ is
\textit{finite}. Most results however can be extended to the case
of countable state-spaces.

To derive a stationary solution of the master equation~(\ref{eq-II-8}),
we must solve the system of $K$ linear equations $\P \bfp =0$;
recall Eq.~(\ref{eq-IV-1}). Since the elements of each column of
matrix $\P$ add to zero, its rows are linearly dependent and, therefore,
the rank of $\P$ will be less than~$K$. As a consequence, $\P \bfp =0$
will have at least one nontrivial solution. Unfortunately, this result
does not tell us how many nontrivial solutions exist and which ones are
valid probability distributions; i.e., which solutions satisfy the
necessary constraints
\beq
0 \leq p_k \leq 1, ~~ \mbox{for $k =1,2,\ldots,K$},
~~ \mbox{and} ~~ \sum_{k=1}^K p_k = 1. \nn
\label{eq-VI-1}
\eeq

In the following, we first consider the dynamic behavior of an
\textit{irreducible} Markovian reaction network. This type of
network is defined by the property that, for any pair
$(\bfx,\bfx')$ of population states, there exists at least
one sequence of reactions that takes the system from state
$\bfx$ to state $\bfx'$ -- these states are said to be
\textit{communicating}. By using a simple graph theoretic
analysis and Kirchhoff's theorem, \textcite{Schn:76} has
shown that an irreducible Markovian reaction network converges to
a \textit{unique} probability distribution $\overline{\bfp}$
at steady-state, which does not depend on the initial probability
distribution $\bfp(0)$, such that $0 < \overline{\bfp} < 1$\footnote{
$a \leq {\bf v} \leq b$ denotes that each element of vector ${\bf v}$
satisfies this relation.} [see also~\textcite{Kamp:07}]. As a
consequence, in an irreducible Markovian reaction network, the
population process can take any value in $\sX$ at steady-state
with \textit{nonzero} probability.

On the other hand, the theory of systems of ordinary differential
equations with constant coefficients implies that, for a given
initial probability distribution $\bfp(0)$, Eq.~(\ref{eq-IV-1})
is satisfied by a \textit{unique} probability distribution $\bfp(t)$,
$t > 0$, which is analytic for all $0 \leq t < \infty$. Since
the elements of each column of matrix $\P$ add to zero,
\beq
\frac{d [\bfe^T \bfp(t)]}{d t}  = \bfe^T \frac{d \bfp(t)}{d t}
= \bfe^T \P \: \bfp(t) = 0, \nn
\label{eq-VI-2}
\eeq
where $\bfe$ is a $K \times 1$ vector with all its elements being
equal to one. This result, together with the fact that $\bfe^T
\bfp(0) =1$, implies $\bfe^T \bfp(t)=1$, for all $t \geq 0$.
Unfortunately, it is not clear whether $0 \leq \bfp(t) \leq 1$,
for every $t > 0$. It turns out however that, for an irreducible
Markovian reaction network, $0 < \bfp(t) < 1$, for every
$t > 0$~\cite{Schn:76}.

If $\lambda_k$, $k=1,2,\ldots,K$, are the eigenvalues of matrix $\P$,
with corresponding right and left eigenvectors $\bfr_k$, $\bfl_k$,
$k=1,2,\ldots,K$, respectively, then the solution to Eq.~(\ref{eq-IV-1})
is given by~\cite{MoleLoan:03}
\beq
{\bfp}(t) = \exp \left ( \P t \right ) \bfp(0) =
\sum_{k=1}^K c_k \: \bfr_k \: e^{\lambda_k t},
\label{eq-VI-3}
\eeq
for $0 \leq t \leq \infty $, where we assume here that the
eigenvalues of $\P$ have the same algebraic and geometric
multiplicity, an assumption satisfied by many Markovian reaction
networks. In this case, the right and left eigenvectors are
biorthogonal (i.e., $\bfl^T_k \bfr_{k'}=0$, for every $k \not= k'$),
which implies that the constants $c_k$ are
given by $c_k = \bfl_k^T {\bfp(0)}/\bfl_k^T \bfr_k$. As a consequence,
we can use the eigenvalues and eigenvectors of $\P$ to analytically
specify the entire mesoscopic behavior of a Markovian reaction network.
See~\textcite{KeelRoss:08} and~\textcite{VellQian:09} for application
of Eq.~(\ref{eq-VI-3}) to problems in epidemiology and computational
biochemistry.

Note that Eq.~(\ref{eq-VI-3}) and the fact that a non-trivial
stationary solution always exists imply that at least one
eigenvalue of $\P$ must be zero. For an irreducible Markovian
reaction network, matrix $\P$ has only one zero eigenvalue,
with the remaining $K-1$ eigenvalues having negative real
parts~\cite{Schn:76}. If we therefore assume that
$\lambda_1=0$, then Eq.~(\ref{eq-VI-3}) implies that the
stationary distribution will be given by $\overline{{\bfp}}
= \bfr_1/ \| \bfr_1 \|$, where~$\bfr_1$ is the eigenvector
corresponding to the zero eigenvalue and $\| \bfr \|$ is the
$\ell_1$-norm of vector $\bfr$. It turns out that the solution
$\bfp(t)$, $t \geq 0$, of Eq.~(\ref{eq-IV-1}) is asymptotically
stable with respect to $\overline{\bfp}$, in the sense that
\beq
\lim_{t \rightarrow \infty} D[\bfp(t), \overline{\bfp}] = 0, \nn
\label{eq-VI-4}
\eeq
where
\beq
D[\bfp,\bfq] := \sum_{k=1}^K p_k \ln \frac{p_k}{q_k}
\label{eq-VI-5}
\eeq
is the Kullback-Leibler distance between the two probability
distributions $\bfp=\{p_k,k=1,2,\ldots,K\}$ and
$\bfq=\{q_k, k=1,2,\ldots,K\}$. As a matter of fact,
$d D[\bfp(t), \overline{\bfp}]/dt \leq 0$, where equality
is archived only at steady-state.

To summarize, for a given initial probability vector $\bfp(0)$, the
master equation associated with an irreducible Markovian reaction
network has a \textit{unique} and strictly positive solution
$0 < \bfp(t) < 1$, $0 < t \leq \infty$. This solution is analytic
for all $0 \leq t < \infty$, converges to a stationary distribution
$0 < \overline{\bfp} < 1$ that does not depend on the initial
probability distribution $\bfp(0)$, and is asymptotically stable
with respect to $\overline{\bfp}$.

It is not in general easy to check whether a Markovian reaction
network is irreducible. However, we often assume that a given
Markovian reaction network is comprised of only reversible
reactions (reactions which can occur in both directions
with nonzero probability). This is a plausible assumption since,
in principle, a transition between two physical states can occur
in the reverse direction as well. In this case, and after
appropriately ordering the states, matrix $\P$ can be cast into
a block diagonal form with diagonal elements $\P^{(1)},\P^{(2)},
\ldots,\P^{(J)}$, for some~$J$, where each submatrix $\P^{(j)}$
is irreducible (when $J=1$, matrix $\P$ is itself irreducible).
The resulting Markovian reaction network is said to be
\textit{completely reducible}~\cite{Kamp:07}. In this case,
the original Markovian reaction network can be decomposed into
$J$ non-interacting subnetworks with non-overlapping state-spaces,
which can be treated independently of each other. Each reaction
subnetwork is characterized by \textit{unique} dynamic and
stationary solutions $\bfp^{(j)}(t)$, $\overline{\bfp}^{(j)}$,
$j=1,2,\ldots,J$, that satisfy the aforementioned properties.
However, the dynamic and stationary solutions of the original
master equation are determined by the initial condition at time
$t=0$. If the master equation is initialized with a population
vector in the state-space of the $\jth$ subnetwork, then its
dynamic and stationary solution will be given by
\beq
\left [
\begin{array}{c}
{\bf0} \\
\vdots \\ [6pt]
{\bfp}^{(j)}(t) \\
\vdots \\
{\bf0}
\end{array}
\right ] \qquad \mbox{and} \qquad \left [
\begin{array}{c}
{\bf0} \\
\vdots \\ [6pt]
\overline{{\bfp}}^{(j)} \\
\vdots \\
{\bf0}
\end{array}
\right ], \nn
\label{eq-VI-7}
\eeq
respectively, where ${\bfp}^{(j)}(t)$ depends on the initial
condition and $\overline{{\bfp}}^{(j)}$ does not.

A question that arises at this point is what happens when the
Markovian reaction network contains irreversible reactions and
matrix $\P$ is not irreducible. To get an idea, let us assume
that, after appropriately ordering the states,
\beq
\P = \left [
\begin{array}{ll}
\P^{(1)}    & \T^{(1)} \\ [6pt]
{\bf0}      & \T^{(2)}
\end{array} \!\!
\right ], \nn
\label{eq-VI-8}
\eeq
where $\P^{(1)}$ and $\T^{(2)}$ are square matrices, $\P^{(1)}$
is irreducible, and at least one element of each column of $\T^{(1)}$
is strictly positive. Note that the nonzero elements of $\T^{(1)}$
correspond to nonreversible reactions. The associated Markovian
reaction network is said to be \textit{incompletely
reducible}~\cite{Kamp:07}. If we denote by ${\bfp}^{(1)}(t)$
and ${\bfp}^{(2)}(t)$ the probability distributions of the state
vectors at time $t$, determined by the partition of the state-space
suggested by matrix $\P$, then the master equation results in the
following two differential equations:
\beqa
\frac{d{\bfp}^{(1)}(t)}{dt} &\!\!=\!\!& \P^{(1)} {\bfp}^{(1)}(t) + \T^{(1)}
{\bfp}^{(2)}(t) \nn \\ [6pt]
\frac{d{\bfp}^{(2)}(t)}{dt} &\!\!=\!\!& \T^{(2)} {\bfp}^{(2)}(t). \nn
\label{eq-VI-9}
\eeqa
Clearly, one can solve the second equation independently from the
first to obtain
\beq
{\bfp}^{(2)}(t) = \exp \! {\bigl \{} \T^{(2)} t {\bigr \}} \:
{\bfp}^{(2)}(0). \nn
\label{eq-VI-10}
\eeq
On the other hand, the dynamic behavior of ${\bfp}^{(1)}$ is now
driven by ${\bfp}^{(2)}(t)$, unless ${\bfp}^{(2)}(0)=0$, in
which case ${\bfp}^{(1)}(t) = \exp \! {\bigl \{} \P^{(1)} t {\bigr \}}
{\bfp}^{(1)}(0)$.
Note however that
\beqa
\frac{d[\bfe^T \bfp^{(2)}(t)]}{dt}
&\!\!=\!\!& \bfe^T \frac{d \bfp^{(2)}(t)}{dt} \nn \\ [6pt]
&\!\!=\!\!& \bfe^T \T^{(2)} {\bfp}^{(2)}(t) \nn \\ [6pt]
&\!\!=\!\!& - \bfe^T \T^{(1)} {\bfp}^{(2)}(t) < 0, \nn
\label{eq-VI-11}
\eeqa
provided that ${\bfp}^{(2)}(t) \not= {\bf0}$, for every
$t \geq 0$, since the elements of each column of matrix $\P$ add
to zero and we have assumed that each column of matrix $\T^{(1)}$
contains at least one element that is strictly positive. Therefore,
${\bfp}^{(2)}(t)$ asymptotically becomes zero as $t \rightarrow
\infty$. As a matter of fact, ${\bfp}^{(2)}(t)$ assigns
probability mass over the \textit{transient} states of the Markovian
reaction network, as opposed to ${\bfp}^{(1)}(t)$ that assigns
probability mass over the \textit{persistent} states. In this case,
and when matrix $\P^{(1)}$ is irreducible, the stationary solution
of the master equation governing an incompletely reducible Markovian
reaction network will be unique and given by the probability vector
\beq
\overline{{\bfp}} = \left [
\begin{array}{l}
\overline{{\bfp}}^{(1)} \\ [6pt]
{\bf0}
\end{array}
\! \right ], \nn
\label{eq-VI-12}
\eeq
where $\overline{{\bfp}}^{(1)}$ is the (unique) solution of the linear
system of equations $\P^{(1)} {\bfp} = {\bf0}$.

In general, the population states in a Markovian reaction network can
be classified into two distinct groups: \emph{transient} and \emph{persistent}.
These states can be uniquely partitioned into non-overlapping sets
$T$ and $P_j$, $j=1,2,\ldots,J$, where $T$ contains all transient
states and $P_j$, $j=1,2,\ldots,J$, are irreducible sets containing
persistent states with the additional property that, for every
$j \not= j'$, each state in $P_j$ does not communicate with any
state in~$P_{j'}$. By appropriately ordering the states, we can
write matrix $\P$ in the form
\beq
\P = \left [
\begin{array}{ccccccc}
\P^{(1)}    & {\bf0}    & \cdot & \cdot & \cdot & {\bf0}    & \T^{(1)}      \\ [6pt]
{\bf0}      & \P^{(2)}  & \cdot & \cdot & \cdot & {\bf0}    & \T^{(2)}      \\
\cdot       & \cdot     & \cdot &       &       & \cdot     & \cdot         \\ [-3pt]
\cdot       & \cdot     &       & \cdot &       & \cdot     & \cdot         \\ [-3pt]
\cdot       & \cdot     &       &       & \cdot & \cdot     & \cdot         \\
{\bf0}      & {\bf0}    & \cdot & \cdot & \cdot & \P^{(J)}  & \T^{(J)}      \\ [6pt]
{\bf0}      & {\bf0}    & \cdot & \cdot & \cdot & {\bf0}    & \T
\end{array}
\right ], \nn
\label{eq-VI-13}
\eeq
where $\P^{(j)}$ is a square irreducible matrix that characterizes how
probability mass is dynamically distributed among the persistent states
in $P_j$, $\T^{(j)}$ is a matrix that tells us how probability mass is
transferred from the transient states in $T$ to persistent states in $P_j$,
and $\T$ is a square matrix that characterizes how probability mass is
dynamically distributed among the transient states in~$T$. In this case,
if the Markovian reaction network is initialized by a persistent state
in~$P_j$, then the stationary solution will be given by the probability
vector
\beq
\bfg_j = \left [
\begin{array}{c}
{\bf0} \\
\cdot  \\ [-3pt]
\cdot  \\ [-3pt]
\cdot  \\
\overline{{\bfp}}^{(j)} \\
\cdot  \\ [-3pt]
\cdot  \\ [-3pt]
\cdot  \\
{\bf0}
\end{array}
\! \right ], \nn
\label{eq-VI-14}
\eeq
where $\overline{{\bfp}}^{(j)}$ is the unique stationary distribution
of the $\jth$ irreducible Markovian reaction subnetwork characterized
by matrix $\P^{(j)}$. However, if the network is initialized with the
$\ith$ transient state in $T$, then the stationary distribution
$\overline{\bfp}_i$ (which now depends on $i$) will be given by a
convex combination of the stationary distributions~$\bfg_j$ above,
with mixing coefficients $\mu_{ij}$; i.e., we have that
\beq
\overline{\bfp}_i = \sum_{j=1}^J  \mu_{ij} \bfg_j ,
\label{eq-VI-15}
\eeq
where
\beq
\mu_{ij} \geq 0 \qquad \mbox{and} \qquad \sum_{j=1}^J \mu_{ij} =1. \nn
\label{eq-VI-16}
\eeq
As a matter of fact, Eq.~(\ref{eq-VI-15}) simply expresses the fact that
the probability of a Markovian reaction network initialized with the
$\ith$ transient state in $T$ to reach a persistent population state
$\bfx$ in $P_j$ at steady-state equals the probability $\mu_{ij}$ that
the system will reach a persistent state in $P_j$ at steady-state
multiplied by the probability that this state will be $\bfx$. It
can be shown that
\beq
\mu_{ij} = - \sum_{i' \in T} \sum_{j' \in P_j} [ \T^{(j)} ]_{j'i'}
[\T^{-1}]_{i'i}, \nn
\label{eq-VI-17}
\eeq
where $[ \T^{(j)} ]_{j'i'}$ is the $(j',i')$ element of matrix $\T^{(j)}$ and
$[\T^{-1}]_{i'i}$ is the $(i',i)$ element of the inverse of matrix $\T$.

To summarize, a fundamental property of the master equation~(\ref{eq-II-8})
associated with a Markovian reaction network is that, when this
equation is initialized with a persistent state, its solution
converges to a unique stationary distribution that assigns
positive probability only to the persistent states that
communicate with the initial state. On the other hand, if the
Markovian reaction network is initialized with a transient
state, then its stationary distribution will be a convex
combination of the distinct stationary distributions obtained
by initializing the system with persistent states chosen from
each individual irreducible set.

%
%

\vspace{-12pt}

\section{Potential energy landscape}

\vspace{-6pt}

To better understand what might happen at steady-state, let us
assume that the master equation~(\ref{eq-II-8}) has a unique
stationary solution $\overline{p}_{\ss \bf X}(\bfx) :=
\lim_{t \rightarrow \infty} \px(\bfx;t)$ that is independent
of the initial state. The probability distribution
$p_{\ss \widetilde{\bfX}}(\widetilde{\bfx};t)$ of the population
density process $\widetilde{\bfX}(t;\Omega)=\bfX(t)/\Omega$ will
be given by $p_{\ss \widetilde{\bfX}}(\widetilde{\bfx};t)
= \Omega \px(\Omega \widetilde{\bfx};t)$ and will depend on
the size parameter $\Omega$ in general. Let us define the
function
\beq
V(\widetilde{\bfx};\Omega) := - \frac{1}{\Omega} \ln
\frac{\overline{p}_{\ss \widetilde{\bfX}}(\widetilde{\bfx})}
{\overline{p}_{\ss \widetilde{\bfX}}(\widetilde{\bfx}_*)},
\label{eq-VII-1}
\eeq
where $\overline{p}_{\ss\widetilde{\bfX}}(\widetilde{\bfx}) :=
\lim_{t \rightarrow \infty} p_{\ss \widetilde{\bfX}}(\widetilde{\bfx};t)$
is the steady-state distribution of the population density process
and $\widetilde{\bfx}_*$ is a state at which the stationary
probability distribution $\overline{p}_{\ss \widetilde{\bfX}}
(\widetilde{\bfx})$ attains its maximum value. Note that
$V(\widetilde{\bfx};\Omega) \geq 0$. Moreover, and as a
consequence of Eq.~(\ref{eq-VII-1}), we have that
\beq
\overline{p}_{\ss \widetilde{\bfX}}(\widetilde{\bfx}) = \frac{1}{\zeta(\Omega)}
\exp {\Bigl \{} \! - \Omega V(\widetilde{\bfx};\Omega) {\Bigr \}},
\label{eq-VII-2}
\eeq
where
\beq
\zeta(\Omega) := \sum_{\bfu} \exp {\Bigl \{} \! - \Omega V(\bfu;\Omega) {\Bigr \}} .
\label{eq-VII-3}
\eeq
In this case, $\overline{p}_{\ss \widetilde{\bfX}}(\widetilde{\bfx})$ is a
Gibbs distribution with ``potential energy'' function $V(\widetilde{\bfx};
\Omega)$, ``temperature'' $1/\Omega$, and partition function $\zeta(\Omega)$.
Clearly, $V(\widetilde{\bfx};\Omega)$ assigns minimum (zero) potential to
the states of maximum probability at steady-state and infinite potential
to the states of zero probability.

We will now assume that, around the thermodynamic limit, the potential
function $V(\widetilde{\bfx};\Omega)$ is an analytic function of
$\Omega^{-1}$. Then, a Taylor series expansion with respect to
$\Omega^{-1}$ around zero (i.e., around the thermodynamic limit)
results in
\beqa
V(\widetilde{\bfx};\Omega) &\!\!=\!\!& V(\widetilde{\bfx};\infty)
+ \frac{1}{\Omega} \frac{\partial V(\widetilde{\bfx};\infty)}
{\partial \Omega^{-1}} + \cdots \nn \\ [6pt]
&\!\!=\!\!& V_0(\widetilde{\bfx}) + \frac{1}{\Omega}
V_1(\widetilde{\bfx}) + \cdots ,
\label{eq-VII-4}
\eeqa
where
\beqa
V_0(\widetilde{\bfx}) := V(\widetilde{\bfx};\infty) = - \lim_{\Omega
\rightarrow \infty} \frac{1}{\Omega} \ln \frac{\overline{p}_{\ss \widetilde{\bfX}}
(\widetilde{\bfx})}{\overline{p}_{\ss \widetilde{\bfX}}(\widetilde{\bfx}_0)}
\geq 0 , \nn
\label{eq-VII-5}
\eeqa
and
\beqa
V_1(\widetilde{\bfx}) := \frac{\partial V(\widetilde{\bfx};\infty)}
{\partial \Omega^{-1}} . \nn
\label{eq-VII-6}
\eeqa
As a consequence of Eqs.~(\ref{eq-VII-2})--(\ref{eq-VII-4}), and for
sufficiently large $\Omega$, we have that
\beq
\overline{p}_{\ss \widetilde{\bfX}}(\widetilde{\bfx}) \simeq \frac{1}{\zeta(\Omega)}
\exp {\Bigl \{} \! - \Omega V_0(\widetilde{\bfx}) - V_1(\widetilde{\bfx})
{\Bigr \}}, \nn
\label{eq-VII-7}
\eeq
where the partition function is now given by
\beq
\zeta(\Omega) = \sum_{\bfu} \exp {\Bigl \{} \! - \Omega V_0(\bfu) - V_1(\bfu)
{\Bigr \}} . \nn
\label{eq-VII-8}
\eeq
Therefore, the stationary probability distribution $\overline{p}_{\ss \widetilde{\bfX}}
(\widetilde{\bfx})$ satisfies a large deviation principle with
rate function~$V_0(\widetilde{\bfx})$~\cite{Touc:09}. It turns out that,
if $\bfchi(t)$ satisfies the macroscopic equations~(\ref{eq-IV-46}), then
\beqa
\frac{d V_0(\bfchi(t))}{dt} = \sum_{n \in \sN} \frac{\partial
V_0(\bfchi(t))}{\partial \chi_n(t)}
\frac{d \chi_n(t)}{dt} \leq 0, \nn
\label{eq-VII-9}
\eeqa
with equality if and only if
\beqa
\frac{d \chi_n(t)}{dt} = 0, \quad \mbox{for~every~$n \in \sN$}. \nn
\label{eq-VII-10}
\eeqa
Therefore, the solution $\bfchi(t)$ of the macroscopic
equation~(\ref{eq-IV-46}) produces a downhill motion in the
value of the potential energy function $V_0$ (which is a
Lyapunov function for the macroscopic system) until it
asymptotically reaches a stable stationary state.

For a given value of $\Omega$, we can view the multidimensional
surface $V_0(\widetilde{\bfx};\Omega)$ as a \textit{potential
energy landscape}~\cite{AoKwonQian:07,WangXuWang:08,ZhouQian:11,%
WangZhanXu-etal:11}. The stable stationary states of Eq.~(\ref{eq-IV-46})
correspond to \textit{potential wells} (basins of attraction) associated
with the (local or global) minima of $V_0$, which are separated
by barriers corresponding to hills (unstable states) and saddles
(transitional states -- states on the potential energy surface from
which stable states are equally accessible). Which minimum will be
reached depends on the initial condition that must be a point in
the potential well of that minimum. It turns out that, once the
macroscopic system, given by Eq.~(\ref{eq-IV-46}), reaches a
stable stationary state, it will stays there forever. As a
consequence, the macroscopic system will only move downhill
on the potential energy landscape.

It is now not difficult to see that Eqs.~(\ref{eq-VII-2})--(\ref{eq-VII-4})
imply that
\beqa
\lim_{\Omega \rightarrow \infty} \overline{p}_{\ss \widetilde{\bfX}}(\widetilde{\bfx})
&\!\!=\!\!& \lim_{\Omega \rightarrow \infty} \frac{\exp {\Bigl \{} \!
- \Omega V_0(\widetilde{\bfx}) {\Bigr \}} \exp {\Bigl \{} \!
- V_1(\widetilde{\bfx}) {\Bigr \}}}{\sum_{\bfu} \exp{\Bigl \{} \!
- \Omega V_0(\bfu) {\Bigr \}} \exp {\Bigl \{} \! - V_1(\bfu) {\Bigr \}}} \nn \\ [6pt]
&\!\!=\!\!& \left \{
\begin{array}{ll}
{\displaystyle \frac{\exp \{ -V_1(\widetilde{\bfx})\}}
{\sum_{\bfu \in \sG_*} \exp \{ -V_1(\widetilde{\bfu}) \}}},
& ~\mbox{if~~$\widetilde{\bfx} \in \sG_*$} \\ [12pt]
~~~~~~~~~~~0, & ~\mbox{otherwise,}
\end{array} \right . \nn
\label{eq-VII-11}
\eeqa
where $\sG_*$ is the set of all \textit{ground states} (global minima)
of the potential energy landscape $V_0$ (i.e., the set of all states
at which $V_0=0$). As a consequence, only the ground states have a
non-negligible probability to be observed as $\Omega$ becomes large
because $\overline{p}_{\ss \widetilde{\bfX}}(\widetilde{\bfx})$
decays exponentially with $\Omega$, for every $\widetilde{\bfx}
\notin \sG_*$. This implies that, in the thermodynamic limit, the
master equation~(\ref{eq-II-8}) will asymptotically converge
almost surely to a ground state of the potential energy function~$V_0$,
independently of the initial state. The particular ground state is
chosen with probability determined by the values of the potential
energy function~$V_1$ over the ground states of $V_0$. On the other
hand, the macroscopic equation~(\ref{eq-IV-46}) will reach a
minimum of $V_0$, which may or may not be a ground state in $\sG_*$
depending on the initial condition. If the macroscopic equation has
a unique asymptotically stable stationary solution that is independent
of the initial condition, then $V_0$ will have only one (global)
minimum. In this case, and as we mentioned before, the master
equation~(\ref{eq-II-8}) will converge almost surely to the same
state in the thermodynamic limit. However, if $V_0$ contains more
than one minimum, then the stationary solution of the master
equation~(\ref{eq-II-8}) may be different from the stationary
solution predicted by the corresponding macroscopic
equation~(\ref{eq-IV-46}). As a consequence,
\beq
\lim_{\Omega \rightarrow \infty} \lim_{t \rightarrow \infty}
p_{\ss \widetilde{\bfX}}(\widetilde{\bfx};t) \not=
\lim_{t \rightarrow \infty} \lim_{\Omega \rightarrow \infty}
p_{\ss \widetilde{\bfX}}(\widetilde{\bfx};t) \nn
\label{eq-VII-12}
\eeq
in general. This distinct difference between the stationary behavior
of the master equation (left-hand side of inequality) and macroscopic
equation (right-hand side of inequality) is known as \textit{Keizer's
paradox}~\cite{Keiz:87,VellQian:07,Qian:10}.

At finite sizes $\Omega$, the modes of the stationary probability
distribution $\overline{p}_{\ss \widetilde{\bfX}}(\widetilde{\bfx})$
(i.e., the most probable states) correspond to the minima of the
potential energy landscape $V(\widetilde{\bfx};\Omega)$. If
$\widetilde{\bfx}'$ is a minimum of $V_0(\widetilde{\bfx})
+ \Omega^{-1}V_1(\widetilde{\bfx})$, then $V_0(\widetilde{\bfx})
\geq V_0(\widetilde{\bfx}') - \Omega^{-1} \left [ V_1(\widetilde{\bfx})
- V_1(\widetilde{\bfx}' ) \right ]$, for every $\widetilde{\bfx} \in
\sW(\widetilde{\bfx}')$, where $\sW(\widetilde{\bfx}')$ is a local
neighborhood of $\widetilde{\bfx}'$ for which the inequality is
satisfied. For large enough $\Omega$, such that
\beq
\max_{\widetilde{\bfx}'} \max_{\widetilde{\bfx}
\in \sW(\widetilde{\bfx}')} \left \{ \frac{ V_1(\widetilde{\bfx})
- V_1(\widetilde{\bfx}')}{V_0(\widetilde{\bfx}')} \right \} \ll \Omega
< \infty, \nn
\label{eq-VII-13}
\eeq
where the first maximum is taken over all minima of $V_0(\widetilde{\bfx})
+\Omega^{-1}V_1(\widetilde{\bfx})$ which are not ground states of
$V_0(\widetilde{\bfx})$, we have $V_0(\widetilde{\bfx}')
\lesssim V_0(\widetilde{\bfx})$, for every $\widetilde{\bfx} \in
\sW(\widetilde{\bfx}')$, and therefore $\widetilde{\bfx}'$ will
approximately be a local minimum of the potential energy landscape $V_0$.
In this case, the peaks of the stationary probability distribution
$\overline{p}_{\ss \widetilde{\bfX}}(\widetilde{\bfx})$ will correspond
to stationary states of the macroscopic equation~(\ref{eq-IV-46}).\footnote{Note
that the converse of this result is not necessarily true. There might be
stationary states of the macroscopic equation that, for a given value
of~$\Omega$, do not introduce peaks in the stationary probability
distribution. To see this, recall that, in the limit as $\Omega
\rightarrow \infty$, the only peaks present in the stationary
probability distribution are the ones associated with the global
minima of~$V_0$.} For this reason, we can refer to the peaks in
$\overline{p}_{\widetilde{\ss \bf X}}(\widetilde{\bfx})$ as
\textit{macroscopic} modes.

At smaller values of $\Omega$, the stationary probability distribution
$\overline{p}_{\ss \widetilde{\bfX}}(\widetilde{\bfx})$ will be given
by Eqs.~(\ref{eq-VII-2}) and~(\ref{eq-VII-3}). The modes will now
depend on the fluctuation size parameter~$\Omega$ and will be
determined by the minima of the potential energy landscape
$V(\widetilde{\bfx};\Omega)$. However, a state that minimizes the
potential energy function $V_0$ may not necessarily minimize $V$,
in which case at least some modes of the probability distribution
$\overline{p}_{\ss \widetilde{\bfX}} (\widetilde{\bfx})$ will not be
predicted by the corresponding macroscopic equation. Since these
modes show up at small system sizes, in which appreciable stochastic
fluctuations may be present in the system due to ``intrinsic noise,''
we refer to them as \textit{noise-induced} modes. Recent
literature has documented the presence of noise-induced modes in
biochemical reaction networks and their importance in modeling
system behavior not accounted for by their macroscopic
counterparts~\cite{ArtyDasKard-etal:07,ArtyMathSamo-etal:09,%
QianShiXing:09,BishQian:10,ZhanGeQian:10,Qian:10,Qian:11}.

We finally note that, if a Markovian reaction network is at a
stable state $\widetilde{\bfx}_1^s$ at time $t_0$, then it may
switch to a another stable state~$\widetilde{\bfx}_2^s$ at
time $t_0 < t < \infty$ with probability $\widetilde{p}_{\ss \Omega}
(\widetilde{\bfx}_2^s;t)$. However, $\lim_{\Omega \rightarrow \infty}
\widetilde{p}_{\ss \Omega}(\widetilde{\bfx}_2^s;t) =
\delta(\widetilde{\bfx}_2^s\!\! -\bfchi(t))$, where $\bfchi(t)$ is the
solution of the macroscopic equations~(\ref{eq-IV-41}), initialized
with $\widetilde{\bfx}_1^s$. Since $\widetilde{\bfx}_1^s$ is a minimum
of the potential energy function $V_0$, the  macroscopic system
will be in state $\bfchi(t) = \widetilde{\bfx}_1^s$ at time~$t$.
Hence, $\lim_{\Omega \rightarrow \infty} \widetilde{p}_{\ss \Omega}
(\widetilde{\bfx}_2^s;t) =0$. As a consequence, the probability of
switching from a stable state to another stable state tends (in general
exponentially) to zero as the system size increases to infinity. At
finite system sizes $\Omega$, switching among stable stationary states
becomes possible, but the probability of switching is very small for
large $\Omega$; i.e., switching among stable stationary states are
\textit{rare} events~\cite{Qian:11,ZhouQian:11}. As a matter of fact,
we can approximate the waiting time for switching by an exponential
distribution~\cite{Aldo:82} with rate parameter that tends to zero
as $\Omega \rightarrow \infty$.

%
%

\vspace{-12pt}

\section{Macroscopic (thermodynamic) behavior}

\vspace{-6pt}

We can view a Markovian reaction network as a thermodynamic system
that absorbs energy, produces entropy, and dissipates heat~\cite{Prig:78,%
Schn:76,OonoPani:98,AndrGasp:04,Gasp:04,AndrGasp:07,SchmSeif:07,%
Ross:08,Ge:09,Qian:09,EspoBroe:10a,EspoBroe:10b,GeQian:10,Qian:10,%
RossVill:10,SantQian:11,ZhanQianQian:12a}. This perspective can
provide important insights into functional properties of Markovian
reaction networks (such as robustness and stability) and can lead
to a better understanding of the relationship between the mesoscopic
(unobservable) and macroscopic (observable) behavior of such
networks~\cite{Gasp:04,HanWang:08,Ross:08,Qian:09,VellQian:09,%
Qian:10,RossVill:10,GeQianQian:12b}.

In this section, we consider an \emph{irreducible} Markovian
reaction network comprised of $M/2$ pairs of \textit{reversible}
reactions $(2m-1,2m)$, $m=1,2,\ldots,M/2$, where $2m-1$ is the
forward reaction and $2m$ is the corresponding reverse reaction.
This does not forbid us to consider irreversible reactions,
since an irreversible reaction can be thought of as being
reversible with negligible propensity in the reverse direction.
As we mentioned in Section~VI, the reaction network is
characterized by a unique population probability distribution
$p_{\ss \bfX}(\bfx;t)$ that is analytic for all $t \geq 0$
and converges to a stationary distribution
$\overline{p}_{\ss \bfX}(\bfx)$, which does not depend on the
initial distribution $p_{\ss \bfX}(\bfx;0)$. By following our
discussion in Section~VII, we can define the energy of state
$\bfx$ by
\beq
E(\bfx) := - \frac{1}{\Omega} \ln \overline{p}_{\ss \bfX}(\bfx),
\quad \mbox{for~~$\bfx \in \sX$},
\label{eq-VIII-1}
\eeq
where $\Omega > 0$ is an appropriately chosen size parameter.

Our discussion in the following is purely mathematical in nature and
can be applied to any physical or nonphysical Markovian reaction
network. However, direct connection to thermodynamics can be made
in certain physical systems, such as biochemical reaction networks,
which may exchange matter, work, and heat through a well-defined
boundary that separates the system with its surroundings~\cite{DillBrom:11}.
In this case, we must take the size parameter $\Omega$ to be the
inverse of $\kb T$, where $\kb$ is the Boltzmann constant and $T$
is the system temperature. Since the exact value of $\Omega$ is
not important here, we set $\Omega=1$ for simplicity.

By viewing a Markovian reaction network as a thermodynamic system,
we can define three fundamental quantities: the internal energy, entropy,
and Helmholtz free energy. The \textit{internal energy} $U(t)$ is the
average energy of the system at time $t$ over all states, given by
\beq
U(t) := \sum_{\bfx \in \sX} E(\bfx) p_{\ss \bfX}(\bfx;t) , \nn
\label{eq-VIII-2}
\eeq
for $t \geq 0, \nn$, whereas, the entropy is defined by
\beq
S(t) := - \sum_{\bfx \in \sX} p_{\ss \bfX}(\bfx;t)
\ln p_{\ss \bfX}(\bfx;t),
\label{eq-VIII-3}
\eeq
for $t \geq 0$. Moreover, the Helmholtz free energy is given by
\beq
F(t) := U(t)- S(t) = \sum_{\bfx \in \sX} p_{\ss \bfX}(\bfx;t) \ln
\frac{p_{\ss \bfX}(\bfx;t)} {\overline{p}_{\ss \bfX}(\bfx)},
\label{eq-VIII-4}
\eeq
for $t \geq 0$.
The Helmholtz free energy measures the energy available in a thermodynamic
system to do work under constant temperature and volume. Note that $F(t)$
coincides with the Kullback-Leibler distance (or relative entropy) of the
probability distribution $p_{\ss \bfX}(\bfx;t)$ from the steady-state
probability distribution $\overline{p}_{\ss \bfX}(\bfx)$ [recall
Eq.~(\ref{eq-VI-5})]. Therefore, the Helmholtz free energy provides a
measure of how far a Markovian reaction network is from steady-state at
time $t$. Note that $F(t) \geq 0$ and $dF(t)/dt \leq 0$, for every
$t \geq 0$, with equality only at steady-state~\cite{Schn:76,CoveThom:91,Qian:09}.

\vspace{-12pt}

\subsection{Balance equations}

\vspace{-6pt}

From Eqs.~(\ref{eq-II-8}) and~(\ref{eq-VIII-3}), we can show the
following \textit{entropy balance} equation:
\beq
\frac{d S(t)}{dt} = \sigma(t) - h(t),
\label{eq-VIII-5}
\eeq
for $t > 0$, where
\beq
\sigma(t) \!:=\! \frac{1}{2} \sum_{m=1}^{M/2} \sum_{\bfx \in \sX}
{\Bigl [} \rho^+_m (\bfx;t) \sA^+_m(\bfx;t) + \rho^-_m (\bfx;t)
\sA^-_m(\bfx;t) {\Bigr ]} \! ,
\label{eq-VIII-6}
\eeq
and
\beqa
&& \hspace{-10mm} h(t) := \frac{1}{2} \sum_{m=1}^{M/2} \sum_{\bfx \in \sX}
\left \{ \rho^+_m(\bfx;t) \ln \! \left [ \frac{\pi_{2m-1}
(\bfx - \bfs_{2m-1})} {\pi_{2m}(\bfx)} \right ] \right . \nn \\
&& \hspace{17mm} \left . + \rho^-_m(\bfx;t)
\ln \! \left [ \frac{\pi_{2m}(\bfx + \bfs_{2m-1})}
{\pi_{2m-1}(\bfx)} \right ]\right \} \!.
\label{eq-VIII-7}
\eeqa
In these equations,
\beq
\begin{array}{l}
\rho^+_m(\bfx;t) := \pi_{2m-1}(\bfx - \bfs_{2m-1}) \px(\bfx-\bfs_{2m-1};t) \\ [4pt]
~~~~~~~~~~~~~~~~~ - \pi_{2m}(\bfx) \px(\bfx;t) \\ [9pt]
\rho^-_m(\bfx;t) := \pi_{2m}(\bfx + \bfs_{2m-1}) \px(\bfx+\bfs_{2m-1};t) \\ [4pt]
~~~~~~~~~~~~~~~~~ - \pi_{2m-1}(\bfx) \px(\bfx;t) , \nn
\end{array}
\label{eq-VIII-8}
\eeq
where $\rho^+_m(\bfx;t)$ is the \textit{net flux} of the $\mth$ pair
of reversible reactions reaching state $\bfx$ from state $\bfx-\bfs_{2m-1}$
and $\rho^-_m(\bfx;t)$ is the net flux of the same pair of reactions
reaching state $\bfx$ from state $\bfx-\bfs_{2m}$ (note that $\bfs_{2m}
=-\bfs_{2m-1}$). Moreover,
\beq
\hspace{-12pt}
\begin{array}{l}
{\displaystyle \sA^+_m(\bfx;t) := \ln \! \left [ \frac{\pi_{2m-1}(\bfx - \bfs_{2m-1})
\px(\bfx-\bfs_{2m-1};t)}{\pi_{2m}(\bfx) \px(\bfx;t)} \right ] } \\ [18pt]
{\displaystyle \sA^-_m(\bfx;t) := \ln \! \left [ \frac{\pi_{2m}(\bfx + \bfs_{2m-1})
\px(\bfx+\bfs_{2m-1};t)}{\pi_{2m-1}(\bfx) \px(\bfx;t)} \right ]}
\end{array}
\label{eq-VIII-9}
\eeq
are the \textit{affinities} corresponding to the net fluxes $\rho^+_m(\bfx;t)$
and $\rho^-_m(\bfx;t)$, respectively. Note that
\beq
\begin{array}{c}
\rho^-_m(\bfx;t) = - \rho^+_m(\bfx+\bfs_{2m-1};t), \\ [9pt]
\sA^-_m(\bfx;t) = - \sA^+_m(\bfx+\bfs_{2m-1};t), \nn
\end{array}
\label{eq-VIII-10}
\eeq
and
\beq
\frac{\partial \px(\bfx;t)}{\partial t} ~ = \! \sum_{m=1}^{M/2}
\! {\Bigl [} \rho^+_m(\bfx;t) + \rho^-_m(\bfx;t) {\Bigr ]}, \nn
\label{eq-VIII-11}
\eeq
for $t > 0$.
Therefore, $[\rho^+_m(\bfx;t) + \rho^-_m(\bfx;t)]dt$ quantifies the
change [increase, when $\rho^+_m(\bfx;t) + \rho^-_m(\bfx;t)>0$, or
decrease, when $\rho^+_m(\bfx;t) + \rho^-_m(\bfx;t)<0$] in the
probability mass of the population process within the infinitesimally
small time interval $[t,t+dt)$ due to the $\mth$ pair of reversible
reactions. These changes in probability mass are driven by the
affinities $\sA^+_m(t)$ and $\sA^-_m(t)$, which can be viewed as
\textit{thermodynamic forces} that drive a Markovian reaction network
away or towards the state of \textit{thermodynamic equilibrium}
in which all net fluxes are zero.

Equation~(\ref{eq-VIII-5}) provides an expression for the rate of
entropy change in a Markovian reaction network. The term $\sigma(t)$
quantifies the rate of entropy production, whereas, the term $h(t)$
quantifies the rate of entropy loss due to heat dissipation. For
this reason, $\sigma(t)$ is called the \textit{entropy production
rate}, whereas, $h(t)$ is called the \textit{heat dissipation rate}.
Equation~(\ref{eq-VIII-6}) shows that $\sigma(t)$ is a sum of terms
$1/2 \sum_{\bfx \in \sX} {\bigl [} \rho^+_m (\bfx;t) \sA^+_m(\bfx;t)
+ \rho^-_m(\bfx;t) \sA^-_m(\bfx;t) {\bigr ]}$, each term quantifying
the contribution of a pair of reversible reactions to the net rate
of entropy production. Similarly, Eq.~(\ref{eq-VIII-7}) shows that
$h(t)$ is a sum of terms $1/2 \sum_{\bfx \in \sX} {\bigl \{}
\rho^+_m(\bfx;t) \ln [\pi_{2m-1} (\bfx - \bfs_{2m-1})/\pi_{2m}(\bfx)]
+ \rho^-_m(\bfx;t) \ln [\pi_{2m}(\bfx - \bfs_{2m})/\pi_{2m-1}(\bfx)]
{\bigr \}}$, each term quantifying the contribution of a pair of
reversible reactions to the net rate of heat dissipation. Therefore,
a reaction with some non-zero net flux must produce entropy and
dissipate heat.

By differentiating Eq.~(\ref{eq-VIII-4}) with respect to $t$ and by
using Eqs.~(\ref{eq-II-8}),~(\ref{eq-VIII-6}) and~(\ref{eq-VIII-9}),
we can show the following balance equations for the Helmholtz free
energy and internal energy:
\beq
\frac{d F(t)}{dt} = f(t) - \sigma(t),
\label{eq-VIII-12}
\eeq
for $t > 0$, and
\beq
\frac{d U(t)}{dt} = f(t) - h(t),
\label{eq-VIII-13}
\eeq
for $ t > 0$, where
\beq
f(t) := \frac{1}{2} \sum_{m=1}^{M/2} \sum_{\bfx \in \sX} {\Bigl [}
\rho^+_m(\bfx;t) \bar{\sA}^+_m(\bfx) + \rho^-_m(\bfx;t)
\bar{\sA}^-_m(\bfx) {\Bigr ]}, \nn
\label{eq-VIII-14}
\eeq
with $\bar{\sA}^+_m(\bfx)$ and $\bar{\sA}^-_m(\bfx)$ being the
affinities of the $\mth$ pair of reversible reactions at steady-state;
i.e., $\bar{\sA}^+_m(\bfx) := \lim_{t \rightarrow \infty} \sA^+_m(\bfx;t)$
and $\bar{\sA}^-_m(\bfx) := \lim_{t \rightarrow \infty} \sA^-_m(\bfx;t)$.
Equation~(\ref{eq-VIII-12}) quantifies the change in Helmholtz free
energy due to the Markovian reaction network being away from
thermodynamic equilibrium at steady-state  [quantified by the first
term on the right-hand-side of Eq.~(\ref{eq-VIII-12})] or reduction in
Helmholtz free energy due to entropy production [quantified by the
second term on the right-hand-side of Eq.~(\ref{eq-VIII-12})].
The term $f(t)$ quantifies the rate of energy (i.e., power) supplied
to the Markovian reaction network in order to keep it away from
thermodynamic equilibrium. For this reason, we refer to $f(t)$ as
the \textit{``motive'' power}.\footnote{This quantity is also referred
to as ``housekeeping'' heat rate~\cite{OonoPani:98,SchmSeif:07,Ge:09,%
EspoBroe:10a,GeQian:10}. However, we prefer to call $f(t)$ the
``motive'' power, since it represents the energy flow per unit time
required to keep the Markovian reaction network away from thermodynamic
equilibrium.} It turns out that $0 \leq f(t) \leq \sigma(t)$, for
every $t \geq 0$.\footnote{We can show the first inequality by
using the fact that the right-hand side of the master equation~(\ref{eq-II-8})
is zero at steady-state and that $\ln x \leq x-1$, for $x > 0$
[see~\textcite{Ge:09}]. The second inequality is due to
Eq.~(\ref{eq-VIII-12}) and the fact that $d F(t)/dt \leq 0$.} Note
that $f(t)$ is a sum of terms $1/2 \sum_{\bfx\in \sX} {\bigl [}
\rho^+_m(\bfx;t) \bar{\sA}^+_m(\bfx) + \rho^-_m(\bfx;t)
\bar{\sA}^-_m(\bfx) {\bigr ]}$, each term quantifying the contribution
of a pair of reversible reactions to the net ``motive'' power.
Therefore, a reaction with non-zero (forward or reverse) flux and
corresponding non-zero affinity at steady-state will supply motive
power to the Markovian reaction network.

Equation~(\ref{eq-VIII-12}) shows that reactions in a Markovian
reaction network can increase the Helmholtz free energy by adding
``motive'' energy to the system, whereas, they can reduce the
Helmholtz free energy due to entropy production. Moreover,
\beq
\sigma(t) =  f(t) + \left | \frac{d F(t)}{dt} \right | , \nn
\label{eq-VIII-15}
\eeq
for $t > 0$,
which implies that entropy production comes from two sources: from
supplying motive power $f(t)$ to sustain the reaction network away
from thermodynamic equilibrium and from a spontaneous change
$|dF(t)/dt|$ in Helmholtz free energy due to relaxation towards
the steady-state~\cite{NicoPrig:77}. On the other hand,
Eq.~(\ref{eq-VIII-13}) expresses the first-law of thermodynamics
(energy conservation): a change $\Delta U(t) = U(t+dt)-U(t)$ in
internal energy within the infinitesimal time interval $[t,t+dt)$
must equal the amount of motive energy $f(t)dt$ added to the system
minus the dissipated heat $h(t)dt$. Note that $\sigma(t) \geq 0$,
for every $t \geq 0$, with equality if and only if $\sA^+_m(t)
= \sA^-_m(t)=0$, for every $m=1,2,\ldots, M/2$.\footnote{This
is a direct consequence of the fact that $(x_1-x_2)\ln(x_1/x_2)
\geq 0$, for any values of $x_1$ and $x_2$, with equality if
and only if $x_1=x_2$.} This is in agreement with the second
law of thermodynamics, which postulates that the rate of entropy
production must always be nonnegative. Finally, Eqs.~(\ref{eq-VIII-5})
and~(\ref{eq-VIII-12}) imply that
\beq
0 \leq \overline{\sigma} = \overline{h} = \overline{f},
\label{eq-VIII-16}
\eeq
where $\overline{\sigma} := \lim_{t \rightarrow \infty} \sigma(t)$,
and similarly for $\overline{h}$ and $\overline{f}$. This result
says that, at steady-state, the amount of motive power supplied
to the system must be equal to the rate of heat dissipation, in
agreement with the first law of thermodynamics. In addition, the
rate of heat dissipation must be equal to the rate of entropy
production. Finally, the steady-state entropy production, heat
dissipation and motive power must all be nonnegative, in agreement
with the second law of thermodynamics.

\vspace{-12pt}

\subsection{Thermodynamic equilibrium}

\vspace{-6pt}

A Markovian reaction network reaches thermodynamic equilibrium
at steady-state if and only if $\bar{\sA}^+_m=\bar{\sA}^-_m=0$,
for every $m=1,2,\ldots,M/2$, which is equivalent to the
following \textit{detailed balance} equations:
\beqa
\pi_{2m-1}(\bfx - \bfs_{2m-1}) \overline{p}_{\ss \bfX}(\bfx-\bfs_{2m-1})
&\!=\!& \pi_{2m}(\bfx) \overline{p}_{\ss \bfX}(\bfx) \nn \\ [9pt]
\pi_{2m}(\bfx + \bfs_{2m-1}) \overline{p}_{\ss \bfX}(\bfx+\bfs_{2m-1})
&\!=\!& \pi_{2m-1}(\bfx) \overline{p}_{\ss \bfX}(\bfx), \nn
\label{eq-VIII-17}
\eeqa
for every $m=1,2,\ldots,M/2$, $\bfx \in \sX$. In this case, $f(t)=0$,
for every $t \geq 0$, which implies that
\beq
\frac{d U(t)}{dt} = - h(t) \quad \mbox{and} \quad
\frac{d F(t)}{dt} = - \sigma(t), \nn
\label{eq-VIII-18}
\eeq
for $t > 0$.
Moreover, Eq.~(\ref{eq-VIII-16}) results in $\overline{\sigma} =
\overline{h} = \overline{f} =0$, which shows that a Markovian
reaction network that reaches thermodynamic equilibrium at
steady-state will not produce entropy or dissipate heat. It
turns out that a Markovian reaction network must be
\textit{reversible} at thermodynamic equilibrium, which means
that the stationary behavior of the population process will be
indistinguishable if the direction of time is reversed. This
behavior may not be desirable, since many Markovian reaction
systems (e.g., biochemical reaction networks) are irreversible
with respect to time. As a matter of fact, entropy production,
heat dissipation, and irreversibility with respect to time are
fundamental properties which are necessary for the formation of
order in many physical systems~\cite{Prig:78}. Therefore,
and in many cases of interest, a Markovian reaction network
must not reach thermodynamic equilibrium in order to be
useful. We can make sure that this is the case by including
nonreversible reactions that transfer mass through the
boundary of the system with its surroundings, thus breaking
detailed balance.

Despite the aforementioned drawbacks, Markovian reaction
networks that reach thermodynamic equilibrium have been
extensively used to model population dynamics. For this
type of networks we can use (at least in principle) a simple
iterative procedure to calculate the steady-state probability
distribution. This is possible because \textit{any} state
$\bfx \in \sX$ can be reached from a given state $\bfx_0 \in \sX$
through at least one ordered chain of reactions $(m_1,m_2,
\ldots,m_L)$. In this case, detailed balance implies
that~\cite{Hake:74}
\beq
\overline{p}_{\ss \bfX}(\bfx) = \overline{p}_{\ss \bfX}(\bfx_0)
\prod_{l=1}^L \frac{\pi_{m_l}(\bfx_0+\sum_{l'=1}^{l-1} \bfs_{m_{l'}})}
{\pi_{m^*_l}(\bfx_0+\sum_{l'=1}^{l} \bfs_{m_{l'}})} \:,
\label{eq-VIII-19}
\eeq
for every $\bfx \not= \bfx_0$,
where $m^*_l$ is the index of the opposite reaction to reaction $m_l$
(i.e., $m^*_l=2m$, if $m_l=2m-1$, and $m^*_l=2m-1$, if $m_l=2m$).
After this procedure is completed for all $x \in \sX$, we can
calculate $\overline{p}_{\ss \bfX}(\bfx_0)$ in Eq.~(\ref{eq-VIII-19})
by setting the sum of all probabilities $\overline{p}_{\ss \bfX}(\bfx)$
equal to $1-\overline{p}_{\ss \bfX}(\bfx_0)$.

\vspace{-12pt}

\subsection{Cycles and affinities}

\vspace{-6pt}

A useful representation of the state-space $\sX$ of a Markovian reaction
network is by means of a graph~$G$ whose nodes are the population states
$\bfx \in \sX$ and whose edges connect pairs of population states
$(\bfx,\bfx+\bfs_m)$ when $\pi_m(\bfx),\pi_{m^*}(\bfx+\bfs_m)>0$~\cite{Schn:76}.
Clearly, an edge connecting two states $\bfx$, $\bfx'$ indicates that
these states can ``reach'' each other using a pair $(m,m^*)$ of
reversible reactions. Note that there might be a number of distinct
edges (corresponding to different reversible reactions) connecting a
given pair of nodes, in which case $G$ is a multi-graph.\footnote{For
example, the reversible reactions $X_1 \rightleftarrows X_2$ and
$X_1 + X_3 \rightleftarrows X_2 + X_3$ are characterized by the same
net stoichiometry and will therefore connect the same pair of nodes in $G$.}
An ordered chain $(m_1,m_2,\ldots,m_L)$ of reactions will produce a
path $(\bfx_0,\bfx_0+\bfs_{m_1},\ldots,\bfx_0 + \sum_{l=1}^L\bfs_{m_l})$
in $G$ of length $L$, provided that each reaction can occur with positive
probability. In the particular case when $\sum_{l=1}^L \bfs_{m_l}=0$,
the reactions $(m_1,m_2,\ldots,m_L)$ will produce a \textit{cycle}
in $G$ of length $L$ that ends in the same state as the starting state.
In the following, we will denote by $\sC$ the set of all cycles in $G$
with $L \geq 2$.\footnote{Here we use an equivalence class of cycles
over cyclical shifts, which implies that a cycle produced by reactions
$(m_1,m_2,\ldots,m_L)$ with starting state~$\bfx_0$ is equivalent to
the cycle produced by reactions $(m_2,\ldots,m_L,m_1)$ with starting
state $\bfx_0+\bfs_1$.}

It is clear from Eq.~(\ref{eq-VIII-19}) that the propensity functions of
a Markovian reaction network that reaches thermodynamic equilibrium
must satisfy the following conditions:
\beq
\sP(C) := \prod_{l=1}^L \frac{\pi_{m_l}(\bfx_{l-1})}
{\pi_{m^*_l}(\bfx_l)}= 1, \nn
\label{eq-VIII-20}
\eeq
over a cycle $C=(\bfx_0,\bfx_1,\ldots,\bfx_L)$ produced by
reactions $(m_1,m_2,\ldots,m_L)$, where $\bfx_l := \bfx_0 +
\sum_{l'=1}^l \bfs_{m_{l'}}$. These are known as \textit{Kolmogorov
cyclic conditions}~\cite{JianQianQian:04}. Equivalently,
\beq
\sA(C):= \sum_{l=1}^L \ln \frac{\pi_{m_l}(\bfx_{l-1})\px
(\bfx_{l-1};t)}{\pi_{m^*_l}(\bfx_l)\px(\bfx_{l};t)} = \ln \sP(C) = 0, \nn
\label{eq-VIII-21}
\eeq
for every $t \geq 0$,
where $\sA(C)$ is the \textit{net} affinity around cycle $C$. In addition
to being necessary, the previous conditions are also sufficient for a
Markovian reaction network to reach thermodynamic equilibrium [see
Theorem 2.2.10 in \textcite{JianQianQian:04}]. Therefore, care must be
taken when dealing with this type of Markovian reaction networks, since
their propensity functions must be appropriately constrained.

Research effort has been recently focused on developing
techniques for enforcing such conditions in the thermodynamic
limit of mass-action systems governed by the macroscopic
equations~(\ref{eq-IV-46}). In this case, the corresponding
constraints are known as \textit{Wegscheider conditions}.
The proposed methods have been developed for performing sensitivity
analysis~\cite{ZhanDempGout:09} or parameter
estimation~\cite{ColqDownBeat-etal:04,LiebKlip:06,YangBrunHlav-etal:06,%
JenkZhonGout:10,JenkGout:11}. However, further work is needed to deal
with the Kolmogorov-Wegscheider conditions in the Markovian setting
discussed in this review.

The Kolmogorov cyclic conditions are usually highly redundant, since a
cycle can often be decomposed into smaller cycles. In this case the
Kolmogorov cyclic condition imposed on the larger cycle is implied
by the conditions imposed on the smaller cycles. To address this
issue, we can derive a minimal set of Kolmogorov cyclic conditions that,
when satisfied, they imply the remaining conditions. As a matter of fact,
it has been shown by~\textcite{Schn:76} that the net affinity$\sA(C)$
of a cycle $C \in \sC$ is given by
\beq
\sA(C) = \sum_{k=1}^K \alpha_k(C) \sA(C^\dag_k),
\label{eq-VIII-22}
\eeq
where $ \alpha_k(C)$ is an appropriately defined constant that takes
integer values and $\{C^\dag_1,C^\dag_2,\ldots,C^\dag_K\}$ is a
(non-unique) set of cycles, known as \textit{fundamental cycles}.

The fundamental cycles and the associated values of $\alpha$ can be
determined by a simple procedure based on graph
theory~\cite{Schn:76}.\footnote{We refer the reader to the excellent
book by~\textcite{Dies:97} for an introduction to graph theory.} Given
the graph $G$, we can define a maximal tree $T(G)$ of $G$ such that:
(i)~$T(G)$ is a covering subgraph of $G$ [i.e., $T(G)$ shares all
vertices of $G$ and an edge of $T(G)$ must be an edge of $G$],
(ii)~$T(G)$ is connected, and (iii)~$T(G)$ contains no circuit (i.e.,
no cyclic sequence of edges). An edge $e_k$ of $G$ is said to be a
\textit{chord} of $T(G)$ whenever it is not an edge of $T(G)$. If
$T_k(G)$ is the graph $T(G)$ with the chord $e_k$ included as an
edge, then this subgraph of $G$ would include exactly one circuit
$\widetilde{C}_k$, which is obtained from $T_k(G)$ by removing all
edges that are not part of the circuit. The set $\{\widetilde{C}_1,
\widetilde{C}_2,\ldots,\widetilde{C}_K\}$ consists of circuits,
known as \textit{fundamental circuits}. Adding an arbitrary
orientation to the fundamental circuit $\widetilde{C}_k$ defines
the fundamental cycle $C^\dag_k$ used in Eq.~(\ref{eq-VIII-22}).

On the other hand, given a cycle $C \in \sC$ and a fundamental cycle
$C^\dag_k$,
\beq
\alpha_k(C) = \sigma_{\! e_k}(C) \: \sigma_{\! e_k}(C^\dag_k), \nn
\label{eq-VIII-23}
\eeq
with $e_k$ being the chord used to produce the circuit $\widetilde{C}_k$ (and
thus $C^\dag_k$) and
\beq
\sigma_e(D) := \left \{
\begin{array}{rl}
 i, & \mbox{if cycle $D$ contains $i$ copies of an} \\
    & \mbox{edge $e$ in the same orientation} \\
    & \mbox{as in graph $G$} \\ [6pt]
-i, & \mbox{if cycle $D$ contains $i$ copies of an} \\
    & \mbox{edge $e$ in the opposite orientation} \\
    & \mbox{as in graph $G$} \\ [6pt]
 0, & \mbox{if cycle $D$ does not contain edge $e$} ,
\end{array} \right . \nn
\label{eq-VIII-24}
\eeq
where the graph $G$ is assigned an orientation to its edges in the
direction of the forward reaction (i.e., the reaction that
corresponds to an odd value of $m$).

Equation~(\ref{eq-VIII-22}) shows that the affinity around a
cycle~$C$ can be written as a linear combination of the affinities
around the fundamental cycles $C^\dag_k$. This result demonstrates
that a necessary and sufficient condition for a system to reach
thermodynamic equilibrium is that the affinity of each fundamental
cycle must be zero. This provides a reduced and more manageable
set of conditions than the Kolmogorov cyclic conditions over all
possible cycles.

The net affinity of a Markovian reaction network that does not
reach thermodynamic equilibrium must be non-zero over at least
one fundamental cycle. This affinity quantifies the net
thermodynamic force applied to the network due to its interaction
with the surroundings (e.g., due to mass flow through the system
boundary); see~\textcite{AndrGasp:04,AndrGasp:07}
and~\textcite{Schn:76}. In many Markovian reaction networks, such
as those with propensities that follow the mass-action rate law,
there is a number of \textit{global affinities} $\sA_q$, $q =
1,2,\ldots,Q$, which describe the macroscopic coupling of the
system to its surroundings, such that the affinity of each
fundamental cycle $C^\dag_k$ equals $\sA_q$, for some $q$.
If the global affinities are known, then the relationships
$\sA(C^\dag_k)=\sA_q$, along with Eq.~(\ref{eq-VIII-22}),
constitute a more general version of the Kolmogorov cyclic
conditions that must be enforced on the propensity functions
so that the Markovian reaction network does not reach thermodynamic
equilibrium.

\vspace{-12pt}

\subsection{Example: Neural dynamics}

\vspace{-6pt}

We now consider a special case of the neural network model
discussed in Section~III-F, which allows us to numerically
compute the dynamics of the joint population probability
distribution and proceed with illustrating a number of
thermodynamic properties of this model. We will assume that
the~$L$ neurons in the network can be divided into an equal
number of $L/2$ \emph{excitatory} and $L/2$ \emph{inhibitory}
neurons. Moreover, we will assume that all neurons synapse
to all other neurons with excitatory and inhibitory weights
$\nu_E \geq 0$ and $\nu_I \leq 0$, respectively. Finally,
we will consider the case in which the neurons are characterized
by the same decay rate $\gamma$, whereas, their external inputs
take the same value~$h$. Under these assumptions, it may not be
of interest to track the state of individual neurons, since the
excitatory or inhibitory neurons are identical to each other.
Instead, it will be more appropriate to track the dynamics of
the net number
\beq
A(t) := Y_1(t)+Y_2(t), \nn
\label{eq-VIII-25}
\eeq
of active excitatory and inhibitory neurons, where
\beq
Y_1(t) := \sum_{l \in \sE} X_{2l}(t) \qquad \mbox{and} \qquad
Y_2(t) := \sum_{l \in \sI} X_{2l}(t), \nn
\label{eq-VIII-26}
\eeq
with $\sE$ and $\sI$ being the set of excitatory and inhibitory
neurons, respectively.

For convenience, and without loss
of generality, we can take $\sE = \{1,2,\ldots,L/2\}$ and
$\sI = \{L/2+1,L/2+2,\ldots,L\}$. It turns out that $Y_1(t)$
and $Y_2(t)$ can be modeled by a simple Markovian reaction
network comprised of two species~$Y_1$ and~$Y_2$ that denote
active excitatory and inhibitory neurons, respectively, which
interact through the following reactions:
\beq
\begin{array}{l}
Y_1 + Y_2 \rightarrow 2  Y_1 + Y_2 \\ [6pt]
Y_1 \rightarrow \emptyset \\ [6pt]
Y_1 + Y_2 \rightarrow Y_1 + 2 Y_2 \\ [6pt]
Y_2 \rightarrow \emptyset .
\end{array} \nn
\label{eq-VIII-27}
\eeq
These reactions correspond to the activation/deactivation of
an excitatory neuron (first and second reactions) and
the activation/deactivation of an excitatory neuron (third
and fourth reactions). Their propensity functions are given by
[recall Eqs.~(\ref{eq-III-21}) and~(\ref{eq-III-22})]
\beq
\begin{array}{l}
\pi_1(\bfy) = (L/2-y_1) [\phi(\bfy)>0] \tanh[\phi(\bfy)] \\ [6pt]
\pi_2(\bfy) = \gamma y_1                                      \\ [6pt]
\pi_3(\bfy) = (L/2-y_2) [\phi(\bfy)>0] \tanh[\phi(\bfy)] \\ [6pt]
\pi_4(\bfy) = \gamma y_2, \nn
\end{array}
\label{eq-VIII-28}
\eeq
respectively, where $\bfy=[y_1~y_2]^T$, with $y_1$ and $y_2$ taking
values in $\{0,1,\ldots,L/2\}$, whereas, $[a > 0]$ is the Iverson
bracket and $\phi$ is the synaptic input to each neuron, given by
[recall Eq.~(\ref{eq-III-20})]
\beq
\phi(\bfy) = \nu_E y_1 + \nu_I y_2 + h. \nn
\label{eq-VIII-29}
\eeq
The propensity functions of the associated reverse reactions
are taken to be zero.

Despite its simplified nature, the previous model has been
shown by~\textcite{BenaCowaDron-etal:10} to be very effective
for predicting experimentally observed, \emph{in vitro} and
\emph{in vivo}, neural behavior, known as \emph{avalanches}.
This behavior is characterized by irregular and isolated
bursts of neural activity during which many neurons fire
simultaneously. In the following, we use the thermodynamic
principles discussed in this section to explore this
interesting behavior. For ease of computational analysis,
we consider a moderately sized neural network comprised
of $L=100$ neurons. This allows us to numerically compute
the solution of the underlying master equation using the
KSA method discussed in Section~IV-B. We adopt parameter
values used in~\textcite{BenaCowaDron-etal:10} and set
$\gamma=0.1 \mbox{ms}^{-1}$ and $h=0.001$, whereas, we
chose values for the synaptic weights $\nu_E$ and $\nu_I$
so that their sum $\nu_{E}+\nu_{I}$ is kept fixed to a
value of $0.004$. Finally, we assume that all neurons
are initially at rest, in which case, $Y_1(0)=Y_2(0)=0$.

\begin{figure}
\includegraphics[width=2.65in]{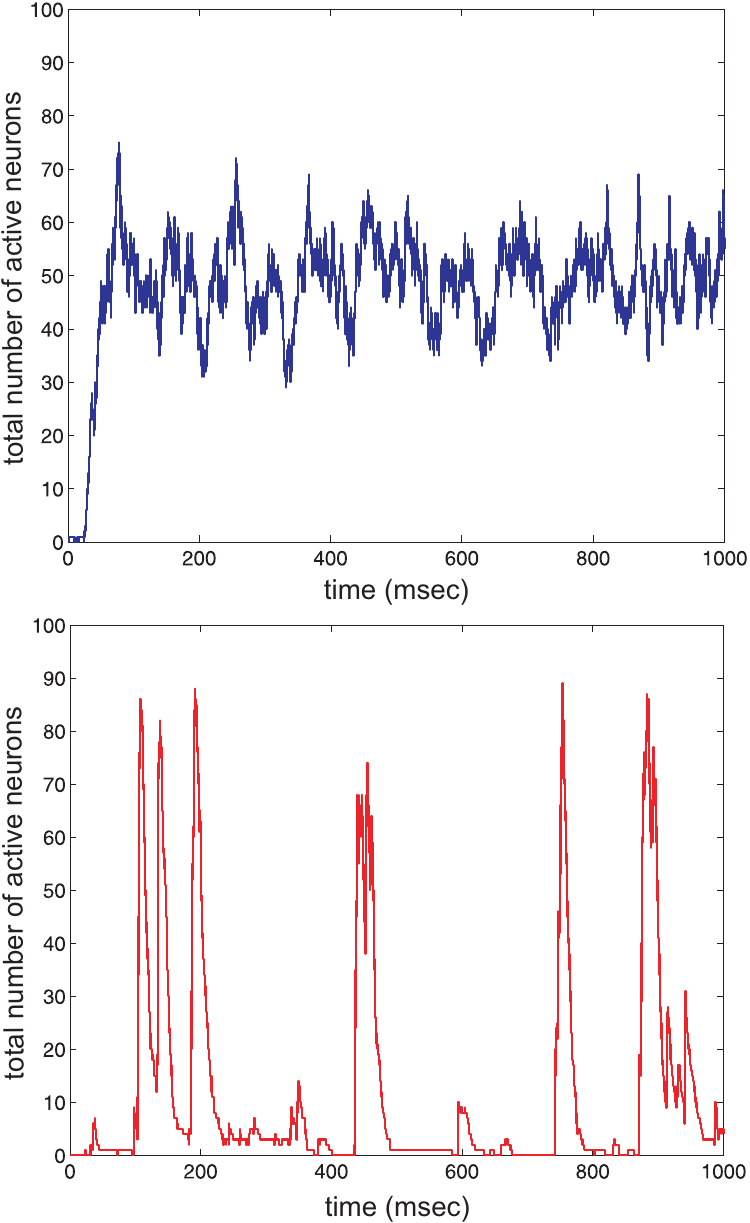}
\caption{Dynamic evolutions of the net number of active neurons in
the neural network example considered in Section~VIII-D drawn from the
underlying master equation using exact sampling. The blue trajectory
indicates that neurons fire asynchronously, whereas, the irregular and
isolated bursts of net neural activity observed in the red trajectory
indicate that neurons fire synchronously resulting in avalanching.}
\label{fig-8}
\vspace{-12pt}
\end{figure}

\begin{figure}
\includegraphics[width=3.4in]{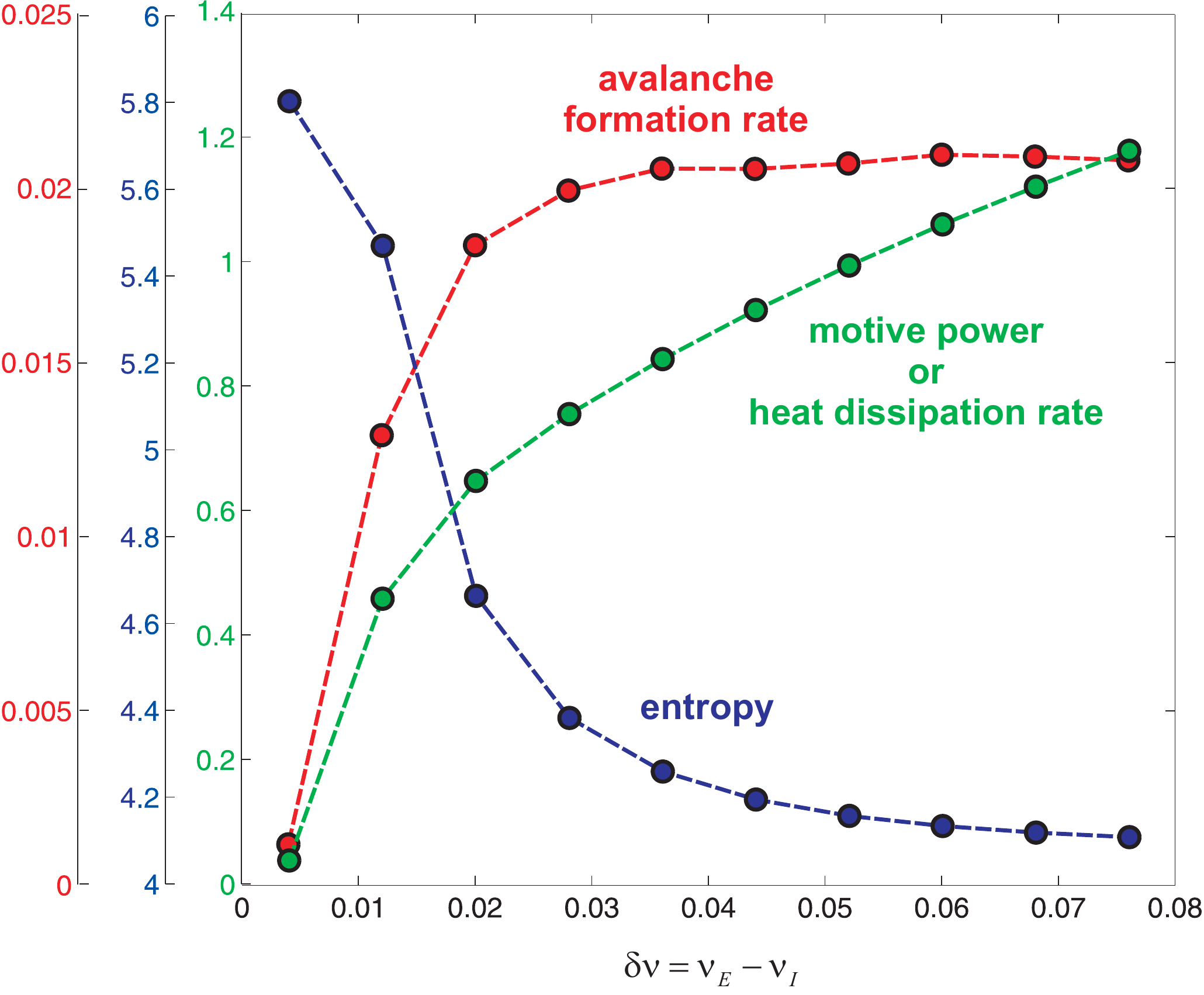}
\caption{Average rate of avalanche formation (red curve) in the
neural network example considered in Section~VIII-D (calculated
from $1,\!000$ trajectories of net neural activity during a
period of $1,\!000$~msec, drawn from the master equation using
exact Monte Carlo sampling), supplied motive power (or heat
dissipation) at steady-state (green curve), and system entropy
at steady-state (blue curve) as a function of the difference
$\delta\nu$ between the excitatory and inhibitory weights. As
expected, increasing the supply of motive power results in
increasing the rate of avalanche formation and decreasing the
system entropy.}
\label{fig-10}
\vspace{-14pt}
\end{figure}

It has been shown by~\textcite{BenaCowaDron-etal:10} that
when the sum $\nu_E+\nu_I$ is kept fixed, then the
difference $\delta\nu := \nu_E-\nu_I$ controls avalanching
(see also Fig.~\ref{fig-10}). More particularly, as $\delta\nu$
increases, the network transitions from asynchronous to
synchronous neural firings that lead to avalanching.
We illustrate this behavior in Fig.~\ref{fig-8}, which
depicts two trajectories of the net number $A(t)$ of
active neurons obtained by sampling the master equation
using exact sampling. The blue trajectory has been obtained
by setting $\nu_E=0.034$ and $\nu_I=-0.00062$, whereas, the
red trajectory has been obtained by setting $\nu_E=0.140$
and $\nu_I=-0.136$. Clearly, the blue trajectory indicates
that neurons fire asynchronously (in this case, $\delta \nu
= 0.00276$), whereas, the red trajectory exhibits avalanching
with neurons firing synchronously, resulting in irregular
and isolated bursts of activity and thus avalanching (in
this case, $\delta \nu = 0.276$, which is $100$ times larger
than the previous value).

\begin{figure}
\includegraphics[width=3.4in]{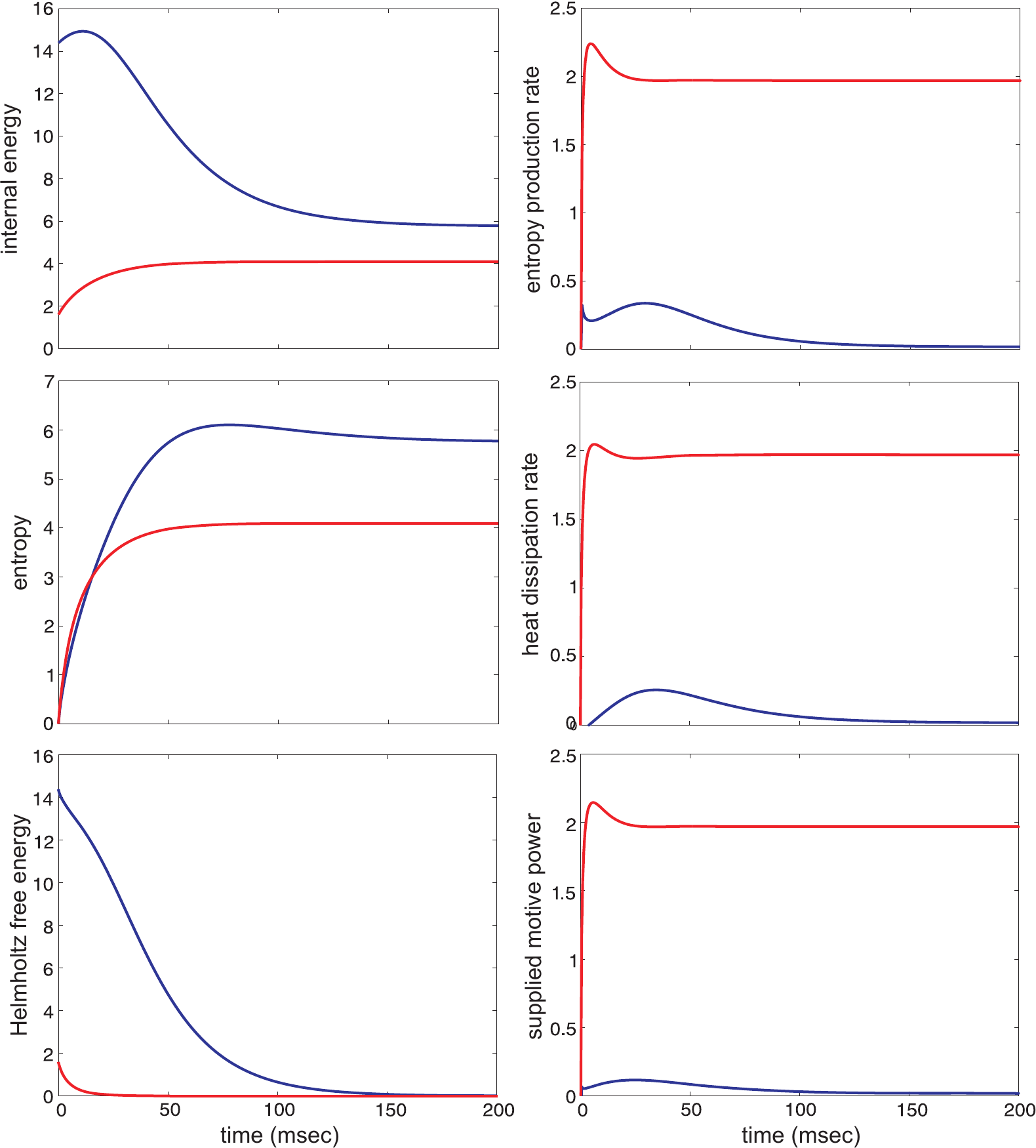}
\caption{Dynamic evolutions of internal energy, entropy, Helmholtz free
energy, entropy production rate, heat dissipation rate, and supplied
motive power in the neural network example considered in Section~VIII-D
for the case of asynchronous (blue lines) and synchronous (red lines)
neural firings leading to avalanching.}
\label{fig-9}
\vspace{-12pt}
\end{figure}

Most biological systems of interest reach a state of homeostasis,
wherein the system is maintained at a given stable operating
point. Mathematically, we can describe this point by a stable
steady state. From a thermodynamic perspective however such a
system must operate away from thermodynamic equilibrium, since
living organisms require transfer of energy and mass with their
surroundings in order to consume nutrients and excrete waste.
To achieve this, nonzero motive power must be supplied to the
system at steady-state, which implies that an equal amount
must be dissipated to the surroundings in the form of heat,
by virtue of the first law of thermodynamics. As a consequence,
the state of homeostasis at which biological systems operate
is often referred to as the \textit{non-equilibrium steady-state}
(NESS).

Naturally, the neural network discussed in this example
must also operate at a NESS [e.g., see~\textcite{StewPlen:08}].
Fig.~\ref{fig-9} shows clearly that this is indeed the case.
This figure depicts the dynamic evolutions of internal energy,
entropy, Helmholtz free energy, entropy production rate, heat
dissipation rate, and supplied motive power, for the two cases
of asynchronous (blue lines) and synchronous neural firings (red
lines). Note that all thermodynamic quantities reach stationarity,
with the entropy production rate, heat dissipation rate, and
supplied motive power converging to the same value in each case,
in agreement with the first and second law of thermodynamics.
The fact that this value is nonzero in both cases shows that
the system operates away from thermodynamic equilibrium at
steady-state regardless of the type of neural firings involved.

The results depicted in Fig.~\ref{fig-9} show that the system
entropy is in general smaller when neurons fire synchronously
than when they fire asynchronously. This indicates an expected
degree of predictability in neural activity when neurons fire
synchronously. On the other hand, the initial Helmholtz free
energy associated with asynchronous neural firings is almost
an order of magnitude larger than that associated with
synchronous firings. In both cases however the Helmholtz
free energy becomes zero at steady-state, as expected. This
indicates that, under constant temperature and volume,
appreciable more work must be done when the neurons fire
asynchronously to reach steady-state than when the neurons
fire synchronously. It is therefore expected that when
neurons fire synchronously, the system will reach
steady-state much faster than when neurons fire asynchronously.
This is clearly verified by the results depicted in Fig.~\ref{fig-9}.
We may therefore postulate that synchronous neural firing is,
among other things, necessary for a neural network to quickly
reach a state of homeostasis~[see also~\textcite{StewPlen:08}].

Fig.~\ref{fig-9} also reveals that the stationary value of the
supplied motive power (as well as the stationary values of the
entropy production and heat dissipation rates) is appreciably
larger in the case of synchronous neural firings than
asynchronous firings. This difference is well predicted by
the theory of dissipative structures~\cite{NicoPrig:77}
according to which self-organization of a system to an
ordered internal state requires that the system is sufficiently
driven by external sources and dissipates appreciable
heat to its surroundings. It is clear from Fig.~\ref{fig-9} that
appreciable amounts of supplied motive power and heat
dissipation is required to achieve avalanching behavior,
which leads to an ordered stationary state, quantified by
a lower entropy. We may therefore conclude that the emergence
of avalanches in a neural network is a consequence of
externally driven self-organization accompanied by appreciable
heat dissipation. This conclusion is further confirmed by
Fig.~\ref{fig-10}, which depicts the (average) rate of
avalanche formation (number of avalanches per unit time, )
the supplied motive power (or heat dissipation), and the
system entropy at steady-state. To calculate the number
of avalanches present in a given trajectory of net neural
activity, we assume that an avalanche occurs within a time
window $[t,t+\tau)$ whenever the following three conditions
are satisfied: ($i$)~$A(t')>0$, for all $t'\in[t,t+\tau)$;
i.e., there is neuronal activity during the time interval
$[t,t+\tau)$, ($ii$)~there exist some $\epsilon>0$ such that
$A(t')=0$, for all $t'\in [t-\epsilon,t)$; i.e.,
there is no neural activity immediately before time $t$,
and ($iii$)~$A(t+\tau)=0$; i.e., there is no
neural activity at time $t+\tau$.

\begin{figure*}
\includegraphics[width=6.25in]{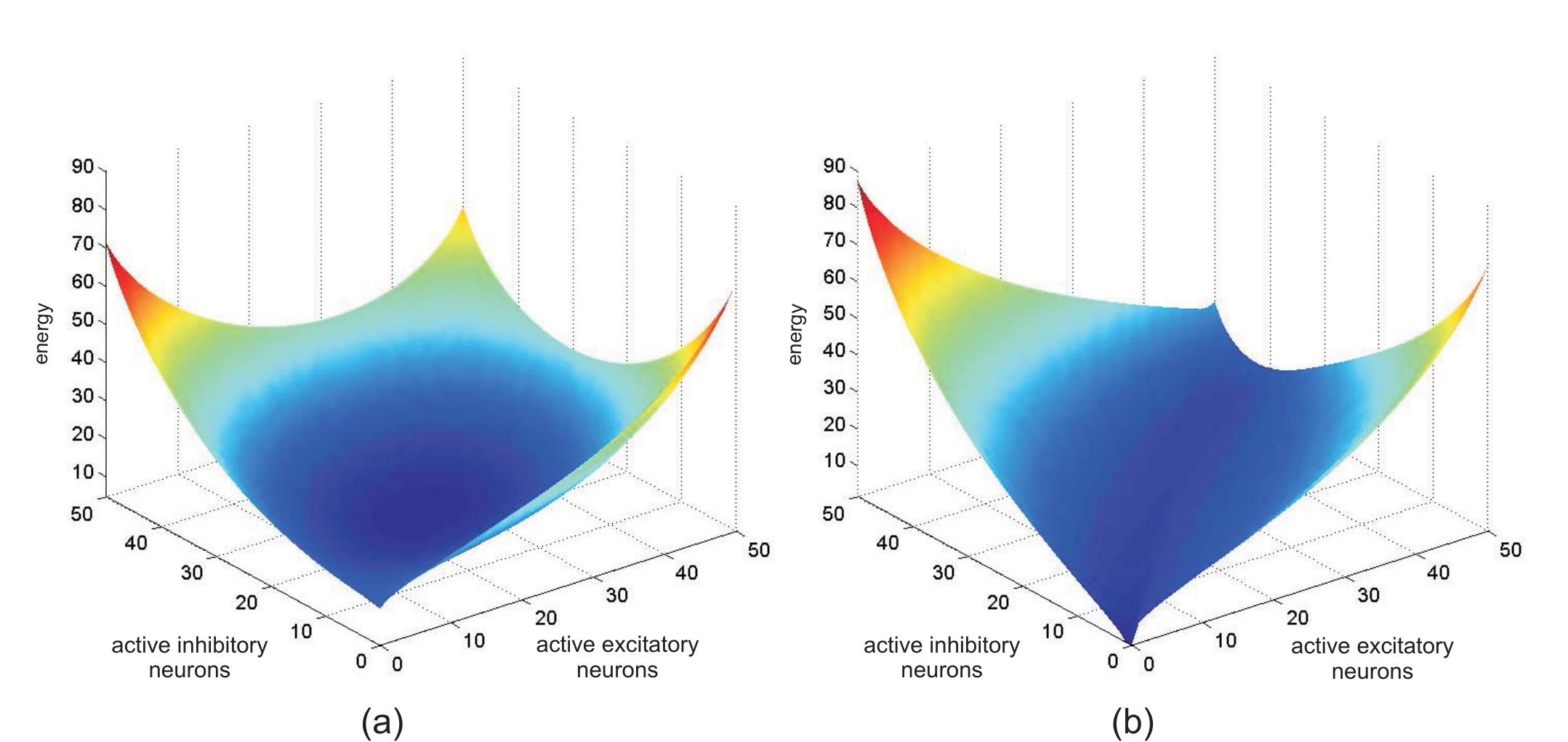}
\caption{Energy landscape of the neural network example considered
in Section~VIII-D when: (a)~neurons fire asynchronously, and (b)~neurons
fire synchronously resulting in avalanching.}
\label{fig-11}
\vspace{-12pt}
\end{figure*}

We conclude by noting that avalanching can be understood
at steady-state through the energy landscape $E(\bfy)$ [see
Eq.~(\ref{eq-VIII-1})]. As a matter of fact, the stationary
dynamics of neural activity may be thought of as a random walk
on this landscape, where the most likely steps follow a path
from higher to lower energy states (i.e., the preference is to
move downhill), with occasional (low probability) jumps from
lower to higher energy states. In Fig.~\ref{fig-11}(a), we
depict the energy landscape when neurons fire asynchronously,
whereas, in Fig.~\ref{fig-11}(b) we depict the energy landscape
when neurons fire synchronously, thus leading to avalanching.
One can see that the ground state (global minimum) of the energy
landscape depicted in Fig.~\ref{fig-11}(a) occurs away from the
origin at which neural activity is zero. As a consequence, the
system spends most time in Gaussian-like fluctuations about the
ground state. The system can jump out of the energy well
surrounding the ground state and reach the origin, but with
very small probability, due to its large width. Therefore,
avalanche formation in this system is a rare event. This
behavior is in agreement with a recent finding that neural
networks may simultaneously support synchronous and
asynchronous dynamics, switching between these two modes
of operation spontaneously~\cite{DeviPesk:08}.

On the other hand, the energy landscape of the system with
synchronous neural firings depicted in Fig.~\ref{fig-11}(b)
contains a valley along the line $Y_1=Y_2$ of equal excitatory
and inhibitory activities, which slopes down to the ground state
that is now located at the origin. Thus, from any point on this
landscape, the most likely trajectory roles downhill until it
reaches the origin. From this point, the dynamics of excitatory
and inhibitory neural activities may again reach the valley by
randomly jumping away in an uphill motion that overcomes the
steep and narrow energy well surrounding the origin. This
mechanism results in avalanching, which can be thought of as
a random sequence of zero (or almost zero) net neural activity
at points in the well surrounding the origin followed by nonzero
net neural activity at points proximal to the valley.

%
%

\vspace{-12pt}

\section{Outlook}

\vspace{-6pt}

The study of Markov processes on complex reaction networks is an
area of research that has been evolving for more than half a
century. Its applicability to many scientific and engineering
disciplines has led to parallel and often independent developments,
which have recently reached a critical mass due to unprecedented
advancements in modern experimental procedures and computational
capabilities. This can be best illustrated by the sharp increase
in the number of articles published on the subject during the last
ten years. For this reason, we believe that the present review is
both appropriate and timely. Our objective has been to provide a
coherent exposure to what has been done so far and illustrate key
methods with simple examples. In the following, we conclude this
review with a brief discussion of some outstanding issues and
problems important to the field. Some old problems require better
solutions, whereas, developing novel methods for the analysis
of Markovian reaction networks can lead to new and important
results. In some cases, the work that needs to be done is at
least as challenging and rewarding as the work done so far.

\vspace{-12pt}

\subsection{Solving the master equation}

\vspace{-6pt}

The tremendous flexibility and generality of Markovian reaction
networks make them an excellent mathematical framework for
studying stochastic processes on complex networks. The coherency
of a single framework means that tools and discoveries made in
one field may be readily ported to distant applications. In
fact, it has been argued by~\textcite{CookSoloWinf-etal:09} that
Markovian reaction networks with mass action propensities can
perform Turing universal computations with arbitrarily small
error, which becomes zero at the thermodynamic limit~\cite{Magn:97}.
This strength however turns out to be one of the most profound
weaknesses of Markovian reaction networks: there will be no
single analytical or even computational method capable of
calculating the exact solution of the underlying master
equation in complete generality using finite resources.
As a consequence, the development of accurate and
computationally feasible techniques for studying the dynamic
behavior of large nonlinear Markovian reaction networks is
still the most important and challenging problem in this
field of research.

One way to deal with this problem is to focus on
specialized structures that may be present in certain
reaction networks and, by exploiting these structures,
develop rigorous approximation techniques tailored to
the specific application at hand. A good example of such
an approach is the IE method discussed in Section~IV-B.
This method calculates the exact solution of the
master equation, up to a desired precision, by exploiting
the structure of the master equation that governs the DA
process. Moreover, it uses the fact that, in some reaction
networks, the sample space associated with the DA process
is bounded whereas its cardinality is not appreciably
larger than the cardinality of the sample space
associated with the population process~\cite{JenkGout:12}.

Another possibility is to move away from estimating the joint
probability distributions of the DA and population processes
and focus on estimating computationally more tractable marginal
probability distributions or statistical summaries. Moment
closure schemes, in conjunction with MaxEnt methods, seem
to be particularly suited in this case. However, much
work is needed for developing appropriate closure schemes and
evaluating the errors introduced by the resulting approximations,
as well as for designing computationally efficient MaxEnt
methods. On the other hand, embedding Monte Carlo steps
within a computationally efficient method for solving the
master equation can prove very useful for guiding correct
implementation. For example, employing a second-order
moment closure scheme will not be appropriate if a small
number of Monte Carlo samples drawn from the master equation
reveals a bistable system behavior. The use of hybrid
techniques, capable of drawing on relative strengths to
mitigate weaknesses of the tools discussed in Sections~IV \&~V,
will likely pave the way to more robust solution methodologies,
especially for stiff reaction networks. Finally,
to deal with the large networks present in many applications,
it is necessary to focus on computational efficiency and, in
particular, on algorithms that can exploit the highly
parallel structure of modern high performance computing
platforms.

\vspace{-12pt}

\subsection{Thermodynamic analysis}

\vspace{-6pt}

Statistical thermodynamics can be effectively used to compactly
describe the macroscopic behavior of a given stochastic system
by a small number of variables. For this reason, it has been
successfully applied in many fields of science and engineering.
For example, a container full of gas molecules can be exhaustively
described by a set of high-dimensional Hamiltonian equations
governing the position and momentum of every single molecule.
However, a small number of statistical thermodynamic summaries,
such as pressure, temperature, entropy, Helmholtz free energy
etc., can be used to provide a more lucid and computationally
tractable description of the system. Compact descriptions
of system dynamics are also possible in the case of Markovian
reaction networks whose analysis, based on statistical
thermodynamics, may be the only amenable method of dealing
with large nonlinear networks. However, development of such
analysis methods are still at their infancy and wide open for
future exploration.

A promising line of inquiry seems to be the potential energy
landscape perspective discussed in Section~VII, which serves
as the starting point for developing the thermodynamic
analysis tools discussed in Section~VIII. Future efforts must
focus on designing accurate and efficient methods for estimating
the potential energy landscape of Markovian reaction networks
and detecting noise-induced modes. The ability to predict
noise-induced modes not present as fixed points in the
macroscopic equations is an important and challenging task.
Since these modes of operation are prominent when appreciable
stochastic fluctuations are present and, often, are perched
near bifurcation points predicted by the corresponding macroscopic
equations~\cite{SamoPlyaArki:05,TurcGarcSuel:08}, it is likely
that their study will lead investigators to focus on the
interface between stochastic processes and bifurcation theory
for nonlinear ODEs.

Aside from the possibility of representing a reaction network
using a small number of variables, conditions on the underlying
propensity functions imposed by the laws of thermodynamics have
shown to be extremely valuable for the modeling and analysis of
such networks. For example, the use of thermodynamic constraints
can alleviate some burden associated with parameter estimation
imposed by the curse of dimensionality and can lead to lower
computational complexity, better estimation performance, and
reduced data overfitting~\cite{ColqDownBeat-etal:04,LiebKlip:06,%
YangBrunHlav-etal:06,JenkZhonGout:10,JenkGout:11}. Moreover,
it can result in physically realizable network models consistent
with the fundamental laws of thermodynamics. The work done so
far on this important subject has focused on reaction networks
with deterministic dynamics. Therefore, extending this line of
research to Markovian reaction networks is an
exciting prospect with potentially fundamental consequences.

\vspace{-12pt}

\subsection{Sensitivity analysis}

\vspace{-6pt}

Often, the main focus of analysis of the dynamic behavior of
a reaction network is a response function that encapsulates
some important system characteristics. In epidemiology,
for example, one may not care so much about the specific details
of the population dynamics, but would rather focus on the
total number of individuals infected by a disease over a
given period of time. Another example would be the case of
cell signaling, where the detailed interactions of a signaling
pathway are not as important as the total amount of a protein
produced at the ``output'' of the pathway. Sensitivity
analysis is a quantitative approach designed to investigate how
variations in the parameters of a reaction network (e.g.,
in the specific probability rate constants associated with the
propensity functions of a mass action system) affect a response
function of interest~\cite{HeinSchu:96,VarmMorbWu:99,%
SaltRatttara-etal:05,SaltRattAndr-etal:08}.

Many physical and man-made reaction networks are designed to
be robust to random fluctuations (or even failures) in system
components. Although robustness is a highly desirable property,
it results in a small number of parameters having a
disproportionately large influence on the system response.
As a consequence, a robust reaction network can be quite
vulnerable to targeted attacks on influential components,
which can be a blessing or a curse, depending on the particular
situation at hand. For example, development of new drugs
may greatly benefit from this property since, to reduce or
even eliminate the effects of a disease caused by deregulation
of key system responses, it may be sufficient to design a
drug that only inhibits influential reactions that shape these
responses. On the other hand, targeted attacks on national
infrastructure by hackers or terrorists may produce large
scale disruptions with devastating results.

The objective of sensitivity analysis is to determine those
factors in a reaction network that produce no noticeable
variations in system response and identify those factors
that are most influential in shaping that response. Although
this is a powerful analysis technique with important
practical consequences, it comes with
a large computational cost, even in the case of reaction
networks with deterministic dynamics~\cite{ZhanDempGout:09,%
ZhanGout:10,ZhanGout:11}. For this reason, the development
of practical methods for sensitivity analysis of Markovian
reaction networks is still at their infancy~\cite{GunaCaoPetz-etal:05,%
KimDebuNajm:07,PlyaArki:07,DandKham:10,KimSaur:10,%
RathShepKham:10,KomoCostRand-etal:11,WarrAlle:12}.

In the stochastic context, sensitivity analysis involves
computing the solution of the master equation using different
parameter values. As a consequence, the development of efficient
solution methods that can be implemented on parallel computer
architectures, paired with novel sensitivity estimators, will
ensure the feasibility of this type of analysis. We should also
note that it has been recently demonstrated by~\textcite{ZhanDempGout:09}
and \textcite{ZhanGout:11} that, at least for the case of
physical reaction networks with deterministic dynamics,
sensitivity analysis methods must be in agreement with
underlying thermodynamic constraints. As a consequence,
developing accurate, computationally efficient, and
thermodynamically consistent sensitivity analysis methods
for Markovian reaction networks is an important research
activity with significant benefits.

\vspace{-12pt}

\subsection{Statistical inference}

\vspace{-6pt}

In general, there are two fundamentally different types of
parameters associated with a Markovian reaction network model:
the stoichiometric coefficients $\nu_{nm}$ and $\nu'_{nm}$
that determine the structure of the network, and the kinetic
parameters that determine the non-structural portion of the
propensity functions. Some parameter values can be deduced
experimentally or by means of appropriate theoretical and
sometimes heuristic arguments. Most parameters however
must be estimated from available data using statistical
inference techniques. Since the predictive power of a
given model is fundamentally constrained by the accuracy
of its parameterization, inferring the unknown parameter
values in a Markovian reaction network is a problem of
paramount interest and practical importance. Although
this problem has been extensively studied for reaction
networks with deterministic dynamics~\cite{MoleMendBang:03,%
CramSchnMcSh:04,Mari:04}, the statistical inference of Markovian
reaction networks is largely an open research problem. This
problem has been recently investigated by \textcite{GoliWilk:05,%
GoliWilk:06}, \textcite{ReinAltmTimm:06},
\textcite{BoysWilkKirk:08}, \textcite{KomoFinkHarp-etal:09},
\textcite{PoovGuna:10}, \textcite{WangChriMjol-etal:10},
and \textcite{DaigRohPetz-etal:12}, but the resulting
algorithms do not adequately address important
issues, such as curse of dimensionality, thermodynamic
consistency, and computational efficiency. These methods
have been primarily designed for biochemical reaction
networks, but can be easily adopted in other applications
with little or no effort.

In most approaches to statistical inference, it is quite
common to assume known structural parameters and proceed
with estimating the kinetic parameters using noisy and
sparse measurements of system dynamics. This problem,
known as \textit{model calibration}, is much easier than
the problem of estimating the structural parameters,
which is often referred to as \textit{model selection}.

The two most difficult issues associated with model calibration
is the curse of dimensionality and the use of non-convex cost
functions which complicate numerical optimization. Curse of
dimensionality refers to a very fast growth of the volume of
the parameter space in terms of the number of unknown parameters
to be estimated. As a consequence, the problem of finding the
``best'' parameter values becomes increasingly difficult as the
number of unknown parameters increases. This is further
exacerbated by the non-convex optimization problem of finding
these values, which is computationally very difficult to solve
in most cases of interest~\cite{Spal:03}. Therefore, the development
of statistical techniques for accurate and computationally
efficient model calibration of Markovian reaction networks
is a very challenging problem. Possible ways to attack this
problem is to effectively reduce the number of parameters
that must be estimated by incorporating appropriate constraints
[e.g., constraints imposed by the fundamental laws of
thermodynamics~\cite{ColqDownBeat-etal:04,LiebKlip:06,%
YangBrunHlav-etal:06,JenkZhonGout:10,JenkGout:11}] and to
identify a smaller set of ``influential'' parameters whose
values must be estimated with sufficient precision [e.g., by
employing a sensitivity analysis approach~\cite{JenkZhonGout:10}].
This reduction in dimensionality must be combined with fast
algorithms for solving the master equation, with efficient
optimization methods, and appropriately designed experimental
protocols for collecting data with high information content
about the values of the unknown parameters~\cite{JenkGout:10}.

In general, model selection is a more difficult problem. Solving
this problem will require the development of novel hypothesis
testing approaches for comparing between two competing network
models (e.g., an originally proposed signaling network and another
network obtained by adding new reactions) in a rigorous
statistical fashion. This approach however requires that both
models are calibrated before compared to each other
(e.g., by a likelihood ratio test), which substantially adds
to the difficulty of the problem. Another major issue is that
more complex models are expected to be more capable of closely
matching experimental data, but these models may result in
undesirable overfitting. It is therefore necessary to develop
methods that appropriately penalize model complexity so that
the chosen ``optimal'' model is the most parsimonious model
capable of adequately explaining available data. Finally,
all of this must be done while taking into account possible
constraints imposed on the structural and kinetic parameters
of the network (e.g., by prior knowledge on feasible structural
parameter values and by the fundamental laws of thermodynamics).

\vspace{-12pt}

\subsection{Adaptive Markovian reaction networks}

\vspace{-6pt}

An important aspect of many real-life interaction networks is
that the network topology, quantified by the stoichiometric
coefficients $\nu_{nm}$ and $\nu'_{nm}$, is intricately coupled
with the network dynamics. As we mentioned in Section~II-B-3,
for all reaction networks encountered in practice, the propensity
functions depend on the stoichiometric coefficients $\nu_{nm}$ and,
therefore, the topological structure of a reaction network directly
affects its dynamics. However, the opposite is also true: the
dynamics can influence the underlying network topology. In
opinion formation, for example, individuals form their beliefs
based on interactions determined by the underlined social
network, whereas, individuals with similar beliefs tend to
eventually influence each other. In these cases, modeling
dynamics on complex networks by assuming fixed topology
is legitimate only when the time scale of interest is
sufficiently smaller than the time scale of change in
network topology.

Recently, several articles have appeared in the
literature introducing \textit{adaptive} networks that take
into account the interplay between network topology and
dynamics~\cite{EhrhMarsRedo:06,GrosLimaBlas:06,HolmNewm:06,%
ZhouKurt:06,GrosBlas:08,RenWangLi-etal:10,ZhanZhouXu-etal:10,%
YuanZhou:11}. These preliminary works clearly demonstrate that
a number of intriguing properties emerge, not previously
observed in nonadaptive networks: formation of complex topologies,
spontaneous emergence of modular organization, more complex
dynamics than the ones observed in nonadaptive models, and
self-organization towards a highly robust critical behavior
characterized by power-law distributions.

Understanding the coupling between Markovian dynamics and
network topology is a fascinating subject of research that
can eventually lead to surprising new theoretical results
and novel computational methods for the modeling and analysis
of networks with significant impact in many fields. For
example, adaptive Markovian reaction networks may potentially
lead to a better understanding of how cell biology at the
molecular level has evolved towards certain types of reaction
network architectures and motifs, characterized by a high
degree of modularity and robustness. These type of networks
can be represented by master equations with time-dependent
stoichiometric coefficients, whose values are updated by the
system state following well-defined rules. We foresee a rapid
growth in developing such models that can lead to many
fascinating and novel results to come.

%
%

\vspace{-12pt}

\section{Conclusion}

\vspace{-6pt}

The recent burst in the amount of research efforts for studying
random processes on complex interaction networks has been
driven primarily by impressive advances in experimental
techniques for measuring these processes and by a clear
understanding that stochasticity plays a fundamental role
in shaping the transient and steady-state dynamic behavior
of real-life networks. This effort has reinforced the fact
that the theory of Markovian reaction networks is at the
foundation of most modeling and analysis techniques for
studying stochastic dynamical systems on networks with
applications in diverse fields, such as chemistry, biology,
sociology, epidemiology, pharmacology, theoretical neuroscience,
and engineering. The main goal of this review was to summarize,
in a systematic fashion, what is known so far about this
exciting field and briefly discuss some open problems that
need to be solved and new methodologies that must be developed.
We believe that a major effort should be focused on introducing
new concepts and ideas that might lead to novel and practical
tools for the modeling and analysis of stochastic dynamics
on interaction networks encountered in practice. We hope
that this article will be used as a reference by network
scientists across different scientific disciplines and help
to catalyze exciting new developments in these fields.

%
%

\vspace{-12pt}

\section*{Acknowledgments}

\vspace{-6pt}

The authors are grateful to Bence M{\' e}lyk{\' u}ti for reading
the manuscript and for providing helpful suggestions.
This research was supported in part by the DoD High Performance
Computing Modernization Program through the National Defense Science
and Engineering Graduate (NDSEG) Fellowship 32 CFR 168a, and in
part by the National Science Foundation (NSF) Grant CCF-0849907.
The funders had no role in study design, data collection and analysis,
decision to publish, or preparation of the manuscript.

%
%

\vspace{-12pt}

\bibliographystyle{apsrmp}
\bibliography{article}

\end{document}